\documentclass[twocolumn]{aastex701}

\usepackage{array}
\usepackage{graphicx}
\usepackage{amsmath}
\usepackage{amssymb}
\usepackage{txfonts}
\usepackage{makecell}
\usepackage{enumitem}
\usepackage{comment}
\usepackage{color}
\usepackage{colortbl} 
\definecolor{pinegreen}{RGB}{1, 121, 111}
\definecolor{salmon}{RGB}{255,160,122}
\usepackage{natbib}
\usepackage{rotating}
\usepackage{adjustbox}
\usepackage{caption}
\usepackage{supertabular}
\usepackage{multirow}
\usepackage{wrapfig}
\usepackage{tabularx}
\usepackage{hyperref}
\usepackage{totcount}

\newtotcounter{citnum}
\def\oldbibitem{} \let\oldbibitem=\bibitem
\def\bibitem{\stepcounter{citnum}\oldbibitem}

\definecolor{YC}{RGB}{115,80,185}

\newcommand{\Msun}{\ensuremath{\,M_\odot}}
\newcommand{\Rsun}{\ensuremath{\,R_\odot}}

\newcommand{\kms}{\ensuremath{\,\rm{km}\,\rm{s}^{-1}}}

\newcommand{\HeII}{He~\textsc{ii}}
\newcommand{\HeI}{He~\textsc{i}}

\newcommand{\dRVm}{\ensuremath{\Delta\mathrm{RV}_{\rm max}}}

\usepackage{refs}  
\usepackage{booktabs}   
\usepackage{morefloats}

\bibpunct{(}{)}{;}{a}{}{,}

\defcitealias{2023Sci...382.1287D}{DGL+23}
\defcitealias{2023ApJ...959..125G}{GD+23}

\begin{document}

\title{Revealing Optically Hidden Companions to Intermediate-Mass Stripped Stars with Radial Velocities}
\shortauthors{Laroche et al.}
\shorttitle{Orbital Properties of Stripped Stars}

\author[0000-0002-5522-0217]{Alexander Laroche}
\affiliation{David A. Dunlap Department of Astronomy \& Astrophysics, University of Toronto, 50 St George St, Toronto, ON M5S 3H4, Canada}
\affiliation{Dunlap Institute for Astronomy \& Astrophysics, University of Toronto, 50 St George Street, Toronto, ON M5S 3H4, Canada}
\email[show]{alex.laroche@mail.utoronto.ca}

\author[0000-0001-7081-0082]{Maria R. Drout}
\affiliation{David A. Dunlap Department of Astronomy \& Astrophysics, University of Toronto, 50 St George St, Toronto, ON M5S 3H4, Canada}
\affiliation{Dunlap Institute for Astronomy \& Astrophysics, University of Toronto, 50 St George Street, Toronto, ON M5S 3H4, Canada}
\email{maria.drout@utoronto.ca}

\author[0000-0002-6960-6911]{Ylva G\"{o}tberg}
\affiliation{Institute of Science and Technology Austria (ISTA), Am Campus 1, 3400 Klosterneuburg, Austria}
\email{ylva.gotberg@ist.ac.at}

\author[0000-0003-0857-2989]{Bethany Ludwig}
\affiliation{Institute of Astronomy, KU Leuven, Celestijnenlaan 200D, 3001, Leuven, Belgium}
\email{bethany.ludwig@kuleuven.be}

\author[0000-0003-0821-4583]{Dandan Wei}
\affiliation{Institute of Science and Technology Austria (ISTA), Am Campus 1, 3400 Klosterneuburg, Austria}
\email{dandan.wei@ist.ac.at}

\author[0000-0002-9132-6561]{Peter Senchyna}
\affiliation{The Observatories of the Carnegie Institution for Science, 813 Santa Barbara Street, Pasadena, CA 91101, USA}
\email{psenchyna@carnegiescience.edu}

\author[0000-0002-6718-9472]{Mathieu Renzo}
\affiliation{Steward Observatory and Department of Astronomy, University of Arizona, 933 N. Cherry Avenue, Tucson, AZ 85721, USA}
\email{mrenzo@arizona.edu}

\author[0009-0000-1383-8011]{Debasish Dutta}
\affiliation{Institute of Science and Technology Austria (ISTA), Am Campus 1, 3400 Klosterneuburg, Austria}
\email{debasish.dutta@ist.ac.at}

\author[0000-0002-7296-6547]{Anna J. G. O'Grady}
\altaffiliation{NASA Hubble Fellow} 
\affiliation{McWilliams Center for Cosmology and Astrophysics, Department of Physics, Carnegie Mellon University, Pittsburgh, PA, USA}
\affiliation{Department of Physics \& Astronomy, University of Pittsburgh, PA, USA}
\email{aogrady@andrew.cmu.edu}

\author[0009-0002-7518-8944]{Max Pritzkuleit}
\affiliation{Institut für Physik und Astronomie, Universität Potsdam, Karl-Liebknecht-Str. 24/25, D-14476 Potsdam, Germany}
\email{maxpri@astro.physik.uni-potsdam.de}

\begin{abstract}
The recent discovery of a population of intermediate-mass helium stars ($\sim$2--8 $M_\odot$) in the Magellanic Clouds has opened a new observational window into massive binary evolution. These hot, helium-rich stars are thought to be products of binary interaction, and some show no evidence for a luminous companion in their optical spectra. Here, we present multi-epoch radial velocity (RV) measurements for eight of these systems obtained with Magellan/MagE from 2018-2023. Five helium stars exhibit significant RV variability consistent with binary motion, while three do not. The five helium stars with evidence of binarity have orbital periods from 16 hours to 300 days, and eccentricities consistent with zero, implying companion mass lower limits from $\sim$0.6$–$2.9 $M_\odot$. We discuss implications of their orbits for the nature of the unseen companions, and find that low-mass late-pre/early-main sequence stars \mbox{($\lesssim$1.5$–$3.1 $M_\odot$)}, low-mass stripped stars and compact objects are currently viable. We identify an unexpected correlation between surface hydrogen abundance and binarity within this sample: helium stars in binaries are hydrogen-poor \mbox{($X_{\rm H,surf}$ $\sim$ 0.3--0.4)}, whereas helium stars without detected binary motion are nearly hydrogen-free \mbox{($X_{\rm H,surf}$ $\lesssim$ 0.05)}. This bimodality suggests that low RV variability in a subset of the sample is not simply due to inclination, but rather a distinct evolutionary pathway. We consider three possible explanations for these objects: wide binaries, disrupted binaries, and stellar mergers. The binary properties of this first sample of intermediate-mass helium stars is diverse, and provides important empirical constraints for binary evolution and compact object formation.
\end{abstract}
\keywords{
\uat{Binary stars}{154};
\uat{Close binary stars}{254};
\uat{Compact binary stars}{283};
\uat{Compact objects}{283};
\uat{Interacting binary stars}{801};
\uat{Early-type stars}{430};
\uat{Helium-rich stars}{715};
\uat{Helium burning}{716}; 
\uat{Stellar properties}{1624};
\uat{Stellar spectral types}{2051};
\uat{Stellar spectral lines}{1630}
}

\section{Introduction}\label{sec:intro}
\setcounter{footnote}{0} 
Our understanding of massive stars
is inextricably tied to binary interaction \citep[e.g.][]{2012ARA&A..50..107L,2019arXiv190305094B}. In particular, 50--70\% of massive stars are thought to interact at some point in their lives \citep{2012Sci...337..444S,2013A&A...550A.107S,2023ASPC..534..275O,2025NatAs...9.1337S}. One of the diverse outcomes long-predicted by binary evolution theory are intermediate-mass binary-stripped helium stars ($\sim$2$-$8 $M_\odot$). They should originate from stars with $\sim$8$-$25~$M_\odot$ initial masses, in the mass gap between low-mass hot subdwarfs and Wolf-Rayet (WR) stars \citep{1967ZA.....65..251K, 1967AcA....17..355P,
2011ApJ...730...76I}.

Intermediate-mass helium stars are thought to be common \citep{2019A&A...629A.134G,
2021ApJ...908...67S, 2024A&A...683A..37Y, 2025A&A...697A.239H}, long-lived \citep{2002ApJ...573..283P, 2008AIPC..990..230D}, and produce a wide range of astrophysical phenomena. For example, observed populations of wind-stripped WRs are not numerous enough to match the high rate of hydrogen-poor core-collapse supernovae (SN). Intermediate-mass helium stars could therefore be the main progenitors of Type Ibc/IIb SN \citep{2011MNRAS.412.1522S, 2017ApJ...840...10Y,
2019ApJ...885..130S}, which is also supported by the low ejecta masses inferred from modeling of stripped-envelope SN light curves \citep{2011ApJ...741...97D,
2016MNRAS.457..328L}. Envelope-stripping should also lead to a variety of compact object mergers \citep{2007PhR...442...75K}. The evolution to binary neutron stars (NS) should involve two envelope-stripping phases \citep{2017ApJ...846..170T, 2020ApJ...888L..10Y,
2020PASA...37...38V}. In addition to being gravitational wave (GW) progenitors, stripped star plus compact object binaries are (LISA detectable) living GW sources \citep{ 2004MNRAS.349..181N,
2018A&A...618A..14W,
2020A&A...634A.126W,
2020ApJ...904...56G,
2020ApJ...891...45K,
2022A&A...668A..80L}. Finally, stripped stars are UV bright and particularly hot sources of ionizing radiation, potentially playing a key role both in cosmic reionization and in producing the sometimes highly ionized nebular spectra observed in star-forming galaxies across redshift \citep{2016MNRAS.456..485S,
2017A&A...608A..11G,
2019A&A...629A.134G,
2020A&A...634A.134G,
2019MNRAS.488.3492S,
2023MNRAS.522.4430S,
2024MNRAS.527.9480L}. 

Despite their wide ranging implications, a population of stars consistent with expectations for intermediate-mass helium stars was only recently discovered.
\citet[][hereafter DGL+23]{2023Sci...382.1287D} identified a sample of sources in the Magellanic Clouds with optical brightness similar to B-type main sequence (MS) stars, but with excess UV light in their spectral energy distributions \citep[see][for the full photometric sample]{2026ApJ...999...73L}. They then obtained spectra for a sample of 25 objects, and classified them into three broad categories based on their spectral morphologies: Class 1 “Helium star-type” sources are dominated by \HeII~absorption, indicating a hot ($T_{\rm eff}$$\sim$60--100 kK) and hydrogen-poor ($X_{\rm H, surf}$$\sim$0$-$0.4) star, Class 2 “Composite-type” sources show both \HeII~and stronger short-wavelength Balmer absorption, indicating the presence of both a hot star and B-type main sequence (MS) star, and Class 3 “B-type” sources which spectroscopically appear as B-type MS stars, but still exhibit apparent UV excess. These three classes are consistent with expectations for stars stripped in binaries, where the stripped star contributes varying fractions of optical flux relative to its companion. Simultaneously, \citet[][hereafter GD23]{2023ApJ...959..125G} performed spectral fitting of the “Helium star-type” objects. They confirmed that these are hot, compact, and hydrogen-poor stars, with temperatures and luminosities consistent with predictions for $\sim$2--8 M$_\odot$ binary stripped helium stars.

Overall, the sample of stars presented in \citetalias{2023Sci...382.1287D} are distinct from helium stars in the literature to date, probing a unique regime: hot, compact, intermediate-mass helium stars. They are higher mass than observed hot subdwarfs \citep[e.g.][]{2015A&A...576A..44K,
2017ApJ...843...60W,
2022ApJ...926..213K,
2022A&A...666A.182S,
2026A&A...708A.187M}, lower mass than WR stars \citep[e.g.][]{2017MNRAS.464.2066S,
2019A&A...627A.151S}, and more hot and compact than puffed-up stripped stars \citep[e.g.][]{2020A&A...641A..43B, 2021MNRAS.502.3436E, 
2024A&A...692A..90R}.

Helium star masses, as estimated by \citetalias{2023ApJ...959..125G}, are an important step towards characterizing the sample of \citetalias{2023Sci...382.1287D}. However, spectral fitting did not show evidence of binary companions. A complete understanding of these systems requires binary orbits. This is necessary to establish that the helium stars experienced binary stripping \citep[cf.][]{2014ApJ...782....7D}. For example, one of few Galactic, hot intermediate-mass helium stars that currently has a measured orbit is HD 45166. While \cite{2005A&A...444..895S} originally estimated a 1.6 day period, \cite{2023Sci...381..761S} found that it is instead in a 22.5 year orbit. With this new-found long period, it is unlikely that HD 45166 experienced binary envelope stripping; a merger of two lower-mass helium stars was instead proposed by \cite{2023Sci...381..761S}. Thus, orbital properties for the recently identified sample of intermediate-mass helium stars are essential, since they can support or reject binary envelope-stripping.

Orbits can also provide direct constraints on companion masses \citep[e.g.][]{2014Natur.505..378C}. Theoretical expectations for stripped star companions are largely unverified observationally, especially evolutionary pathways which produce compact object binaries \citep[e.g.][]{2017ApJ...840...10Y,2018MNRAS.481.1908K}. 
A sample of orbital properties for stripped stars will directly confront binary theory, and provide critical benchmarks for binary population synthesis \citep[e.g.][]{2015ApJ...805...20S,2018A&A...615A..78G}.

Here we begin this process by presenting a radial velocity (RV) survey for 8 sources from \citetalias{2023Sci...382.1287D}. In this first paper, we consider a majority of the sample with no evidence for a luminous companion in their optical spectra\footnote{As described further in Section~\ref{ssec:sample}, one object from this class, Star 3, will be presented in a companion paper \citep{LudwigStar3}.}, whose helium star surface properties were characterized in \citetalias{2023ApJ...959..125G}. 
This paper is organized as follows: In \secref{sec:obs}, we describe our sample, observations, and data reduction. In \secref{sec:rv}, we present our RV methodology and measurements. In \secref{sec:binary_motion}, we analyze RVs to detect binary motion. In \secref{sec:orbits}, we derive orbital solutions and measure the binary properties of helium stars. In \secref{ssec:context}, we compare our sample of binary properties to other helium stars in the literature. In \secref{sec:stellar-comp}, we consider whether the helium stars orbit stellar companions. We then consider compact object companions in \secref{sec:co-comp}. In \secref{sec:singles}, we discuss the nature of helium stars with non-detections of binary motion, and whether they could be (currently) single stars. Finally, we conclude in \secref{sec:conc}. Our data, results and code will be publicly available on Zenodo.\footnote{Upon acceptance of this manuscript.}

\begin{deluxetable*}{ccccccc}[ht!]
    \caption{Summary of the Intermediate-Mass Helium Stars in Our Radial Velocity Survey.}\label{tab:rv_summary}
    \tablehead{
    \colhead{Name} & \colhead{Location} &\colhead{$N_{\rm RV}/N_{\rm obs}$} & \colhead{\dRVm} &\colhead{$\sigma_{\rm detect}$} &\colhead{Type} &\colhead{Spectral lines} \\ 
    \colhead{} & \colhead{} &\colhead{} & \colhead{[km s$^{-1}$]} &\colhead{} &\colhead{} &\colhead{[\AA]}
    }
    \startdata
    Star 1 & SMC & 22/24 & $42\pm5$ & 9 & Binary & He~\textsc{ii}: 4200, 4542, 4686, 5412, He~\textsc{ii}-H: 4100, 4339, 4860 \\
    Star 2 & SMC & 13/14 & $127\pm4$ & 29 & Binary & He~\textsc{ii}: 4200, 4542, 4686, 5412, He~\textsc{ii}-H: 4100, 4339, 4860 \\
    Star 4 & SMC & 15/16 & $109\pm10$ & 11 & Binary & He~\textsc{ii}: 4200, 4542, 4686, 5412 \\
    \midrule
    Star 5 & LMC & 15/17 & $22\pm7$ & 3 & Non-detection & He~\textsc{ii}: 4200, 4542, 4686, 5412, He~\textsc{ii}-H: 4100, 4339, 4860 \\
    Star 6 & LMC & 16/18 & $61\pm6$ & 10 & Binary & He~\textsc{ii}: 4200, 4542, 4686, 5412, He~\textsc{ii}-H: 4100, 4339 \\
    Star 7 & LMC & 13/14 & $23\pm9$ & 2 & Non-detection & He~\textsc{ii}: 4200, 4542, 4686, 5412, He~\textsc{ii}-H: 4339, 4860 \\
    Star 8 & LMC & 5/5 & $24\pm18$ & 1 & Non-detection & He~\textsc{ii}: 4200, 4542, 5412, He~\textsc{ii}-H: 4100, 4339, 4860 \\
    Star 16 & LMC & 10/11 & $125\pm3$ & 42 & Binary & He~\textsc{i}: 4388, 4713, 5016 \\
    \enddata
    \tablecomments{Name convention from \citetalias{2023Sci...382.1287D,2023ApJ...959..125G}. Location specifies membership in SMC/LMC. $N_{\rm RV}$ ($N_{\rm obs}$) specifies number of high-quality RV measurements (observations). \dRVm~is the measured peak-to-peak RV shift (maximum of \eqref{eq:drv}). $\sigma_{\rm detect}$ is the significance of \dRVm~(\eqref{eq:sig_detect}). Type specifies whether binary motion (see \secref{sec:binary_motion}) is detected (binary) or not (non-detection). Finally, we list the spectral lines used for RV measurements, which includes \HeII, \HeII-H~and \HeI~absorption lines (see \secref{ssec:rv_lines} and \appref{sapp:rv_lines}).} 
\end{deluxetable*}

\section{Sample, Observations and Data Reduction}\label{sec:obs}

\subsection{Sample}\label{ssec:sample}

In this work, we consider 8 stars from \citetalias{2023Sci...382.1287D} which were characterized in \citetalias{2023ApJ...959..125G}: Stars 1, 2, 4, 5, 6, 7, 8 and 16. We select these stars because none show evidence for a luminous companion in their UV-optical photometry or optical spectra. We begin with this sample because they offer the cleanest view of the hot star in each system. This allowed the properties of the hot star to be well measured \citepalias[][see below]{2023ApJ...959..125G} and also allows us to measure robust RVs of the hot star without the need for spectral disentangling. Since none of the stars in our sample showed signatures of a luminous companion in their optical spectra, they are consistent with: (i) having low-mass stellar companions, (ii) having compact object companions, or (iii) being currently single stars. The possibility of compact object companions highlights that some of these systems may be direct progenitors to double compact object binaries.

The targets in our sample are predominantly the “Helium star-type” systems from \citetalias{2023Sci...382.1287D}, which had spectra dominated by HeII and high-ionization metal lines. However, there are two notable exceptions.  First, we include Star 16. This star was initially designated as a “Composite-type” system in \citetalias{2023Sci...382.1287D} due to the presence of stronger short-wavelength Balmer lines in its spectrum. However, \citetalias{2023ApJ...959..125G} demonstrated that, similar to the “Helium star-type” systems, its optical spectrum can be well-fit by a single stripped star (without the need for a luminous companion), simply with a cooler temperature. Second, we do not include Star 3 in this manuscript. Star 3 was a “Helium star-type” system that showed evidence for very large RV variations in the preliminary data presented in \citetalias{2023Sci...382.1287D}. Detailed analysis of Star 3 will be presented in \cite{LudwigStar3}.

As noted above, \citetalias{2023ApJ...959..125G} measured stellar properties for all of the objects in our sample by performing spectro-photometric fitting to a grid of stripped star models. The effective temperatures of the stars in our sample are high ($T_{\rm eff}$$\sim$60$-$100 kK), they are compact  
($\log g$$\sim$4.5$-$5.5), and their surfaces are helium-rich (with mass fractions 
$X_{\rm He, surf}$$\sim$0.6$-$1) and hydrogen-poor ($X_{\rm H, surf}$$\sim$0$-$0.4). They have intermediate bolometric luminosities ($10^3$-$10^5\,L_\odot$; comparable to 5$-$30 $M_\odot$ MS stars; e.g. \citealt{2013A&A...553A..24G}), small radii ($R_{\rm eff}$$\sim$0.5$-$1.5$\Rsun$), and estimated current masses in the range of $\sim$2-8$\Msun$ (for further details on mass constraints for these objects see Section~\ref{sssec:M1}). All but one of the objects have luminosities, temperatures, and radii consistent with expectations for evolutionary models of core-helium burning stripped stars originating from stars with initial masses of $\sim$8$-$25 $M_\odot$. The one exception is Star 16, which is slightly cooler and inflated, and is thus likely contracting or expanding as opposed to core-helium burning \citepalias{2023ApJ...959..125G}.

\subsection{Observations}\label{ssec:obs}

We observed 119 individual epoch spectra of the 8 intermediate-mass helium stars in our sample between 2018 and 2023. Here, we describe these individual optical observations in more detail.

The data used in this work is part of an observational program with the Magellan Echellette (MagE) spectrograph \citep{2008SPIE.7014E..54M} mounted on the Magellan/Baade 6.5m telescope in a search for stars stripped of their hydrogen-rich envelopes via binary interaction (PI: G\"{o}tberg \& Drout, 2018B-2024B). We use the 0.85$\arcsec$ slit, which yields a roughly constant spectral resolution of $\mathcal{R}$$\sim$4100 for each MagE spectrum. For a given epoch we typically obtained 2--4 individual exposures sequentially with exposure times ranging from 10 to 20 minutes each, depending on the brightness of the source. These sequential exposures were stacked (see \secref{ssec:data-red} below) and the typical median signal-to-noise ratios (SNR) obtained for these individual epochs was $\sim$20-45. 

In total, we obtained 5$-$24 individual epochs per star over 2$-$5 years, with a few epochs in each year of observation. Since our sources are in the Magellanic Clouds, observations were predominantly taken in December and January (a smaller subset range from October to February). To accurately calibrate the wavelength solution for our observations, we use a Thorium-Argon (ThAr) lamp. Specifically, we took ThAr exposures throughout the night, either immediately before or after (or both) the sequence of science observations for each target in order to account for potential MagE instrument drift. The impact of when these ThAr exposures were taken on our RV measurements is discussed (along with other possible systematics) in Appendix~\ref{sapp:rv_sys_err}. All epoch spectra will be publicly available on Zenodo.

\subsection{Data Reduction}\label{ssec:data-red}

Reduction of the MagE data followed broadly the same procedure as outlined in the supplemental information of \citetalias{2023Sci...382.1287D}. In brief: we use the standard CarPy reduction tools for MagE \citep{2000ApJ...531..159K, 2003PASP..115..688K} which performs order identification, flat field and bias correction, wavelength calibration, spectral extraction, and stacking of sequential exposures. Typical root mean square (RMS) of the wavelength solution for our observations was $\sim$0.01-0.03 \AA, (corresponding to $\sim$1-2 km/s at the wavelengths of interest from $\sim$4000 to 5000 \AA). We then normalize each order by fitting a low-order polynomial with \texttt{pyraf.continuum} and correct each observation to the heliocentric reference frame with \texttt{pyraf.rvcor} \citep{2012ascl.soft07011S}. 

Our reduction procedure differs from \citetalias{2023Sci...382.1287D} in two ways. First, we do not stitch or average different MagE orders for a given observation. This was for two reasons. First, the precision of wavelength solutions can decrease near the edges of individual orders depending on the number and location of ThAr lines within the order. Second, the SNR of the observations also decreases near the edge of each order. Thus, to avoid possible data degradation being introduced into our analysis, we consider each order independently when measuring RVs. Second, we found that the initial flattening procedure applied by \citetalias{2023Sci...382.1287D} occasionally left features in the continuum. As such, we apply a second iteration of continuum normalization where we fit a cubic spline to all data with SNR$>$10 within a given order. If less than 100 pixels pass the SNR cut, we compute the error weighted mean and adopt a constant continuum correction. However, we note that the individual orders that contain all the spectral lines we use for the RV measurements here all have more than 100 data points that pass the SNR cut.

\section{Radial velocities}
\label{sec:rv}

Measuring binary star RVs is usually performed through cross-correlation of a ‘template’ spectrum with observed spectra \citep[e.g.][]{1979AJ.....84.1511T,
2023ApJ...947...77M,
2023AJ....165..203W}, or by fitting individual lines to measure central wavelengths \citep[e.g.][]{2013A&A...550A.107S,
2017A&A...598A..84A}. 

Previously, \citetalias{2023Sci...382.1287D} used cross-correlation for the stars in our sample to determine RVs for individual epochs when averaging together observations taken across multiple years. Specifically, they applied an iterative approach where multiple observations were shifted and added together and the combined spectrum was then used as a new template for additional rounds of RV refinement \citepalias[see Supplementary Materials of][]{2023Sci...382.1287D}. However, our individual epochs have moderate SNR (ranging from $\sim$10-40, see \tabref{sapp:tab_rv}), making cross correlation between two observed spectra less reliable as noise spikes can influence the result \cite[e.g.][]{2023ApJ...947...77M}. Here, we therefore opt for line fitting where we fit multiple lines simultaneously, while estimating uncertainties. As noted by \cite{2013A&A...550A.107S}, by requiring all lines within an epoch to have the same Doppler shift we increase the precision of our RV measurements. Our method has been optimized for our moderate SNR and resolution regime of SNR$\sim\mathcal{O}(10^1)$ and $\mathcal{R}\sim\mathcal{O}(10^3),$ but can also be applied to higher SNR/resolution. Our RV fitting code will be publicly available. 

\subsection{Spectral line selection}
\label{ssec:rv_lines}

The stars in our sample have few optical spectral lines, typical for hot stars \citep[e.g.][]{2009ssc..book.....G}. The optical spectra largely contain absorption lines, but also weaker metal emission lines. In particular, the spectra are dominated by lines of hydrogen and ionized helium, with weaker highly ionized nitrogen and carbon lines. There are no ‘isolated’ H lines in the spectra, as every Balmer line is blended with a Pickering ($n\to4$) \HeII{} line, which we refer to as ‘He II-H blends.’ However, every other Pickering line is isolated (\HeII\,4200, 4542, 5412 \AA). The $\alpha$-line of the \HeII{} Fowler series ($4\to3$) at 4686 \AA{} is also present in all spectra. 

To include a spectral line in our RV analysis for a given star, we require that it is: (i) present in all epochs, (ii) detected at sufficient SNR in individual epochs (S/N$\gtrsim$10), and (iii) dominated by the helium star (so that RVs are not biased by a companion). With respect to (i) and (ii), for the majority of stars in our sample we select from seven lines depending on their presence and strength: \HeII{} 4200, 4542, 4686, 5412, and \HeII-H 4100, 4339, 4860 (containing H$\delta,$ H$\gamma$ and H$\beta$, respectively). The one exception is Star 16 which, due to its lower temperature, shows strong \HeI~lines in its spectra (e.g. \HeI~4388, 4713, 5016 \AA). These \HeI~lines can yield more precise RV measurements since they are narrower than the Stark (pressure) broadened H and \HeII~lines. While several stars in our sample also show narrow N lines, they have low SNR in individual epochs, and are therefore not included for RV measurements.

With respect to (iii), that a given line not be significantly impacted by a companion, recall that none of the optical spectra for the stars in our sample show evidence for a luminous companion. \citetalias{2023ApJ...959..125G} demonstrated that a companion could contribute at most 10\% of the optical flux before the overall spectral fit would be significantly impacted. However, it is still a potential concern that a low-mass star could contribute just enough optical light to bias RV measurements.  In particular, while we do not expect a low-mass MS companion to exhibit \HeII~lines (they are simply not hot enough), they could---in principle---contribute to \HeII-H blends through Balmer absorption. 

We therefore run two tests to check whether the \HeII-H blends  are helium star dominated and therefore viable to use in our analysis. 
First, we measure RVs from \HeII~and \HeII-H separately, and find a directly proportional relationship. If a companion contributed significantly to \HeII-H blends, we would expect to see a deviation from direct proportionality. We infer that the \HeII-H blends are not significantly biased by a companion (\appref{sapp:rv_compare}). Second, for systems where we confirm binary motion and estimate orbits (\secreftwo{sec:binary_motion}{sec:orbits}), we compare the orbital posterior distributions resulting from different subsets of lines. In all cases, the orbital properties are robust (\appref{sapp:corner_plots}). Based on these results, we choose to include \HeII-H blends in our analysis for most stars since more (unbiased) lines increases RV accuracy.

In \tabref{tab:rv_summary}, we list the spectral lines we use to measure RVs for each star. Given that our line selections, while similar, are not entirely homogeneous across our sample, we present more detailed explanations of individual line selections for each star in \appref{sapp:rv_lines}.

\subsection{Spectral line templates}

Before measuring RVs, we first create line templates: model line profiles which represent the observed shape of a spectral line for a given star, which we will then fit to our data to measure RVs. Since the \HeII-H blends, \HeII~and \HeI~lines in our stars have typical widths of $\sim$ 15, 8 and 4 \AA, respectively, we adopt these windows for the templates (determined by standard deviation from preliminary template fitting). 

To create accurate spectral line templates, we fit Voigt or Gaussian profiles to high SNR stacked line profiles. Given that our observations were taken over multiple years, creating a high SNR stack requires preliminary RV estimates. We therefore fit Voigt profiles for \HeII{} and \HeII-H lines and Gaussian profiles for \HeI~lines to estimate the line centers for each line within an epoch (Voigt profiles account for Stark broadening). We then shift each observed line, individually, to the rest frame of the highest SNR epoch. After interpolating all epochs to the observed wavelength grid of the highest SNR epoch, we compute the error weighted mean of the shifted spectra to create one high-SNR stacked line profile. The associated pixel errors for the stacked profiles are the inverse sum of inverse variances \citep{2007nras.book.....P}. Finally, we fit the high-SNR stacked line profiles (as above) to obtain an accurate line template for each spectral line in Table \ref{tab:rv_summary}. Our spectral line templates are reasonable fits since they are consistent with the high-SNR stacked line profiles, within error.

\subsection{Radial velocity measurement}\label{ssec:rvs}

We determine RVs from our observed spectra by shifting all of the best-fit line templates simultaneously and minimizing $\chi^2$ to determine the best-fit RV. The wavelength ranges around each line center that we fit are  the line widths of the templates (as above).

To determine the measured RV and associated uncertainty for an epoch, we perform Monte Carlo (MC) resampling \citep[e.g.][]{2009PASP..121.1016M}. We draw 1000 random samples from the observed spectral line pixels from Gaussian distributions centered about the observed mean with a standard deviation equal to the observed error per pixel. The best-fit RV per sample is determined by jointly optimizing the RV over all line templates relative to the MC sampled epoch lines, through $\chi^2$ minimization. This yields an RV distribution composed of best-fit RVs for the random samples. After removing 3$\sigma$ outliers, we fit the distribution with a Gaussian to obtain a mean RV and $1\sigma$ uncertainty. In almost all cases, the resulting RV distributions are well-fit by a Gaussian (see \secref{ssec:cuts} for the small minority of cases with anomalous distributions).

Initially, these RVs are measured relative to the highest SNR spectrum of a given star (where our spectral line templates are centered; see \secref{ssec:rv_lines}). We then shift them to absolute rest according to the absolute RV shift of the highest SNR epoch, with rest wavelength line lists \citep[in air:][]{1977ApJ...214..759C,
1994ApJ...420..869U,
2018Galax...6...63V}. RV measurements are tabulated in \appref{app:rv}, and will be publicly available on Zenodo.

We find that our resulting statistical RV uncertainties range from $\sim$1--10 km/s with a mean of $\sim$5 km/s. In \appref{sapp:rv_sys_err}, we investigate several potential sources of systematic uncertainty that can result from our observations and data reduction process, such as the timing of wavelength calibration lamp observations and the spectral normalization process. We find that these systematic uncertainties are small (typically $\sim$0.5 km/s, maximally 4 km/s). Thus, in general, our uncertainties are dominated by statistical (random) errors, and any observed RV variability is unlikely to be driven by systematic effects.

\subsection{Quality cuts}\label{ssec:cuts}

Using the process described above, we obtain RVs for the number of observations listed in \tabref{tab:rv_summary} ($N_{\rm obs}$). In general, we find that our RV measurement technique performs well, but in a small minority of cases struggles due to low data quality, particularly for SNR$\lesssim$10. We employ quality cuts to reject RVs we consider imprecise. For each epoch, we evaluate (i) if the SNR is sufficiently high, and (ii) if the template fit is reasonable.

First, for a given star, we remove RVs with large uncertainties, corresponding to lower SNR epochs, via a statistical test. Following \cite{Tukey1949ComparingIM}, we compute the upper outlier threshold $Q_3$$+$$1.5IQR,$ where $IQR$$=$$Q_3$$-$$Q_1$ is the inter-quartile range and $Q_1$ ($Q_3$) is the 25th (75th) percentile of the RV errors. The upper outlier threshold is the boundary beyond which observed uncertainties are considered outliers relative to the RV error distribution (effectively, a statistical confidence limit). Since the upper outlier threshold is computed for an individual star, it varies across our sample. On the low (high) end, Star 16 (Star 4) has an upper outlier threshold of 4 (13) km/s. We discard RVs with uncertainties larger than the upper outlier threshold.

Second, in a small minority of cases we remove epochs with anomalous RV measurements due to noisy data, by visually inspecting fits and associated RV distributions. While these cuts are subjective, we emphasize that we only eliminate 0-2 anomalous epochs per star (5 out of 119 total epochs).

After applying these cuts, we are left with the $N_{\rm RV}$ measurements per source that we analyze in this work (see \tabref{tab:rv_summary}). In \appref{app:rv}, we provide a quality flag (0 $=$ accepted, 1 $=$ rejected due to error, 2 $=$ rejected due to anomalous) for all RVs, as well as a visualization of all RVs that both pass and fail our quality cuts in \figref{fig:rv_jd}.

\subsection{Updated stacked spectra}\label{ssec:stack}

In addition to epoch-by-epoch RVs, we also produce updated, stacked, normalized spectra for the helium stars in our sample, which will be publicly available on Zenodo. In comparison to the stacked spectra for these stars that were published in both \citetalias{2023Sci...382.1287D} and \citetalias{2023ApJ...959..125G}, our stacked spectra (i) are created from more observations and (ii) use the RVs from our updated methodology. However, the overall spectral morphology is minimally changed. We now describe this process in more detail.

We only include epochs which pass our RV quality cuts. First, we combine the flattened orders together into a single spectrum. We interpolate the orders to a common wavelength grid which matches the natural MagE wavelength spacing ($\sim$0.3--0.6 \AA) and perform error-weighted averaging in regions where orders overlap. The limits of the wavelength grid are set by the global maximum and minimum wavelength across all epochs. Second, we shift each epoch by the absolute RVs determined in \secref{ssec:rvs}. Third, we compute the error weighted mean at each pixel over all epochs, while the final uncertainty per pixel is defined as the inverse sum of the inverse variances, after 3$\sigma$ outlier rejection.

The resulting SNR of our stacked spectra are higher than those of \citetalias{2023Sci...382.1287D} and \citetalias{2023ApJ...959..125G} because we now have more observations. We obtain continuous wavelength coverage from 3800-8000 \AA{} for all spectra. Gaps beyond this wavelength range are due to our original process of only extracting portions of MagE spectra above a certain SNR. We note that beyond 7000 \AA~the SNR is typically low, no telluric correction has been performed, and less care was taken when normalizing. Caution is therefore warranted when using data beyond 7000 \AA. For additional details on our stacked optical spectra, see \appref{sapp:rv_stack}, where we also visualize the spectra.

\begin{figure}[t!]
    \centering
    \includegraphics[width=\columnwidth]{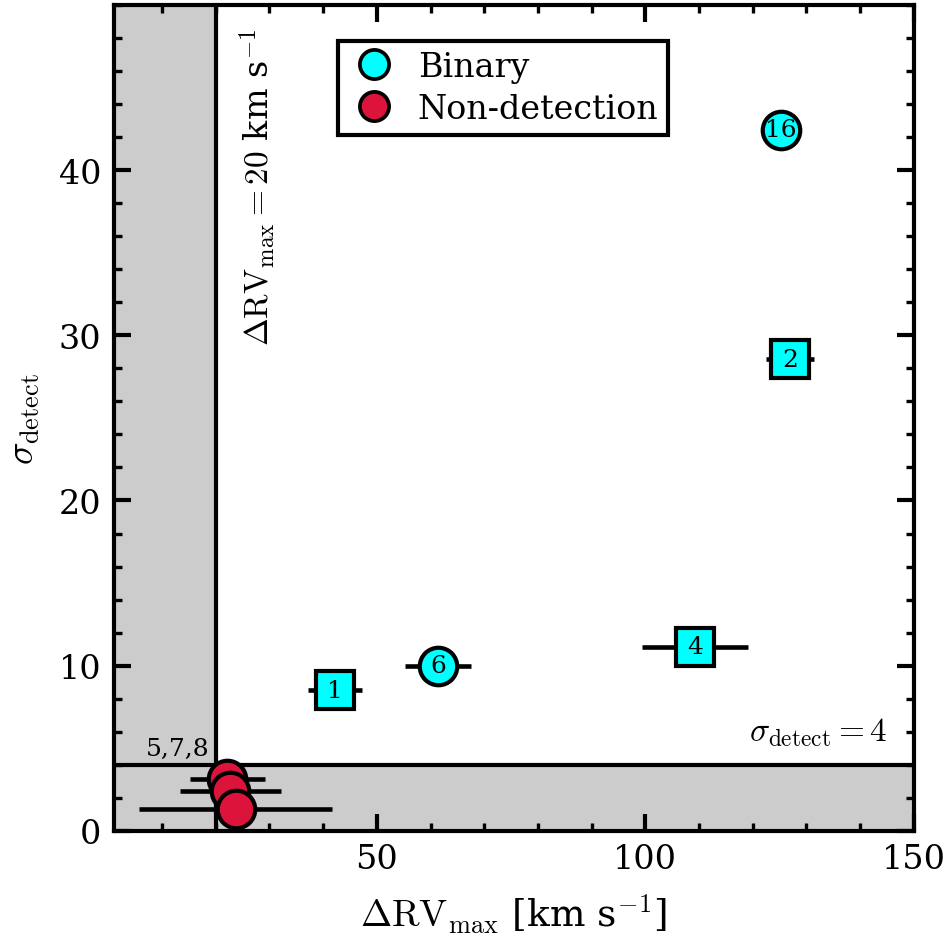}
    \caption{Intermediate-mass helium stars classified according to our binary criteria.
    5 out of 8 sources exhibit binary motion (cyan), whereas 3 sources do not (red). The grey region highlights the zone where binary motion is not detected. Squares (circles) are SMC (LMC) sources.}
    \label{fig:binary_detect}
\end{figure}

\section{Evidence for binary motion}
\label{sec:binary_motion}

Before obtaining orbital solutions, RV measurements of a source can determine whether there is significant evidence for binary motion. This is often possible with a smaller number of epochs than required to obtain an orbit (three is often sufficient, see references in \appref{app:binary-motion}), particularly if the observed RV shifts are large (see below). The stellar properties of the helium stars in our sample are well matched by binary evolution models for binary-stripped stars \citepalias[see \secref{ssec:sample} and][]{2023ApJ...959..125G}, indicating that stars of this type could be expected to exhibit binarity. Indeed, \citetalias{2023Sci...382.1287D} found RV variations for all sources in their sample with more than one epoch of spectroscopy (including all stars in this work).  However, in practice, even single stars can show variations in their measured RVs due to a number of astrophysical effects \citep[e.g. pulsations, winds, magnetism:][]{2001A&A...368..601D,
2019A&A...622A..67B,
2023Ap&SS.368..107B}.  
Thus, it is common to argue for evidence of binary motion using summary statistics. In particular, a combination of the amplitude and significance of RV variability is often used \citep[e.g.][]{2009A&A...507.1585R,
2013A&A...550A.107S,
2023MNRAS.521.4473C}. Here we calculate these quantities across our sample and determine if the resulting RV variability is consistent with binary motion. 

The pairwise RV shift for a given star is:
\begin{equation} \label{eq:drv}
    \Delta\text{RV}(i,j) = |v_i -  v_j|,
\end{equation}
where $v_i$ and $v_j$ are the $i$-th and $j$-th measured RVs. The pairwise shift significance for a given star is:
\begin{equation}\label{eq:sig_detect}
    \sigma_{\rm detect}(i,j) 
    = \frac{\Delta\text{RV}(i,j)}{\sqrt{\sigma_i^2+\sigma_j^2}},
\end{equation}
where $\sigma_i$ and $\sigma_j$ are $i$-th and $j$-th observed RV errors.

Typically, thresholds on $\Delta\text{RV}$ and $\sigma_{\rm detect}$ that support binary motion are based on empirical or theoretical estimates for the amount of RV variability that could be caused by other sources for a specific class of star. Given the lack of observational studies for intermediate-mass helium stars, we choose to adopt the following thresholds to claim significant evidence for binary motion in a given system:
\begin{equation} \label{eq:binary_bool}
\Delta\text{RV}>20\,\text{km}\,\text{s}^{-1}\quad
\text{and}\quad\sigma_{\rm detect} > 4.
\end{equation}
These thresholds mirror those recently adopted by the BLOeM collaboration for OB stars in the SMC \citep[e.g.][]{2025arXiv250202641B}. For further discussion of previous thresholds adopted in the literature, see \appref{app:binary-motion}.

In \figref{fig:binary_detect} and in \tabref{tab:rv_summary}, we present the binary detection results for our stripped star sample. Specifically, we show the maximum peak-to-peak RV shift, $\Delta\rm RV_{\max}$, and its associated $\sigma_{\rm detect}.$ Grey shading and black lines denote regions that fall below the thresholds required for us to claim a detection of binary motion (\eqref{eq:binary_bool}). Based on these thresholds, we detect binary motion in 5 out of 8 of our stars (Stars 1, 2, 4, 6, 16: cyan)\footnote{We note that the fraction of 5/8 would increase to 6/9 if Star 3, which will be published in \cite{LudwigStar3}, is included.}. In contrast, we do not claim a significant detection of binary motion in 3 out of 8 stars (Stars 5, 7, 8: red). In particular, we note that while Stars 5, 7, and 8 exhibit \dRVm$\sim$$20\,\rm km\,s^{-1},$ they all have $\sigma_{\rm detect}<4.$ Thus, in practice, our lower limit on \dRVm{} does not impact our results. It is however possible that these systems could still be in binaries with orbital properties that our thresholds are not sensitive to (see \secref{ssec:single_wide}). 

Thus, we find that most of the intermediate mass helium stars presented in \citetalias{2023Sci...382.1287D} which do not show evidence of luminous companions  \emph{are} in binaries. As we will show in \secref{sec:orbits}, all stars with detections of binary motion are single-lined spectroscopic binaries (SB1s) in well-characterized orbits. Conversely, we hereafter refer to systems for which we do not claim significant evidence of binary motion as ‘non-detections’ (see \tabref{tab:rv_summary}). 

In Sections~\ref{sec:orbits}-\ref{sec:co-comp}, we examine the stars which exhibit RV variability from binary motion. Eventually, in \secref{sec:singles}, we will examine the non-detections. 

\begin{deluxetable*}{lccccccccc}[t!]
    \caption{Summary of Binary Properties for Our Intermediate-Mass Helium Star Sample.}\label{tab:orbit_props}
    \tablehead{
    \colhead{} & \multicolumn{5}{c}{Orbital Properties} &
    \multicolumn{1}{c}{Helium Star} &
    \multicolumn{3}{c}{Optically Faint Companion} \\
    \colhead{Name} & \colhead{$P_{\rm orb}$} &\colhead{$e$} & \colhead{$K_1$} &\colhead{$v_{\rm sys}$} &\colhead{$f$} &\colhead{$M_1$} 
    &\colhead{$M_{2,\rm min}$} &\colhead{$\left<M_2\right>_{\rm inc}$} 
    &\colhead{$M_{2,\rm max}^\star$}  \\
    \colhead{} & \colhead{} &\colhead{} & 
    \colhead{[km s$^{-1}$]} &\colhead{[km s$^{-1}$]} &\colhead{[$M_\odot$]} &\colhead{[$M_\odot$]} &\colhead{[$M_\odot$]} &\colhead{[$M_\odot$]} &\colhead{[$M_\odot$]} 
    }
    \startdata
    Star 1 & $301^{+2}_{-2}$ d & $<0.21$ & $18^{+1}_{-1}$ & $162^{+6}_{-6}$ & $0.18^{+0.03}_{-0.03}$ & $8.45^{+1.04}_{-0.58}$ & $2.86^{+0.18}_{-0.17}\,\left(^{+0.25}_{-0.17}\right)$ & $3.1^{+1.2}_{-0.6}\,\left(^{+1.2}_{-0.6}\right)$ & $3.1^{+0.4}_{-0.3}$ \\
    Star 2 & $16.1702^{+0.0002}_{-0.0002}$ h & $<0.13$ & $65^{+2}_{-2}$ & $161^{+3}_{-3}$ & $0.019^{+0.002}_{-0.002}$ & $3.31^{+1.82}_{-0.18}$ & $0.67^{+0.03}_{-0.03}\,\left(^{+0.17}_{-0.03}\right)$ & $0.7^{+0.3}_{-0.1}\,\left(^{+0.5}_{-0.1}\right)$ & $1.7^{+0.3}_{-0.2}$ \\
    Star 4 & $18.6394^{+0.0003}_{-0.0003}$ h & $<0.20$ & $63^{+3}_{-3}$ & $184^{+7}_{-7}$ & $0.020^{+0.003}_{-0.002}$ & $3.04^{+4.43}_{-0.59}$ & $0.65^{+0.03}_{-0.03}\,\left(^{+0.40}_{-0.03}\right)$ & $0.7^{+0.3}_{-0.1}\,\left(^{+0.6}_{-0.1}\right)$ & $1.5^{+0.1}_{-0.1}$ \\
    \midrule
    Star 5 &  &  &  & $285^{+4}_{-4}$ &  & $4.06^{+1.45}_{-0.54}$ &  &  &  \\
    Star 6 & $8.600^{+0.003}_{-0.003}$ d & $<0.36$ & $30^{+4}_{-2}$ & $273^{+4}_{-4}$ & $0.024^{+0.009}_{-0.005}$ & $3.74^{+0.91}_{-0.94}$ & $0.79^{+0.10}_{-0.08}\,\left(^{+0.12}_{-0.11}\right)$ & $0.8^{+0.4}_{-0.1}\,\left(^{+0.4}_{-0.2}\right)$ & $1.7^{+0.3}_{-0.1}$ \\
    Star 7 &  &  &  & $281^{+6}_{-6}$ &  & $2.91^{+0.51}_{-0.46}$ &  &  &  \\
    Star 8 &  &  &  & $288^{+10}_{-10}$ &  & $2.14^{+1.00}_{-0.58}$ &  &  &  \\
    Star 16 & $2.3252^{+0.0001}_{-0.0001}$ d & $<0.16$ & $76^{+3}_{-4}$ & $276^{+4}_{-4}$ & $0.11^{+0.01}_{-0.01}$ & $0.76^{+1.73}_{-0.35}$ & $0.58^{+0.04}_{-0.04}\,\left(^{+0.43}_{-0.04}\right)$ & $0.6^{+0.3}_{-0.1}\,\left(^{+0.7}_{-0.1}\right)$ & $1.6^{+0.3}_{-0.1}$ \\
    \enddata
    \tablecomments{Orbital period ($P_{\rm orb}$), eccentricity ($e$) and RV semi-amplitude of the helium star ($K_1$) are from the orbital solutions (\secref{sec:orbits}). Eccentricities ($e$) are consistent with zero; we quote 99\% confidence upper limits. For the binary helium stars (Stars 1,2,4,6,16), the systemic velocity ($v_{\rm sys}$) is inferred from the orbit, whereas $v_{\rm sys}$ for the non-detections (Stars 5,7,8) is the error-weighted mean of measured RVs. The binary mass function ($f$) is given by \eqref{eq:bmf}. Helium star masses ($M_1$) originate from \citetalias{2023ApJ...959..125G} as described in Sect.~\ref{sssec:M1}. The minimum companion mass ($M_{2,\rm min}$) and inclination-averaged companion mass ($\left<M_2\right>_{\rm inc}$) are described in Sects. \ref{sssec:M2min} and \ref{sssec:M2inc}. The first set of errors for $M_{2,\rm min}$ and $\left<M_2\right>_{\rm inc}$ come from orbit uncertainty, whereas the second set of errors, in parentheses, come from combined orbit and helium star mass uncertainties.  Finally, the companion mass upper limit for a \emph{luminous} companion ($M_{2,\rm max}^\star$) is determined from optical photometry (\secref{sssec:opt-flux}). SMC/LMC sources are separated by a line (above/below). Best-fit orbital parameters are presented in \tabref{tab:detailed_orbits} of \appref{sapp:orbits}. }
\end{deluxetable*}

\section{Orbital solutions and masses}
\label{sec:orbits}

Having assessed evidence for binary motion in our sample of intermediate-mass helium stars, we will now determine orbits with our measured RVs. We use \texttt{The Joker} \citep{2017ApJ...837...20P}\footnote{\url{https://thejoker.readthedocs.io/en/latest/}}, which determines orbits in two steps: First, rejection sampling broadly surveys the orbital parameter space. Second, Markov Chain Monte Carlo (MCMC) derives orbital posterior distributions. 

The orbital parameter space that RVs can probe depends on the number of epochs, epoch spacing, observing baseline, and RV precision. For all 5 helium stars in binaries, we obtain 10-22 usable RVs (see \tabref{tab:rv_summary} and \appref{app:rv}). The total baseline for these measurements is 3-6 years (see \tabref{tab:stack}) and the smallest epoch spacing ranges from 3 hours to 1 day. We find that this is sufficient to constrain orbital periods up to several hundred days, provided inclinations are not prohibitively low. However, we are likely insensitive to wider orbits ($\gtrsim10^3$ days; see \secref{ssec:single_wide} for further discussion of the types of orbital periods we are insensitive to as a function of companion mass).

\subsection{Rejection sampling}\label{ssec:theJoker}

We first use \texttt{The Joker}'s rejection sampling to measure the plausible orbital parameter space for each star given its observed RVs. Initially, we consider our full set of 8 stars, whether or not we found significant evidence for binary motion in \secref{sec:binary_motion}. \texttt{The Joker} models RVs with six main free parameters: orbital period ($P_{\rm orb}$), eccentricity ($e$), pericenter argument ($\omega$), mean anomaly ($M_0$), RV semi-amplitude ($K_1$), and systemic velocity ($v_{\rm sys}$). For each star, we generate 1,000,000 parameter samples, adopting the following priors \citep[similar to ][]{2020ApJ...895....2P}:

The logarithm of the orbital period ($P_{\rm orb}$) is drawn from a uniform distribution: 
\begin{equation}
p(\ln P_{\rm orb}) = \mathcal{U} (\ln P_{\min}, \ln P_{\max}),
\end{equation}
where $p$ is the prior distribution, and we set the minimum (maximum) period $P_{\min}$$=$$10^{-1}$ ($P_{\max}$$=$$10^3$) days. The eccentricity ($e$) is drawn from a $\beta$ distribution, from 0 to 1: 
\begin{equation}
p(e) = \dfrac{\Gamma (a+b)}{\Gamma (a) \, \Gamma (b)} \, e^{a-1} \, (1-e) ^{b-1},
\end{equation}
where $a$$=$0.867 and $b$$=$3.03 following \citet{2013MNRAS.434L..51K}.
Both the pericenter argument ($\omega$) and the 
mean anomaly ($M_0$) are drawn from uniform distributions:
\begin{equation}
p(\omega) = \mathcal{U}(0,2\pi);\quad p(M_0) = \mathcal{U}(0,2\pi).
\end{equation}
The RV semi-amplitude ($K_1$) is drawn from a normal distribution:
\begin{equation}
p(K_1) = \mathcal{N}(K_1 | \mu_{K_1}, \sigma_{K_1}),
\end{equation}
where $\sigma_{K_1}$$=$$\sigma_{K,0} (P_{\rm orb}/P_0)^{-1/3} \, (1-e^2)^{-1/2},$ $P_0$$=$1 year and $\sigma_{K,0}$$=$100 km s$^{-1}$ for Stars 2, 4, 6, 16, and 30 km s$^{-1}$ for Stars 1, 5, 7, 8 due to the difference in magnitude of their $\Delta\rm RV_{\rm max}$ (see Table \ref{tab:rv_summary}). Thus, $K_1$ is drawn from a Gaussian prior with mean $\mu_{K_1}$ and variance $\sigma_{K_1},$ with hyper-parameters $\sigma_{K,0}$ and $P_0.$ We did not find that our results are sensitive to these hyper-parameters.
Finally, we draw the systemic velocity from a normal distribution:
\begin{equation}
p(v_{\rm sys}) = \mathcal{N}(0,\sigma_{v_{\rm sys}}), 
\end{equation}
where we set $\sigma_{v_{\rm sys}}$$=$30 \kms. Note that we center $p(v_{\rm sys})$ at 0 \kms~since we fit RVs relative to the highest SNR epoch (in \S\ref{ssec:orb_params} we transform $v_{\rm sys}$ for our final orbital solutions to the rest frame). 

Finally, when performing rejection sampling for Star 16 we allow for an additional uncertainty of 2 \kms~with \texttt{The Joker}'s jitter term. We apply this to Star 16 since it has smaller RV uncertainties from narrow \HeI~lines (see \secref{ssec:rv_lines}). We do not utilize this term for any other star.

\texttt{The Joker} sifts through each randomly drawn orbit and uses the accept-reject method \citep{book:NeumannCollectedWorksVolV} to determine which orbits are good fits to the observed RVs. If the accepted samples are unimodal in the orbital parameter space, we consider the orbit to have been constrained. 
This is the case for all five of the helium stars that passed our criteria for detection of binary motion in  \secref{sec:binary_motion}. We therefore continue with posterior sampling for these stars in \secref{ssec:orbit_mcmc}. Conversely, if the orbital parameter space is degenerate, we consider the orbit unconstrained. This is the case for three stars identified as ``non-detections'' in \secref{sec:binary_motion}, and we therefore do not continue with posterior sampling for these stars. Overall, our analysis totally agrees with the independent RV significance test applied in \secref{sec:binary_motion}; all sources with significant RV variation converge to a reasonable orbital solution. This also supports our conclusion that the non-detections do not exhibit detectable RV shifts from binary motion. We present the rejection samples for the helium stars in binaries, as well as the non-detections, in \appref{sapp:corner_plots}. For the remainder of this section, we will focus on the helium stars in binaries.

\subsection{Posterior sampling}\label{ssec:orbit_mcmc}
We continue with \texttt{The Joker}'s MCMC method to measure orbital properties and uncertainties. We initialize four chains from the rejection samples and sample 10,000 steps after a 2000 step tuning period. Four chains are initialized to cover the region of the target distribution with significant probability. Initialization from the rejection samples avoids low probability regions. The tuning period ensures that the chains adapt to the posterior geometry and stabilize. This setup balances computational efficiency with adequate exploration of the parameter space for robust posterior estimation.

In \appref{sapp:corner_plots}, we present orbital posteriors for Stars 1, 2, 4, 6 and 16. We also compare inferred orbital properties when fitting the RVs measured from different spectral line combinations. 
If a luminous companion was contributing a non-negligible fraction of the observed Balmer absorption, then the derived orbital properties should systematically change if we include/exclude \HeII-H~blends. However, this is not the case for the systems considered here. Regardless, of the line combination we use, we find that the derived orbital properties are robust. Thus, we do not find any evidence of companion RV motion, and only identify RV motion from the helium star, as expected for SB1s. This strengthens the interpretation that the companions are either optically faint low-mass stars or optically dark compact objects.

\begin{figure*}[ht!]
\centering
\includegraphics[width=\textwidth]{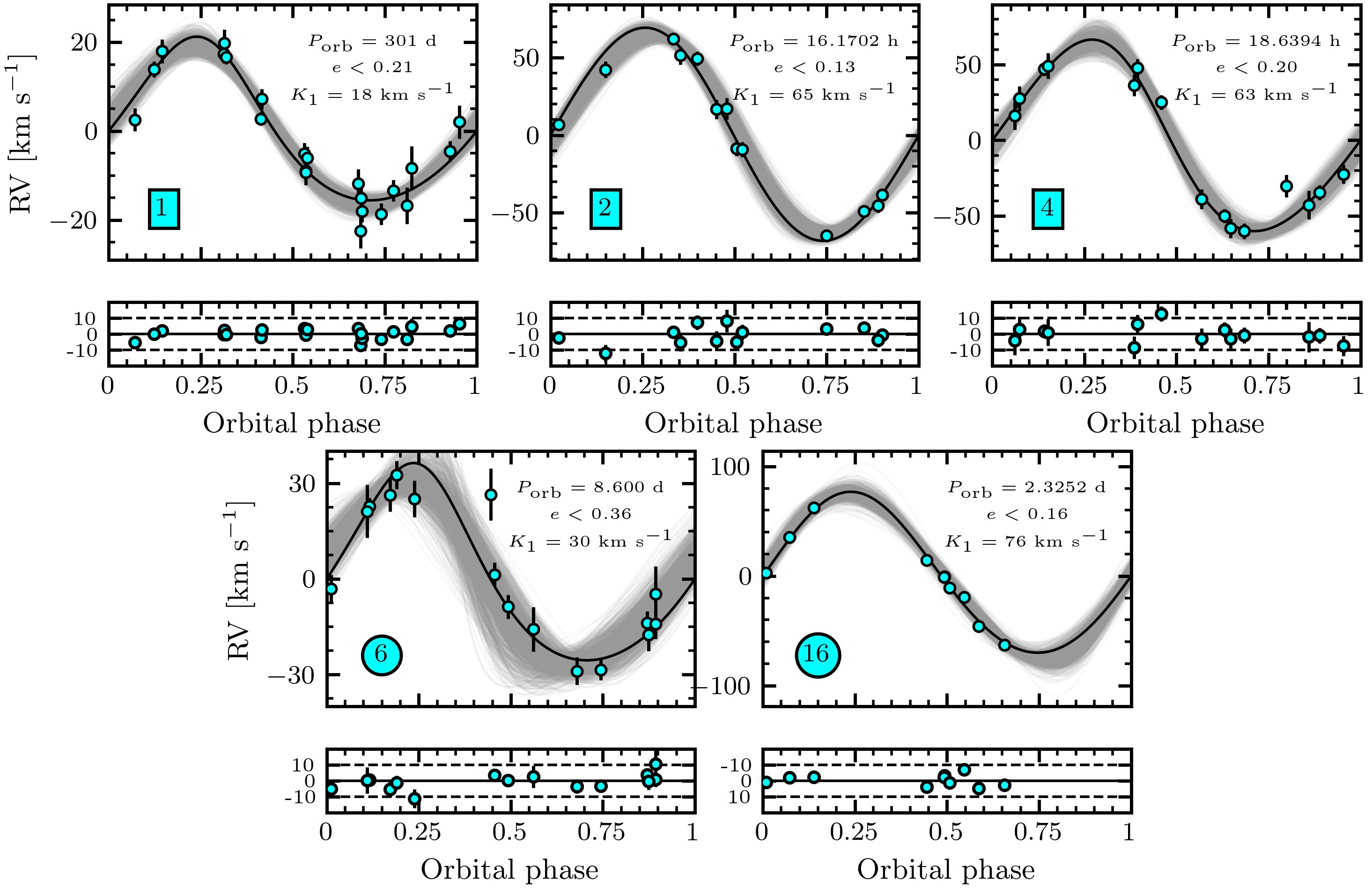}
\caption{Helium stars in binaries have orbital periods ranging from hours to almost 1 year. Phase-folded RV curves are presented, with 10$^3$ posterior samples (grey) about the maximum a posteriori orbit (MAP, black, \tabref{tab:detailed_orbits}). Systemic velocities are subtracted. RVs are cyan circles. Residuals are largely confined to $\pm$10 km s$^{-1}$. The orbital period $P_{\rm orb},$ 99\% upper limit on eccentricity $e$, and RV semi-amplitude $K_1$ are quoted in each sub-panel. Zero orbital phase is defined as RV $=$ 0.}
\label{fig:orbits}
\end{figure*}

\subsection{Derived Orbital Properties}\label{ssec:orb_params}

In \figref{fig:orbits}, we present the phase-folded orbital solutions resulting from our posterior sampling. We show the maximum a posteriori orbit (MAP: orbit which maximizes the posterior distribution given the RV data, black line), with 1000 posterior samples (grey lines) to represent uncertainty. Our RV data span the majority of the orbital phase, indicating we have sufficient phase coverage. Below each orbit we show fit residuals, which are almost exclusively within 10 \kms~(dashed black lines) of the MAP orbit (solid black line). This 20 \kms~scatter is effectively equivalent to \dRVm{} for the three systems for which we do not find significant evidence of binary motion (see \tabref{tab:rv_summary} and \appref{app:rv} for details). Individual RV measurements for each system are shown as cyan points, and RVs are centered at $v=0$ for visualization. In \tabref{tab:orbit_props}, we report our orbital period ($P_{\rm orb}$), eccentricity ($e$), RV semi-amplitude ($K_1$), and systemic velocity ($v_{\rm sys}$) constraints for Stars 1, 2, 4, 6 and 16.

For $P_{\rm orb}$ and $K_1$, we report the posterior median, with 1$\sigma$ errors corresponding to 68\% confidence intervals. $P_{\rm orb}$ ranges from $<1$ day to $\lesssim 1$ year for the stars in our sample: Stars 2 and 4 have very short periods ($\sim$16 and $\sim$18 h), Stars 6 and 16 have short periods of several days ($\sim$2 and $\sim$9 d), and Star 1 has a longer $\sim$300 day period. In all cases, $P_{\rm orb}$ is well constrained due to the long baseline of our observations. $K_1$ ranges from 18--76 km s$^{-1}$ for our sample. In all cases, the measured $K_1$ values are close to \dRVm/2 (in \tabref{tab:orbit_props}) indicating that our our measured \dRVm~are representative of $K_1$.

For $v_{\rm sys}$, we also report the posterior median and 68\% confidence intervals. However, \texttt{The Joker} values were measured relative to the RV of the highest SNR spectrum of a given source (see \secref{sec:rv}). We therefore combine these values with the absolute measured velocities presented in Section \ref{ssec:rvs} to provide the true systemic velocity. We note that we also provide systemic velocities for the 3 stars with non-detections of binary motion in \tabref{tab:rv_summary}. These values simply correspond to the error-weighted mean of the measured RVs. Overall, these range from $\sim$160--190 km s$^{-1}$ for stars in the SMC (Stars 1, 2, 4) and from $\sim$270--290 km s$^{-1}$ for stars in the LMC (Stars 5, 6, 7, 8, 16). These values are consistent with expectations for SMC and LMC membership \citetext{see Fig.~S7 in \citetalias{2023Sci...382.1287D}; 
\citealp{2008MNRAS.386..826E}; 
\citealp{2015A&A...584A...5E,2015A&A...574A..13E}}.

We choose to fit for $e$, instead of fixing orbits to be circular, to allow for the possibility of detecting non-zero $e$. However, this necessarily adds an extra degree of freedom to orbital fitting. As such, and because $e$ has a one-sided distribution, observations of circular orbits can sometimes be well fit to non-zero $e$. We therefore implement the revised Lucy-Sweeney (LS) test to assess whether circular orbits are consistent with the data \citep{1971AJ.....76..544L,
2013A&A...551A..47L}. First, we compute $\left<e\right>\pm\mu,$ where $\left<e\right>$ is the median posterior eccentricity, and $\mu$ is the standard error. If $\left<e\right>/\mu>3.06,$ a non-zero $e$ is accepted. This corresponds to the detection threshold for non-zero $e$, for a confidence level of 99\% \citep{1971AJ.....76..544L}. Otherwise, if $\left<e\right>/\mu<3.06,$ the LS test concludes the orbit is consistent with circular. We adopt a 99\% upper limit on $e$ by interpolating Table A.2 in \cite{2013A&A...551A..47L}. In all cases, $\left<e\right>/\mu<3.06$, implying that we do not detect significant evidence for non-zero $e.$ As such we report 99\% upper limits on $e$ in Table \ref{tab:orbit_props}, since our orbits are consistent with circular. For Stars 1, 2, 4, and 16 these are all $e\lesssim0.2$. For Star 6, our constraint is slightly weaker ($e <0.36$ at 99\% confidence), due to lack of RV observations at 0.3--0.4 of the orbital phase. Additional observations at this orbital phase could improve the $e$ constraint. See \appref{app:orbits} for the $e$ posterior distributions of each system.

In \tabref{tab:detailed_orbits} of \appref{sapp:orbits}, we also list the period, eccentricity, pericenter argument, mean anomaly, semi-amplitude, systemic velocity and reference time for the best-fit orbit (MAP). Since we find that all orbits are consistent with circular, we also provide best-fit orbital parameters assuming $e=0$ in \tabref{tab:detailed_orbits}.

\subsection{Companion mass constraints from orbits}\label{ssec:m2}

\begin{figure*}[ht!]
    \centering
    \includegraphics[width=\textwidth]{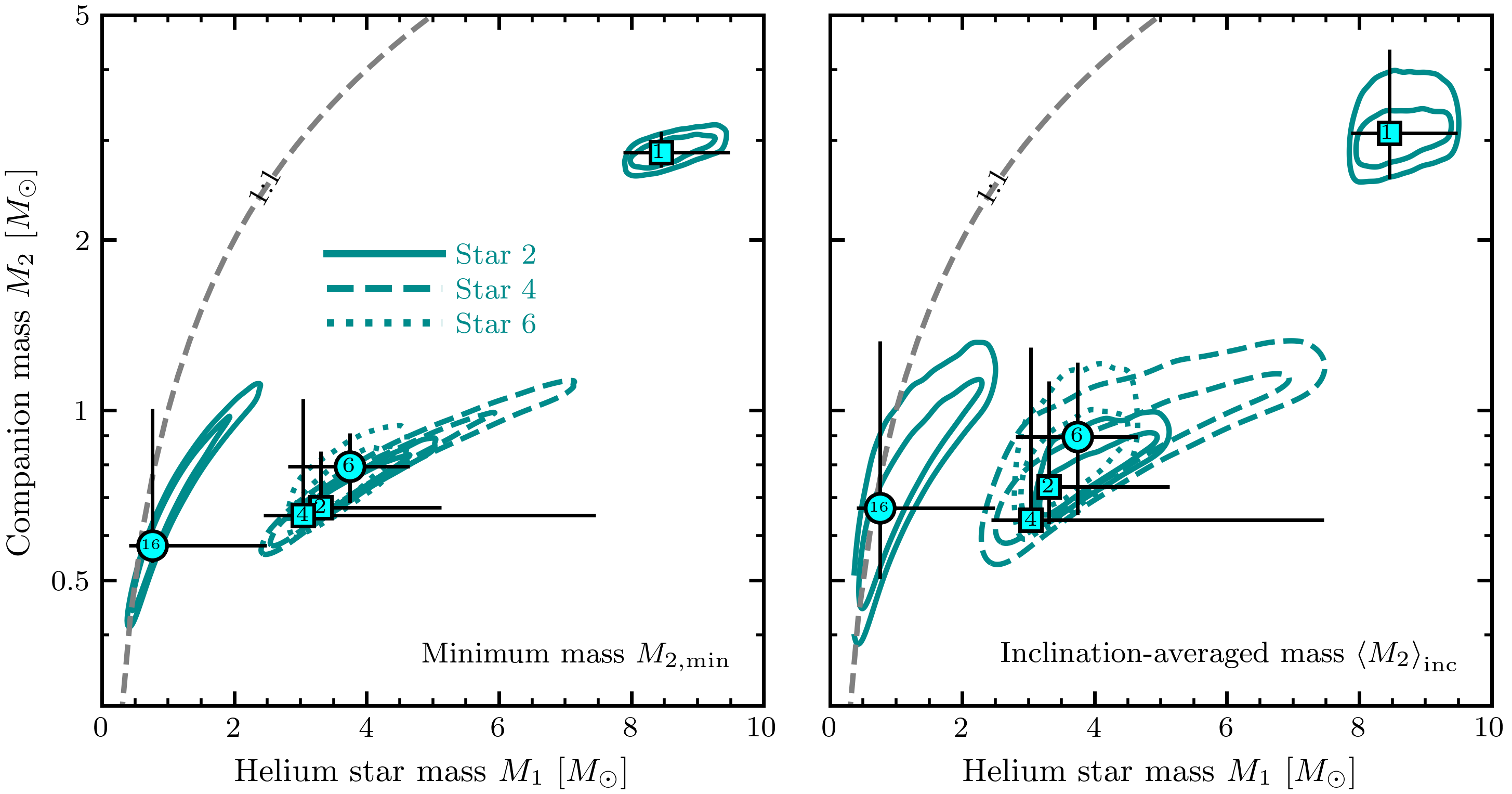}
    \caption{Binary mass constraints for our binary helium star sample. Helium star masses originate from \citetalias{2023ApJ...959..125G} (see \secref{sssec:M1} for details). For the optically dark companions, we present both the minimum companion mass ($M_{2,\rm min}$, left) and inclination-averaged companion mass ($\left<M_2\right>_{\rm inc}$, right), with uncertainties which are described in \secref{ssec:m2} (numbered cyan points). For each star, we show 0.5$\sigma$ and 1$\sigma$ contours which highlight the degeneracy between $M_1$ and $M_2$: a more massive helium star implies a more massive companion. $\left<M_2\right>_{\rm inc}$ is a probabilistic mass estimate, which is therefore larger than the $M_{2,\rm min}$ lower bound, which assumes the system is edge-on. $M_1$ axis is linear scale, $M_2$ axis is log scale, and the grey dashed line corresponds to $q=1.$}
    \label{fig:m2}
\end{figure*}

We have just  demonstrated that Stars 1, 2, 4, 6, and 16 are in binaries by obtaining orbital solutions. However, as discussed in \secref{ssec:sample}, we selected these stars because they do not show evidence for significant optical flux contribution from a companion in the works of \citetalias{2023Sci...382.1287D} and \citetalias{2023ApJ...959..125G}. In addition, we demonstrated that we do not find evidence for a systematic shift in either the measured RVs or orbital solutions based on which spectral lines we used in our analysis (Appendix \ref{sapp:rv_compare} and \ref{sapp:corner_plots}), supporting that the companions are optically faint or dark. To better understand these companions, here we use the orbital solutions to constrain the companion masses.

\subsubsection{Binary Mass Function}\label{sssec:massfunc}
Combining orbital properties with a helium star mass estimate yields an inclination angle dependent constraint on an optically invisible companion through the binary mass function:
\begin{equation}\label{eq:bmf}
    \frac{(M_2\sin i)^3}{(M_1+M_2)^2} 
    = \frac{P_{\rm orb} K_1^3}{2\pi G}(1-e^2)^{3/2} \equiv f,
\end{equation}
where $M_1$ is the stripped star mass, $M_2$ is the companion mass and $i$ is the inclination angle. In \tabref{tab:orbit_props}, we present the median $f$  with 1$\sigma$ errors (defined as the 68\% confidence interval) for all 5 helium stars in binaries in our sample, derived directly from \texttt{The Joker}'s posteriors. These values span $f\approx0.02-0.2\,M_\odot.$ These $f$ constraints are critical since, combined with a mass estimate for the helium star, they can constrain the mass of an optically hidden companion.

\subsubsection{Assumed Helium Star Masses}\label{sssec:M1}

Using the results of their spectro-photometric fitting, \citetalias{2023ApJ...959..125G} derived two constraints on the masses of the helium stars in our sample. In particular, they measured both a spectroscopic mass ($M_{1,\rm spec}$) computed from surface gravity and effective radius (which itself is derived from the effective temperature and bolometric luminosity), and an evolutionary mass ($M_{1,\rm evol}$) computed by comparing the derived bolometric luminosity to a mass-luminosity relation from evolutionary models for stripped stars halfway through core-helium burning \citepalias[see Figure 9 in][]{2023ApJ...959..125G}.
In general, $M_{1,\rm spec}$ and $M_{1,\rm evol}$ are in agreement, while $M_{1,\rm spec}\lesssim M_{1,\rm evol}$. 

For all stars whose bolometric luminosities and effective radii match expectations for the long-lived core-helium burning phase \citepalias[stars 1, 2, 4, 5, 6, 7, and 8; see][]{2023ApJ...959..125G}, we adopt the evolutionary mass as our fiducial mass estimate. In particular, follow-up work by \cite{GotbergPopSynth} shows that regardless of convective overshoot, stellar winds, metallicity, or amount of leftover hydrogen, the bolometric luminosity scales tightly with the stellar mass for stripped stars. The luminosity also varies minimally during the majority of the long-lasting central helium burning phase 
\citetext{see e.g. Fig.~5 of \citetalias{2023ApJ...959..125G}; 
\citealp{1967ZA.....65..251K}}. We, therefore, consider the evolutionary masses reliable for hot helium stars that are undergoing central helium burning. For the stars in our sample with detected binary motion, the helium star in Star 1 is the most massive with $M_1\approx8\,M_\odot,$ while Stars 2, 4 and 6 are $\sim$3--4 $M_\odot$ helium stars.

In contrast to the other stars in our sample, Star 16 is somewhat inflated: its effective radius is approximately twice that expected for a core-helium burning stripped star of its bolometric luminosity. It is therefore likely undergoing a brighter expansion or contraction phase. Hence, the evolutionary mass would overestimate the true mass (see Sec. 4.2.5 in \citetalias{2023ApJ...959..125G}). We therefore adopt the spectroscopic mass as the primary mass estimate, only for Star 16. This value is lower than the other stars in our sample with $M_1\approx0.8\,M_\odot$.

We recognize that the adopted $M_1$ can impact the derived companion masses. As such, in \tabref{tab:detailed_m2} of \appref{sapp:m2} we present the masses described below assuming both $M_{1,\rm spec}$ and $M_{1,\rm evol}$ for all five of the helium stars in binaries in our sample. We also provide approximate analytic formulae for arbitrary $M_1$.

\subsubsection{Minimum companion masses}\label{sssec:M2min}

For a given star, using the above $f$ and $M_1$, we can numerically solve for the minimum companion mass $M_{2,\min}$. This is the mass of the companion assuming the system is edge-on ($i=90^\circ$). If the system is not edge-on, the intrinsic RV shifts of the binary are larger than the observed $K_1,$ which would mean that the companion is more massive.

The ranges of $M_1$ values from \citetalias{2023ApJ...959..125G} are not calibrated likelihood uncertainties; rather, they reflect the spread of models which provide an acceptable fit to the spectroscopic data, and are thus conservative. We therefore adopt the best-fit $M_1$ mass in our fiducial $M_{2,\min}$ estimate. Then, to account for $M_1$ uncertainty, we sample $M_1$ from a truncated, split-normal distribution centered on the best-fit $M_1,$ with asymmetric widths set by the quoted upper and lower errors on $M_1$ \citep[similar to e.g.][]{2025RASTI...4af052D}. This prescription provides a smooth approximation of the allowed $M_1$ range, while accounting for the best-fit $M_1$ being most probable. Note that since we demonstrated in \secref{ssec:orb_params} that all of our stripped star orbits are consistent with circular, we set $e=0$ for our companion mass calculations. Setting $e\neq0$ would lower our $M_{2,\rm min}$ by 1$-$10\%. 

In \tabref{tab:orbit_props}, we present the median $M_{2,\rm min}$ associated with the best-fit $M_1$ from \citetalias{2023ApJ...959..125G}, corresponding to the most probable minimum companion mass. We quote two uncertainties. The first is obtained by fixing $M_1$ to its best-fit value and sampling the orbital posterior; this captures the orbital uncertainty alone, which is typically small ($\sim$0.01--$0.1\,M_\odot$). The second, shown in parentheses, is obtained by sampling both the orbital posterior and the above $M_1$ distribution; this captures the combined orbital and helium-star mass uncertainty. Note that for Stars 1 and 6, the dominant source of error comes from the orbit, since both sets of errors are comparable; the uncertainty in $M_1$ is small relative to the orbit. Conversely, for Stars 2, 4, and 16, the uncertainty in $M_1$ dominates, producing larger upper errors when the allowed $M_1$ range is propagated.

In the left panel of \figref{fig:m2}, we show $M_1$ and $M_{2,\rm min}$. The capped error bars show the orbital uncertainty at fixed $M_1$, while the uncapped error bars include the additional uncertainty from the adopted $M_1$ range. Dark cyan contours show 0.5 and 1$\sigma$ regions. Stars 2, 4, 6, and 16 have $M_{2,\rm min}\simeq0.6$--$0.8\,M_\odot$, whereas Star 1 has a substantially larger minimum companion mass, $M_{2,\rm min}\approx2.9\,M_\odot$. There is clearly degeneracy between $M_1$ and $M_{2,\rm min};$ a more massive helium star implies a more massive companion. Still, assuming the best-fit $M_1$, $M_{2,\rm min}$ is sharply constrained.

\subsubsection{Inclination-averaged companion masses}
\label{sssec:M2inc}

While $M_{2,\rm min}$ is the most conservative estimate of a binary companion mass, since it assumes an edge-on inclination ($i=90^\circ$), we can also examine the companion mass, considering a range of inclinations. Since we do not currently have observations at our disposal which can constrain inclination (such as light curve eclipses), we adopt a random inclination distribution. In particular, the probability density function for $i$ assuming isotropy is $P(i)\,di\propto\sin i\, di.$ The expectation value for the inclination of a binary is therefore $\left<i\right>\approx 57^\circ.$ We also compute the ‘inclination-averaged companion mass’ ($\left<M_2\right>_{\rm inc}$) which randomly samples over the isotropic inclination angle distribution when computing $M_2$ through \eqref{eq:bmf}. Including prior information on the inclination angle distribution yields a probabilistic estimate for $M_2.$ 

It is important to consider that the companion mass distribution, when sampling over isotropic inclinations, exhibits a sharp rise at lower masses, corresponding to near edge-on inclinations, and an extended high-mass tail, corresponding to low inclinations. As such, in \tabref{tab:orbit_props}, we quote the mode $\left<M_2\right>_{\rm inc}$ associated with the best-fit $M_1$ from \citetalias{2023ApJ...959..125G}. The mode captures the peak in $\left<M_2\right>_{\rm inc}$ distribution, whereas the median is biased towards low-inclinations. Analogous to $M_{2,\rm min},$ we quote two uncertainties: the uncertainty associated with the orbit, and the uncertainty when also accounting for the $M_1$ range (in parentheses). Again, because of the skewed distribution which arises from sampling inclinations, uncertainties reflect the highest-probability density interval (HPDI) containing 68\% of the $\left<M_2\right>_{\rm inc}$ distribution.

In the right panel of \figref{fig:m2}, we show $M_1$ vs. $\left<M_2\right>_{\rm inc}$ on the right. $\left<M_2\right>_{\rm inc}> M_{2,\rm min},$ and $\left<M_2\right>_{\rm inc}$ uncertainties are larger than $M_{2,\min},$ since $\left<M_2\right>_{\rm inc}$ marginalizes over the isotropic inclination angle distribution. In other words, assuming an isotropic inclination angle distribution, we expect that a companion is intrinsically more massive than $M_{2,\rm min}.$ In particular, Stars 1, 2, 4, 6, and 16 have $\left<M_2\right>_{\rm inc}=3.1$, 0.7, 0.7, 0.8, 0.6, respectively, and $\left<M_2\right>_{\rm inc}$ is 0.1--0.2 $M_\odot$ larger than $M_{2,\rm min}.$ However, the upper uncertainties on $\left<M_2\right>_{\rm inc}$ are much larger than $M_{2,\rm min}$ due to lower inclinations being sampled. 

\section{Helium star binaries in observational context}
\label{ssec:context}

Given that we have now measured the orbital properties of five intermediate-mass helium stars with optically invisible companions, a natural first step is to contextualize them by comparing to other well-studied binary populations that contain helium stars \citep[e.g., systems with subdwarfs or WR stars:][]{2004Ap&SS.291..299M,
2017A&A...605A.109V,
2022A&A...664A..93D,
2023A&A...674A..88D}. 


\subsection{Summary of comparison sample}\label{sssec:obs-summary}

The comparison sample of helium stars in binaries that we consider contains a wide-range of objects, including: (i) lower mass objects with $M_1\lesssim2\,M_\odot$ (mostly hot subdwarfs), (ii) the higher mass regime with $M_1>8\,M_\odot$ (exclusively WRs), and (iii) puffed-up stripped stars which have large radii relative to expectations for hot stripped stars\footnote{We recognize that the ‘intermediate’ mass range for helium stars is somewhat arbitrary. In particular, our 2--8 $M_\odot$ definition reflects the dearth of helium stars more massive than 2 $M_\odot$, and the approximate (Milky Way) mass threshold for WR stars, above 8 $M_\odot$. While it is reasonable to consider lower mass helium stars from 1--2 $M_\odot$ as intermediate-mass, we retain the 2--8 $M_\odot$ range of \citetalias{2023Sci...382.1287D}, implying $\sim$8--25 $M_\odot$ progenitors, both for consistency and simplicity.}. We note that many systems were drawn from the Post Mass Transfer Catalog \citep{2026arXiv260531290V}\footnote{\url{https://binary-observations.github.io/post_mt_catalog/}}.

In the low-mass regime, we consider observed samples of hot subdwarfs in binaries. Specifically, we consider the sdB+MS and the sdB + white dwarf (WD) binaries from \cite{2017A&A...605A.109V} and \cite{2022A&A...666A.182S,2023A&A...673A..90S}. Both samples contain subdwarfs with an (assumed) canonical sdB mass of \mbox{$M_{\rm sdB}=0.47\,M_\odot$}, while the former contain low-mass MS companions and the latter have both MS and WD companions. We also consider the very recent work of \cite{2026arXiv260827673D}, who presented a 500 pc volume-limited sample of short-period sdO/B+MS and sdO/B+WD binaries. Importantly, subdwarf masses from \cite{2026arXiv260827673D} are SED-fitted masses from \cite{2026A&A...707A...6D}, not the canonical subdwarf mass. In addition, we consider a set of more massive hot subdwarfs ($\sim$0.5--2 $M_\odot$) that orbit Be companions \citep{2023AJ....165..203W, 
2024ApJ...962...70K,
2025A&A...694A.208K}, HD 49798: a low-mass X-ray binary (LMXB) with a $\sim$1.5 $M_\odot$ subdwarf orbiting a compact object \citep{2023MNRAS.523.3043R}, and HD 45166: a quasi-WR ($q$WR) helium star straddling the low/intermediate-mass helium star boundary \citep{2023Sci...381..761S,2025A&A...695L..20D}.

In the high-mass regime, we consider a subset of seven Galactic WR+O binaries
and five WR+O binaries in the Magellanic Clouds which previous authors have argued likely underwent binary-stripping, as opposed to wind-stripping \citep{2006A&A...447..667F,
2016A&A...591A..22S,
2017MNRAS.464.2066S,
2018A&A...616A.103S,
2019A&A...627A.151S}. When selecting stars from these works, we omit WR binaries which only have minimum mass estimates. In addition, we include Cyg X-3, a Galactic lower mass $\sim$$10\,M_\odot$ WR, which orbits an undetermined compact object \citep{2002A&A...392..161S,
2013MNRAS.429L.104Z}. 

Finally, we consider puffed-up stripped stars with large radii, which suggest they may not be completely contracted. Since our helium stars are H-poor, with $X_{\rm H, surf}\lesssim0.4,$ we only consider puffed-up stripped stars with $X_{\rm H, surf}<0.5$, to match our sample. HR 6819 is a $X_{\rm H, surf}\sim 0.45,R\sim4\,R_\odot$ low-mass stripped star orbiting a Be companion \citep{2021MNRAS.502.3436E, 2025A&A...694A.208K}. Similarly, HIP 15429 is a low-mass stripped star + Be binary with $X_{\rm H, surf}\sim0.3$ and $R\sim 9\,R_\odot$ \citep{2025A&A...701A...9M}. The only intermediate-mass, H-poor puffed-up stripped star is 2dFS 163: a $\sim 4$$M_\odot$ stripped star with $X_{\rm H, surf}\sim0.3$ and $R\sim6\,R_\odot$ \citep{2024A&A...692A..90R}.

\subsection{Masses}\label{sssec:M1_M2_obs}

\begin{figure}[t!]
    \centering
    \includegraphics[width=\columnwidth]{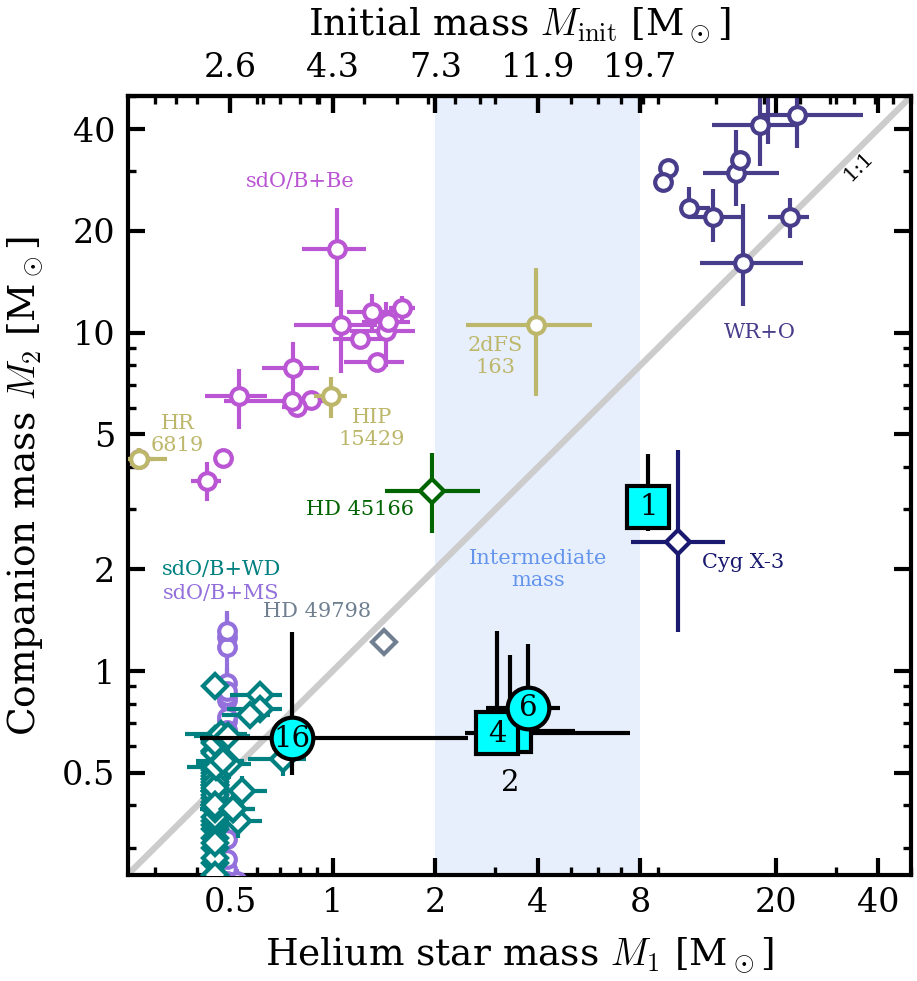}
    \caption{Binary component masses $M_1$ (helium star) and $M_2$ (companion) for helium star binaries (cyan) in comparison to observed helium star binaries: subdwarfs with Be, MS and WD companions (pink circles, light purple circles, teal squares), puffed-up stripped stars with $X_{\rm H, surf}<0.5$ (yellow circles: HR 6819, HD 45166, 2dFS 163), WRs with O-star companions (dark purple circles), HD 49798 (dark grey diamond), HD 45166 (green diamond), Cyg X-3 (dark blue diamond). The light blue shading highlights the intermediate-mass regime, while the grey line denotes $q=1.$ We slightly offset the sdB+MS and sdB+WD binaries which assume the the canonical subdwarf mass, for visibility. The initial-stripped star mass relation of \cite{2025A&A...697A.239H} at LMC metallicity ($Z=0.006$) is shown above.}
    \label{fig:M1_M2_He}
\end{figure}

We compare the component masses of our binaries to the comparison sample in \figref{fig:M1_M2_He}. Specifically, we plot the values of $M_1$ versus $\left<M_2\right>_{\rm inc}$ for our sample alongside $M_1$ and $M_2$ estimates for the helium star binaries from the literature described above. Our stripped star binaries are plotted as numbered cyan squares (SMC) and circles (LMC), while the comparison samples are plotted with various colours (see \figref{fig:M1_M2_He} caption for details). From examining Figure~\ref{fig:M1_M2_He}, we can draw two important conclusions:

First, as previously noted, intermediate-mass helium stars ($\sim$2-8 $M_\odot$, shaded blue) in binaries are observationally rare \cite[e.g.][]{2026PASP..138b4202B}. Prior to the sample presented here, few helium stars with $X_{\rm H, surf}<0.5$ had been observed: 2dFS 163, and (towards the limits) HD 45166 and Cyg X-3.

Second, the optically dark companions to the helium stars in our sample are lower mass than the companions in most previously observed helium star binaries. In particular, if we consider objects in our comparison sample above the canonical subdwarf mass\footnote{i.e. excluding the sdO/B+MS and sdO/B+WD samples, some of which are placed this diagram approximately because the companion masses depend on the assumed subdwarf mass. This is not true for the sample of \cite{2024A&A...686A..25D,2026A&A...707A...6D,2026arXiv260827673D}, who employ SED fitting to estimate the subdwarf mass. Lastly, most WD/MS companion masses presented are minimum masses.}
essentially all of them have mass ratios $q=M_2/M_1\geq1$ (i.e. the companions masses exceed the helium star masses; they are located above the 1:1 line shown in Figure~\ref{fig:M1_M2_He}). In contrast, the helium star binaries in our sample have $q \lesssim 1$. The one notable exception is Cyg X-3, which is thought to contain a compact object and has $q \sim 0.2$ \citep{2013MNRAS.429L.104Z}, similar to the objects in our sample. Whether or not the preference for $q \lesssim 1$ in our sample is intrinsic to (a subset of) the intermediate-mass stripped star population, or driven by selection effects, is an open question which will be addressed in future work \citep[but see also][]{2026PASP..138b4202B}. 

We emphasize that the aforementioned helium star binary samples were discovered through several detection methods and therefore have a variety of selection effects. For example, the sdO/B+Be binaries were identified by searching for companions to Be stars, which preferentially identifies brighter stripped stars in nearby systems. The hot subdwarf population is biased towards short-period systems due to identifiable signatures in their light curves, like reflection effects for cool companions, or tidal deformations and Doppler boosting, in the case of a close WD companion \citep{2022A&A...666A.182S,
2023A&A...673A..90S,
2024A&A...686A..25D,
2026A&A...707A...6D,
2026arXiv260827673D}. With respect to the intermediate-mass helium stars, the UV selection method implemented by \cite{2026ApJ...999...73L} preferentially selects optically faint/dark companions \citep{2026PASP..138b4202B}. Further samples will be necessary to map the full distribution of component masses, while spectroscopic follow-up of the \cite{2026ApJ...999...73L} sample is ongoing.

For completeness, we reproduce \figref{fig:M1_M2_He}, including all helium star binaries which we excluded from our comparison sample, in \figref{fig:M1_M2_He_all} of \appref{app:obs}, such as puffed-up stripped stars with $X_{\rm H,surf}\geq0.5$ and WRs with minimum mass estimates.

\subsection{Periods}\label{sssec:Porb_M1_obs}

\begin{figure}[t!]
    \centering
    \includegraphics[width=\columnwidth]{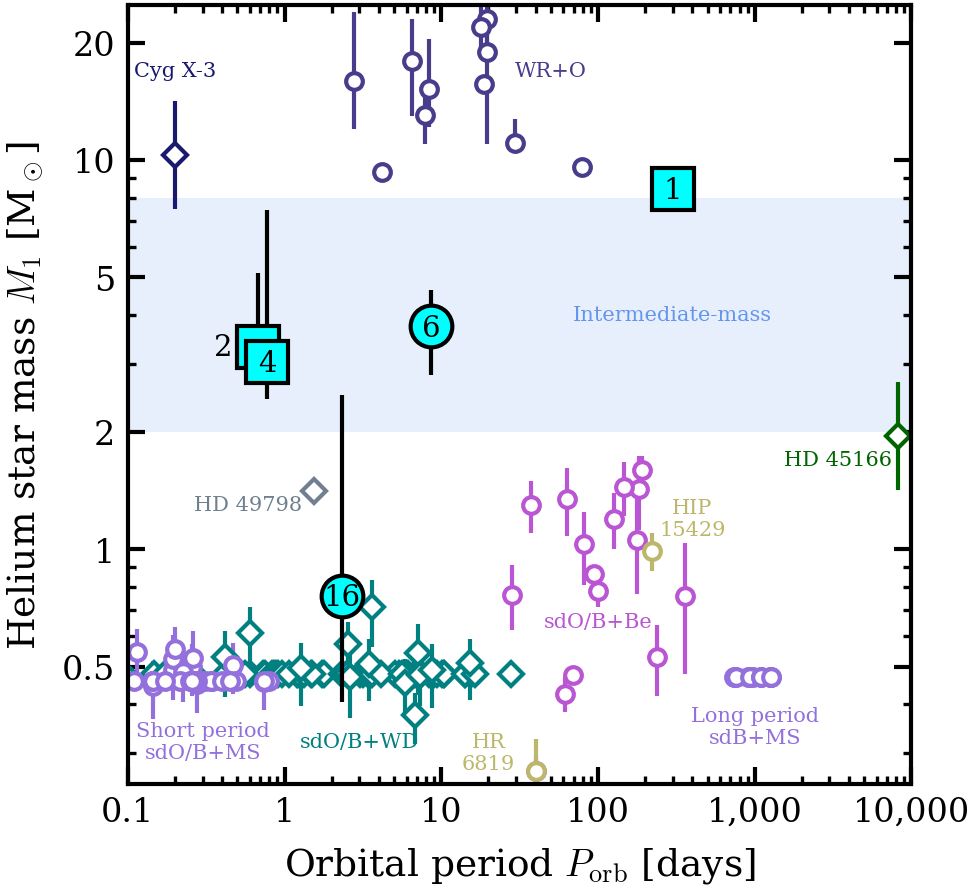}
    \caption{Helium star mass $M_1$ (cyan) as a function of orbital period $P_{\rm orb}$ in comparison to observed helium star binaries. All helium star binaries with period estimates from \figref{fig:M1_M2_He} are included. Our sample spans almost 3 orders of magnitude in $P_{\rm orb}$, across much of the range of previously observed helium star binaries, from sub-day to $\lesssim$ 1 year. sdB+MS and sdB+WD binaries which assume the canonical subdwarf mass are offset, for visualization.}
    \label{fig:M1_Porb_He}
\end{figure}

In \figref{fig:M1_Porb_He}, we show both our binaries and the comparison sample in a plot of helium star mass versus orbital period. Symbols are color-coded in the same manner as in \figref{fig:M1_M2_He} (described in Section~\ref{sssec:M1_M2_obs}). Overall, this plot---which spans 5 orders of magnitude---emphasizes the incredibly wide range of orbital periods observed for both the systems in our sample as well as the comparison sample. 

In the literature, orbital periods found for different subsets of helium stars have been linked to their evolutionary history. For example, the sdO/B+MS star periods (purple circles) are bimodal: short period sdO/B+MS have sub-day periods, while long period sdB+MS cluster at 1000 days. The short period sdO/B+MS binaries are thought to form through common envelope evolution (CEE), whereas the long period sdB+MS binaries likely underwent stable mass transfer \citep[e.g.][]{2002MNRAS.336..449H}. The intermediate-mass helium star \mbox{HD 45166} is even more extreme, with an anomalously wide period of $P_{\rm orb}\lesssim10,000$ days. This long period has been used to argue that this helium star may not have formed through traditional stripping but is instead the result of a merger in a triple system \citep[c.f.][]{2023Sci...381..761S}. Finally, multiple subsets of our comparison sample are found at intermediate periods including the sdO/B+Be binaries, puffed-up stripped stars, and WR+O binaries. The sdO/B+Be stars in particular are thought to originate from (conservative) stable mass transfer \citep{2023AJ....165..203W,2025ApJ...990L..51L}.

Overall, the binaries in our sample span almost three orders of magnitude in $P_{\rm orb}$, from sub-day to $\lesssim1$ year. This covers much of the range of previously observed helium star binary periods, many of which likely experienced binary envelope-stripping.  While a larger sample of intermediate-mass helium star binary periods is required to obtain a more complete distribution, this already points to the potential for a diverse set of evolutionary histories even within the current set of objects presented here. 

For completeness, as above, we reproduce \figref{fig:M1_Porb_He}, including all helium star binaries which we excluded for the aforementioned reasons, in \figref{fig:M1_M2_He_all} of \appref{app:obs}. 

\section{Possible binary companions I: \\ Living stars}
\label{sec:stellar-comp}

\begin{figure*}
    \centering
    \includegraphics[width=\textwidth]{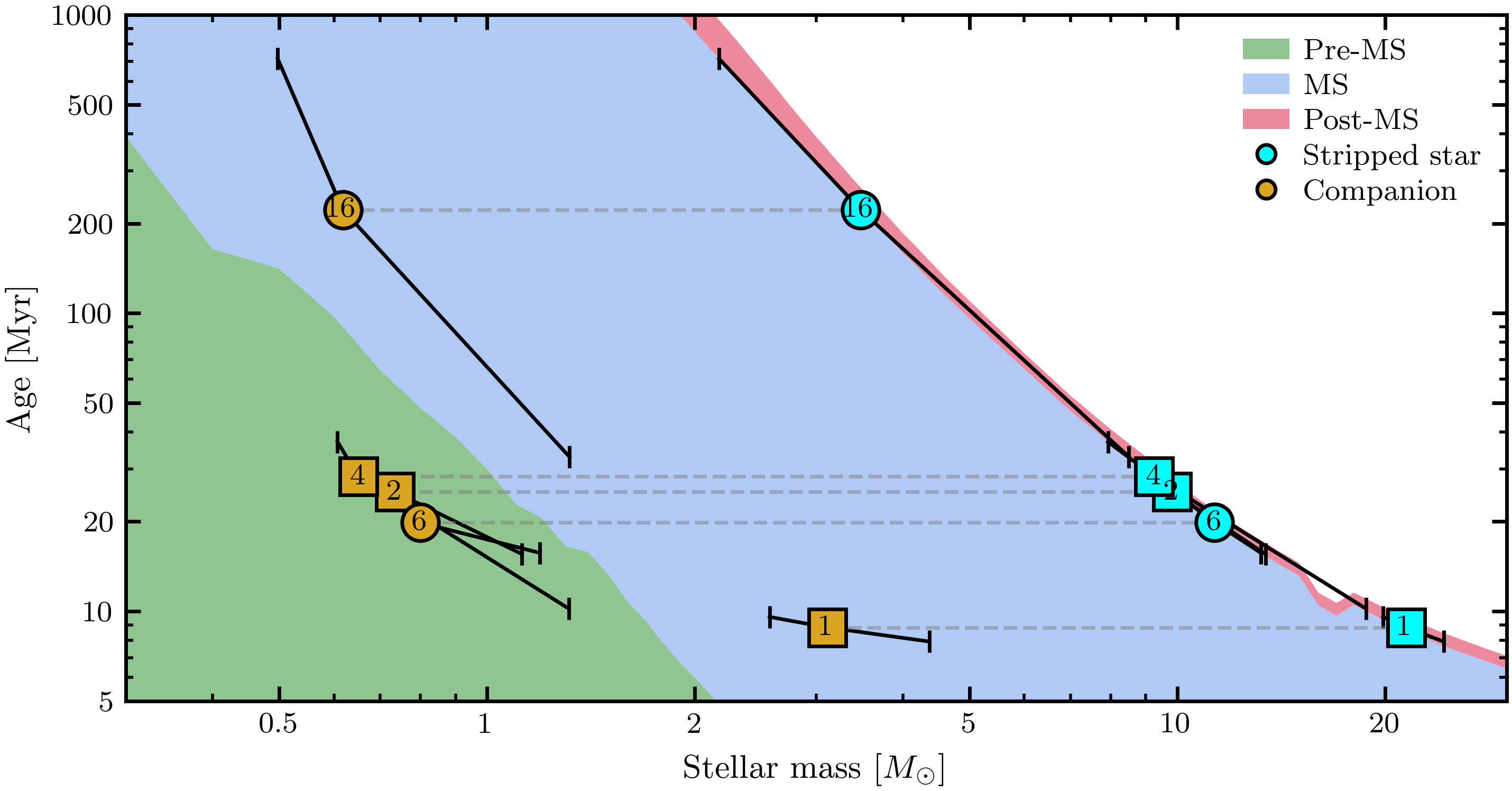}
    \caption{System age constraints for the helium stars in binaries, assuming they orbit stellar companions. We estimate the system age to be the TAMS age of the primary (cyan, squares: SMC, circles: LMC), assuming envelope-stripping after the main sequence (see \secref{sssec:ncmt}). The evolutionary phase of the secondary (orcher) can then be inferred from its mass, assuming it is coeval (green: pre-MS, blue: MS, red: post-MS). Given our system age constraints and companion mass estimates, we infer that stellar companions should be late pre-MS or young MS stars. We visualize evolutionary phases using MIST models at LMC metallicity (which are similar to SMC metallicity). Lastly, error bars are bent due to the logarithmic axis scaling.}
    \label{fig:M-age}
\end{figure*}

In the sections above, we have established that a subset of the sample of intermediate mass helium stars from \citetalias{2023Sci...382.1287D} \emph{for which no luminous companion was visible in the optical spectra} are indeed in binaries. However, given that these companions are optically dark, their nature remains uncertain. Here we will consider the possibility that the binaries in our sample orbit (living) stellar companions, while in Section~\ref{sec:co-comp} we will consider the possibility that they orbit compact objects. We use ‘living’ to refer to active nuclear burning or pre-MS, while excluding WD/NS/black hole (BH).

We emphasize that because the companions are optically dark, this already suggests that any living stellar companions are low mass for their optical flux to be sub-dominant relative to the stripped star. A low-mass stellar companion orbiting a stripped star (massive enough to reach core collapse) would imply that the binary could be a LMXB progenitor \citep[cf.][]{2023hxga.book..120B,2022abn..book.....C}. While LMXBs are primarily thought to form through isolated binary evolution \citep[e.g.][]{paczynski_1976,
2006csxs.book..623T,
2015ApJ...800...17F}, their formation channels remain uncertain \citep{2003MNRAS.341..385P}, and triple evolution pathways have been proposed \citep{2026PASP..138g4204N,2025ApJ...983..115S}. However, it is not guaranteed that a companion will fill its Roche lobe within a Hubble time. Additionally, the explodability of stripped stars, as a function of mass, is non-trivial \citep{2019ApJ...878...49W, 2020ApJ...890...51E, 2021A&A...645A...5S, 2023ApJ...948..111F,2024RNAAS...8..302K}.

Our main goal for this section is to determine if stellar companions are viable and--if so--which types of stellar companions are permitted.
First, in \secref{sssec:ncmt} we highlight an important result: that a stellar companion would
imply that binary envelope stripping was non-conservative. This statement does not rely on any binary evolution assumptions, and is directly inferred from the component masses. 
We will then constrain viable stellar companions in multiple ways: In \secref{sssec:stellar-age}, we determine which evolutionary stages for unperturbed stellar companions (i.e. those that follow single star evolutionary tracks) are consistent with the inferred ages of our helium star binaries. In \secref{sssec:opt-flux}, we derive upper limits on unperturbed stellar companion masses given their apparent lack of optical flux contribution. In \secref{sssec:roche-lobe}, we use our measured orbital properties to determine which types of stellar companion can physically fit within their Roche lobes. Finally, we discuss the possibility/viability of low-mass helium star companions (i.e.\ hot subdwarfs) in \secref{sssec:sbd-comp}.

\subsection{Non-conservative envelope stripping}
\label{sssec:ncmt}

In Figures \ref{fig:m2} and \ref{fig:M1_M2_He}, we observe that our stripped star binaries have present-day mass ratios $q=M_2/M_1\leq1.$ We can estimate the initial primary mass $M_{1,\rm init}$ of the helium star using the (interpolated) initial-stripped star mass relations of \cite{2025A&A...697A.239H}. For Stars 1, 2, 4, 6 and 16, we infer $M_{1,\rm init}\approx22$, 10, 9, 11, and 3.5 $M_\odot,$ respectively, with uncertainties of several $M_\odot$ due to uncertainty in the relation between  current helium star mass and initial mass \citep[e.g. convective boundary mixing:][]{2011A&A...530A.115B,2011A&A...530A.116B}. This implies that the masses of the envelopes that were removed during stripping are large: for Stars 1, 2, 4, 6, and 16, the stripped envelope mass is approximately $M_{1,\rm env}=M_{1,\rm init}-M_1\approx12$, 7, 6, 8, and 2.7 $M_\odot$, respectively. Notably, the inclination averaged companion masses, $\left<M_2\right>_{\rm inc}$, for the binaries in our sample are much less massive than these estimates for $M_{1,\rm env}$. This implies that very little, if any, accretion onto a stellar companion could have taken place\footnote{A similar argument is made by \cite{2026ApJ...999...73L}, but here we are relying on companion masses derived from orbital solutions, rather than photometric fitting.}. However, we caveat that low/zero accretion refers to a minimal fractional mass gained \emph{and} retained by the companion relative to the total mass lost by the stripped star progenitor \citep{2016ApJ...833..108S}. In principle, material could be transferred and subsequently lost \citep{2005A&A...435.1013P}, or lost from a disk \citep{2023MNRAS.519.1409L}, while relatively small amounts of transferred material can have a significant impact on the companion \citep{1981A&A...102...17P,
2021ApJ...923..277R, 2023ApJ...942L..32R,
2026MNRAS.549g1132H}.

The companion mass estimates we derive clearly indicate that stellar companions would imply highly non-conservative envelope stripping. However, whether the envelope stripping was the result of non-conservative stable mass transfer or common envelope evolution is presently unclear. Still, the initial mass ratios we infer are quite extreme: for fully non-conservative mass transfer we infer that the initial mass ratios are \mbox{$q=\left<M_2\right>_{\rm inc}/M_{1,\rm init}\approx0.1$} for all systems. If mass transfer was partially conservative, the initial companion masses would be lower than we measure now, making the initial mass ratios even more extreme. Such extreme mass ratios are usually considered to lead to the development of a common envelope \citep[e.g.][]{2020ApJ...899..132G,2023A&A...669A..45T,2017MNRAS.465.2092P}. The observed luminosity distribution of stripped-envelope SN companions may favour inefficient mass accretion \citep{2026MNRAS.546f2208Z}. However, the large range of orbital periods do not reflect the expected (short) orbital periods from common envelope ejection \citep[cf.][]{2003MNRAS.341..669H,2021ApJ...920...81S}, and more in-depth binary evolution modeling is ongoing \citep{WeiEvol}.

\subsection{System age constraints on evolutionary phase}
\label{sssec:stellar-age}

The age of a binary is informative for constraining the evolutionary stage of a hypothetical stellar companion. Here we investigate the types of stellar companions that would be permitted for the binaries in our sample based on the system age under the following assumptions: (i) the primary and companion stars are coeval (both stars are born and evolve together, dynamical capture and triple evolution are not considered) and (ii) the companion star is relatively unperturbed. Here we define unperturbed to mean that the companion is unaffected by binary interaction and therefore follows standard single star evolutionary tracks. This assumption is motivated by our finding above that the systems in our sample necessarily went through highly non-conservative mass transfer. We will discuss qualitatively below how our results would change if the companions had accreted a small amount of mass. However, we emphasize that the results of this section explicitly assume that the companions are not stripped themselves, which we will instead address in \secref{sssec:sbd-comp}.

We first note that it is very unlikely that any unperturbed stellar companion is a post-MS star. This is because the (inclination averaged) companion masses inferred in Section~\ref{sssec:M2inc} are \emph{significantly} lower than the inferred initial masses for the helium stars in our sample. As such, these companions would have significantly longer MS lifetimes than the primary stars. However, it is also the case that \emph{pre-MS} lifetimes for lower mass stars are significantly longer than high mass stars. It is therefore an open question whether an unperturbed stellar companion would have yet reached the MS by the time the stripped stars in our sample were formed. A similar ambiguity was noted by \cite{2020arXiv201208531N}, who highlight that any binary companion to a red supergiant that is $\lesssim$3 $M_\odot$ would still be a pre-MS star \citep{Neugent2020}. We note that pre-MS stars during the earliest, embedded phase of their evolution can be heavily dust enshrouded \citep[e.g., Class 0/I protostars:][]{2009ApJS..181..321E}. However, in our case, the pre-MS companions are well into their pre-MS lifetimes, such that they should have dispersed their natal envelope/disk \citep{2011ARA&A..49...67W}. Thus, their brightnesses are likely not impacted by dust enshrouding. Here, we use stellar models to infer the likely evolutionary state for stellar companions to the helium stars in our sample. 

For our analysis, we take the age of a helium star binary to be the age of the primary at core-hydrogen exhaustion, i.e. the terminal-age main-sequence (TAMS). This assumes binary interaction occurred as the primary crossed the Hertzsprung gap, and that the helium main-sequence lifetime is short relative to the main-sequence lifetime, both supported by binary evolution theory \citep[e.g.][]{1969AJ.....74.1095V,2018A&A...615A..78G}. We extract ZAMS and TAMS ages for 0.2$-$30 $M_\odot$ stars from MIST models with initial rotation $v/v_{\rm crit}=0.4$ at LMC/SMC metallicities \citep{2011ApJS..192....3P,2015ApJS..220...15P,2013ApJS..208....4P,2016ApJS..222....8D,2016ApJ...823..102C,2026ApJS..283...64D}. The system age is the MIST-interpolated TAMS age corresponding to $M_{1,\rm init}$ (as inferred in Section~\ref{sssec:ncmt} above). The resulting initial masses and system ages are shown as colored cyan points in \figref{fig:M-age}. Stars 1, 2, 4, 6, and 16 have an inferred system age of $\sim$8, 25, 28, 20, and 200 Myr, respectively, with uncertainties of several Myr due to uncertainty in $M_{1,\rm init}.$ Star 16 has particularly large uncertainties because of its low helium star mass.

With this age estimate, in \figref{fig:M-age}, we infer the evolutionary phase of an unperturbed, coeval stellar companion. Using the inclination-marginalized mass $\langle M_2 \rangle_{\rm inc}$, we compare the system age to the companion’s evolutionary timescales. If the system age is shorter than the ZAMS age of $\langle M_2 \rangle_{\rm inc}$, the companion (numbered orcher) would be a pre-MS star (green); if it falls between the ZAMS and TAMS, a MS star (blue). Our age estimates include uncertainties in the helium star mass, and should be considered qualitative. In all cases, uncertainty in the age of the binary from the exact phase during the core-helium burning is very sub-dominant in comparison to the age uncertainty from the helium star mass. The same is true of convective overshoot: age uncertainty due to uncertainty in the primary mass greatly exceeds the $\sim$10\% variation in stellar lifetime between stellar evolution models with and without overshoot \citep[see Fig. 1 in][]{2017A&A...601A..29Z,2020MNRAS.497.4549A}. As such, for the purposes of broadly constraining the evolutionary phase of a stellar companion, the precise single star models we use should not affect our overall conclusions. 

As expected, we find that none of our systems are old enough for the secondary to have evolved beyond the MS. For our inferred $\langle M_2 \rangle_{\rm inc}$, the MS lifetimes of the companions greatly exceed the system ages. As such, it is very unlikely that helium stars orbit post–MS companions. Stars 1 and 16 have inferred system ages implying that a stellar companion of mass $\langle M_2 \rangle_{\rm inc}$ would be a young MS star: less than $\sim$10\% through core-hydrogen burning. Interestingly, Stars 2, 4, and 6 have system ages which imply companions of mass $\langle M_2 \rangle_{\rm inc}$ would be late pre-MS companions. We find that the helium main-sequence lifetime (ignored in the system ages) is not long enough to create ambiguity between late-pre-MS and early-MS since these phases are much longer. 

To summarize, age estimates for the binaries in our sample imply that any relatively unperturbed stellar companion would likely be either a late pre-MS or young MS star.

\subsection{Optical flux constraints}
\label{sssec:opt-flux}

\begin{figure}
    \centering
    \includegraphics[width=\linewidth]{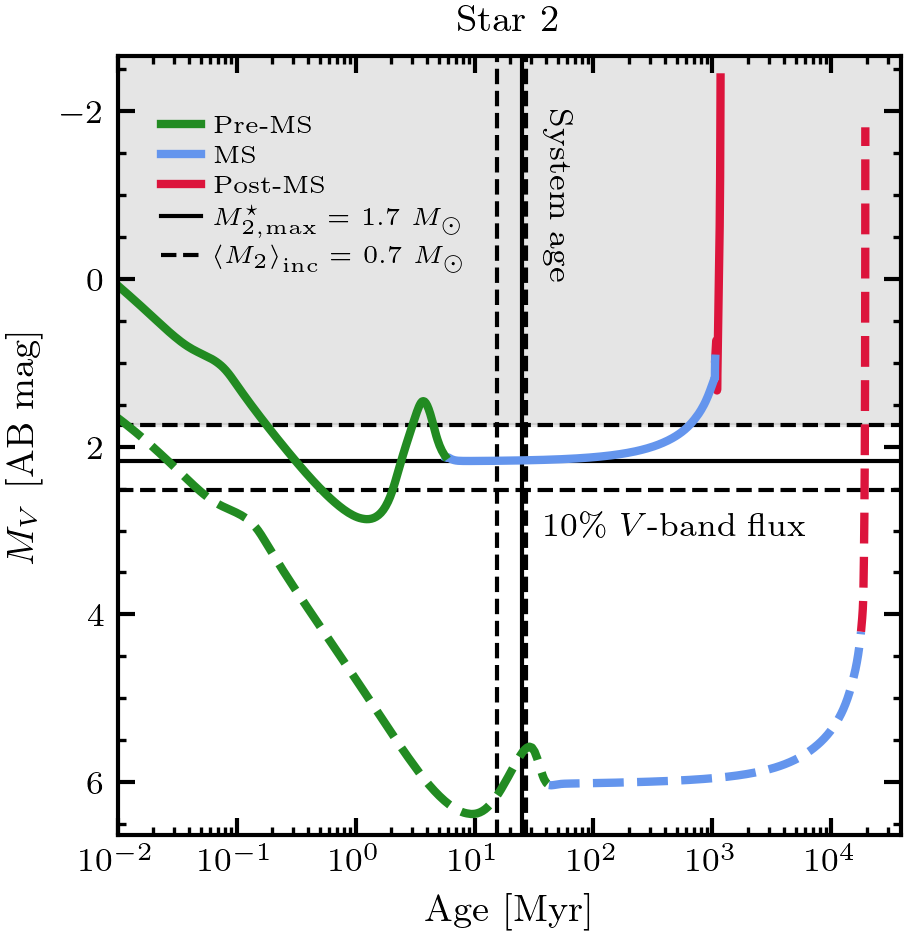}
    \caption{Optical flux contribution of a stellar companion for Star 2 in the V-band, as a function of time. The system age (vertical) and 10\% V-band flux contribution (horizontal) are shown (solid lines), with uncertainties (dashed lines). A stellar companion can be at most a $M_{\rm 2,\rm max}^\star=1.7\,M_\odot$ MS star (solid colored lines), otherwise we would see it in the optical spectrum. On the other hand, a $\left<M_2\right>_{\rm inc}=0.7\,M_\odot$ pre-MS star is faint enough to be optically hidden (dashed colored lines). However, young pre-MS and post-MS stars are too bright. Post-MS tracks terminate at the end of the red giant branch.}
    \label{fig:star2-V}
\end{figure}

The fact that none of the objects in our sample show clear evidence of a stellar companion in their optical spectra allows us to place constraints on the mass of an unperturbed stellar companion under the same assumptions outlined in Section~\ref{sssec:stellar-age}.
As noted in Section~\ref{ssec:sample}, \citetalias{2023Sci...382.1287D} classified most of the objects in our stripped star sample as “Helium star type”. This assessment was based on the equivalent widths of H$\eta$-\HeII~3835 and \HeII~5411. H$\eta$-\HeII~3835 is sensitive to the presence of a MS companion, since short-wavelength Balmer lines should be weak in hot H-poor stars, whereas \HeII\,5411 is sensitive to the helium star, since B-type and cooler MS stars are not hot enough to ionize helium. Specifically, a star was classified as “Helium star type” if EW(H$\eta$-\HeII\,3835) $<1.2$ \AA\, and EW(\HeII\,5411) $>0.2$ \AA. By comparison to a model grid of stripped star plus MS binaries, \citetalias{2023Sci...382.1287D} found that these cuts were consistent with a expectations for a system where a MS companion contributes $<20\%$ of the optical flux. In parallel, \citetalias{2023ApJ...959..125G} performed a few tests investigating the types of MS stars which could be hidden without impacting the quality of their spectral fits (which considered only the contributions of a stripped star). Specifically they showed that removing a 2.2 $M_\odot$ late B-type MS star to the spectrum of Star 6 contributing 10\% (20\%) of the optical flux produced an acceptable (poor) fit to the data. In addition, for Star 5 they found that adding both a 10\% and 20\% optical flux contribution from the same B-star produced poor fits. We note that this assumes both stars are subject to the same attenuation/dust column. 

These results allow us to convert observed V-band magnitudes for our entire sample into approximate upper limits on the mass of an unperturbed stellar companion. We compute the absolute V-band magnitude $M_{\rm V}$ using distances of 49.97 (62.1) kpc for the LMC (SMC) \citep{2013Natur.495...76P, 2020ApJ...904...13G}. We also correct for extinction using $A_{\rm V}$ derived by \citetalias{2023ApJ...959..125G}. After correcting for distance and extinction, we compute the magnitude which corresponds to 10\% of the absolute V-band flux, with associated errors computed by varying distance, extinction and photometric uncertainties. For the following analysis, we assume that a stellar companion cannot be brighter than this limit, otherwise it would be detectable in the optical spectrum. We emphasize that this 10\% limit is conservative, since \citetalias{2023ApJ...959..125G} found that a stellar companion contributing as little as 10\% of the optical flux was (in one of the cases examined) detectable. Future work will examine both spectral fits and UV/IR data to constrain the companion flux contribution on a star-by-star basis, but here we begin this process with this 10\% optical limit.

For a given system, we take the age constraints from Section~\ref{sssec:stellar-age} and retrieve the MIST $M_{\rm V}$ at the system age, as a function of stellar companion mass \citep[derived from theoretical isochrones with bolometric corrections from][]{2016ApJ...823..102C}.  We then interpolate the mass-$M_V$ relation to determine the maximum mass of a stellar companion before it exceeds the 10\% optical flux contribution limit. We obtain associated uncertainties by varying both the system age and the V-band photometry within the upper and lower uncertainties.

We present an example of this V-band flux analysis for Star 2 in \figref{fig:star2-V}, where we show the absolute V-band magnitude of two potential stellar companions as a function of age. We constrain the age of Star 2 to be approximately 25 Myr (vertical line), and a stellar companion cannot be brighter than $\sim$2 mag in the V-band to satisfy our 10\% V-band flux constraint (horizontal line). Given these age and photometry constraints, we find that an unperturbed stellar companion cannot be more massive than $M_{2,\rm max}\approx$ 1.7 $M_\odot.$ However, we note that such a 1.7 $M_\odot$ stellar companion would be young, since it would expand and brighten as it evolves along the MS. Additionally, we observe that young pre-MS and post-MS companions would exceed our 10\% V-band flux constraint by several mags, providing additional support that these evolutionary stages are disfavored. We also plot the optical flux contribution for the most likely companion star mass $\left<M_2\right>_{\rm inc}$ (assuming isotropic inclinations). At our inferred age for the system, an unperturbed stellar companion of mass $\left<M_2\right>_{\rm inc}\approx0.9$ $M_\odot$ is 3 mags dimmer than our 10\% V-band flux constraint: such a stellar companion would be completely optically hidden. In \appref{sapp:opt-flux}, we present analogous figures for our analysis of Stars 1, 4, 6 and 16.

The resulting unperturbed stellar companion upper limits $M_{2,\rm max}^\star$ are presented in \tabref{tab:orbit_props} for each binary helium star in our sample. For Star 1, an unperturbed stellar companion cannot be more massive than $M_{2,\rm max}\approx3.1$ $M_\odot,$ whereas for Stars 2, 4, 6 and 16 unperturbed stellar companions are limited to $M_{2,\rm max}\approx1.5-1.7$ $M_\odot.$ We can compare these stellar companion mass upper limits we just derived to the companion mass constraints from orbital solutions. In all cases, $M_{2,\rm min}<M_{2,\rm max}^\star$. In other words, we find that unperturbed stellar companions are allowed within both our orbital solutions and V-band flux constraints. If we instead consider inclination-averaged companion masses, Stars 2, 4, 6 and 16 also have  $\left<M_2\right>_{\rm inc}<M_{2,\rm max}^\star.$ Thus, given the most likely companion masses $\left<M_2\right>_{\rm inc}$, it is possible that unperturbed stellar companions are present because they would not contribute sufficiently to the total light output to be detectable in the optical spectra. For Star 1, there is a narrow window from $M_2\approx2.9-3.1\,M_\odot$ where a stellar companion is viable.

\subsection{Roche lobe constraints}
\label{sssec:roche-lobe}

\begin{figure}
    \centering
    \includegraphics[width=0.97\linewidth]{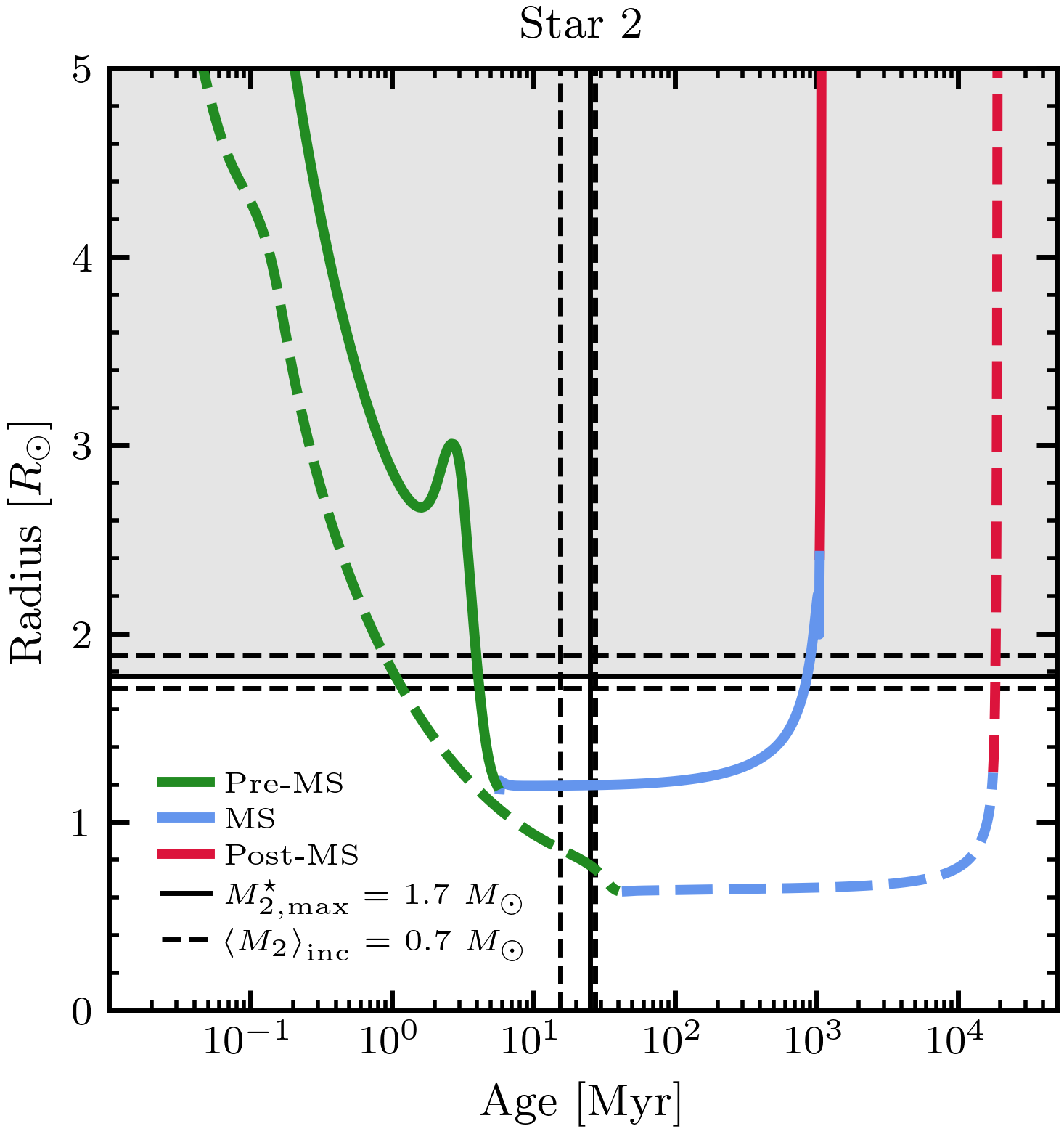}
    \caption{Radius of a stellar companion for Star 2, as a function of time. The system age (vertical) and Roche lobe radius (horizontal) are shown (solid lines), with errors (dashed lines). A $M_{\rm 2,\rm max}^\star=1.7\,M_\odot$ MS star fits within its Roche Lobe, as does a $\left<M_2\right>_{\rm inc}=0.7\,M_\odot$ pre-MS star. However, young pre-MS and post-MS stars would overflow their Roche lobes. Post-MS tracks extend to the end of the red giant branch.}
    \label{fig:star2-R}
\end{figure}

Finally, the Roche lobe radius can provide an additional constraint for a stellar companion, which can be estimated using the 
approximation of \citet{1983ApJ...268..368E}:
\begin{equation}
\frac{R_{L,2}}{a} =
\frac{0.49\, q^{2/3}}
{0.6\, q^{2/3} + \ln\!\left(1 + q^{1/3}\right)},
\end{equation}
where $q = M_2/M_1$ is the binary mass ratio, and $a$ is the orbital separation. Since our stripped star binaries are consistent with being detached post-mass transfer systems, a stellar companion should not overflow its Roche Lobe. However, in all cases we find that the radii of an unperturbed stellar companion with mass $M^\star_{2,\rm max}$ at the system age is within $R_{L,2}.$ In other words, the Roche Lobe constraint does not further restrict the allowed unperturbed stellar companions.

We present an example of our Roche lobe analysis for Star 2 in \figref{fig:star2-R}, where we show the companion radius as a function of age. Star 2 is approximately 25 Myr old (vertical line) and the Roche lobe of the secondary is $R_{L,2}\approx1.8$ $R_\odot.$ Given our age and Roche lobe radius constraint, the most massive unperturbed stellar companion allowed by optical photometry, an early-MS star with $M_{2,\rm max}=$ 1.7 $M_\odot,$ is contained within its Roche lobe while being on the MS, with a filling factor of $\sim$0.6. A late pre-MS star with $\left<M_2\right>_{\rm inc}\approx0.9\,M_\odot$ is also contained within its Roche lobe. Both young pre-MS and post-MS stars would overflow their Roche Lobes, which again disfavors these evolutionary stages. 

In \appref{sapp:roche-lobe}, we present analogous Roche lobe constraints for Stars 1, 4, 6 and 16. In particular, Star 1,  4, 6, and 16 have $R_{L,2}=120$, 1.9, 10, and 4 $R_\odot$, respectively, which for an unperturbed stellar companion with mass $M_{2,\rm max},$ at the system age, would have a Roche lobe filling factor $\sim$0.01, 0.6, 0.1, and 0.3.

\subsection{Hot subdwarf companions}
\label{sssec:sbd-comp}

Finally, we will briefly discuss the possibility that the companions could be low-mass helium stars (i.e.\ hot subdwarfs) instead of relatively unperturbed stars. Hot subdwarfs are core helium-burning objects with typical masses $\sim$0.4$–$0.6 $M_\odot$ \citep{2009A&A...504L..13Z} thought to form through binary interaction \citep[e.g.][]{2002MNRAS.336..449H,2003MNRAS.341..669H}. In principle, subdwarfs could evade detection in the optical due to their low luminosities, $\log_{10}(L/L_\odot)\lesssim3$, relative to intermediate-mass helium stars, $\log_{10}(L/L_\odot)\sim3-5$ \citepalias[see Fig. 8 in][]{2023ApJ...959..125G}. While this scenario is disfavored for Star 1 because of its large companion mass, it is plausible for Stars 2, 4, 6 and 16. A low-mass stripped star companion below $\sim$1 $M_\odot$ would have an optical flux ratios of at most $\sim$0.01--0.1 across our sample, depending on the brightness of the intermediate-mass helium star. As such, they could be optically hidden. Moreover, hot subdwarfs will certainly fit within their Roche lobes since subdwarfs below $\sim$1 $M_\odot$ have effective radii $R_{\rm eff}\lesssim0.4\,R_\odot$ \citep[according to the models of][]{2018A&A...615A..78G}.

The minimum companion masses for Stars 2, 4, 6 and 16 are above the canonical subdwarf mass of 0.47 $M_\odot$ \citep[e.g.][]{2012A&A...539A..12F}. Nevertheless, more massive sdOs are known \citep[e.g.][]{2012MNRAS.427.2180N,2022A&A...666A.182S}, and therefore cannot be ruled out. However, double helium star binaries appear to be quite rare, observationally. Few double hot subdwarf binaries are known: PG 1544+488 \citep{2004A&A...418..275A,2014MNRAS.440.2676S} and HE 0301-3039 \citep{2004Ap&SS.291..351L,2007A&A...462..269S}, while some additional candidates were recently found \citep{2026MNRAS.546f2202J}. Some binary evolution channels for NS-WD mergers predict an intermediate evolutionary phase involving an intermediate-mass + low-mass helium star \citep[e.g.][]{2018A&A...619A..53T}. Similarly, double core common envelope, which produces a double helium star binary, has been proposed as double neutron star (DNS) merger pathway \citep{2018MNRAS.481.4009V,
2020PASA...37...38V}. Such scenarios would likely require a mass-ratio reversal, in which the initially more massive star evolves first to produce the low-mass helium star companion, followed by stripping of the initially lower-mass secondary. It is therefore plausible that the companions are low-mass stripped stars, and this scenario requires further investigation.

\section{Possible binary companions II: \\ Compact objects}
\label{sec:co-comp}

\begin{figure*}
    \centering
    \includegraphics
    [width=\textwidth]{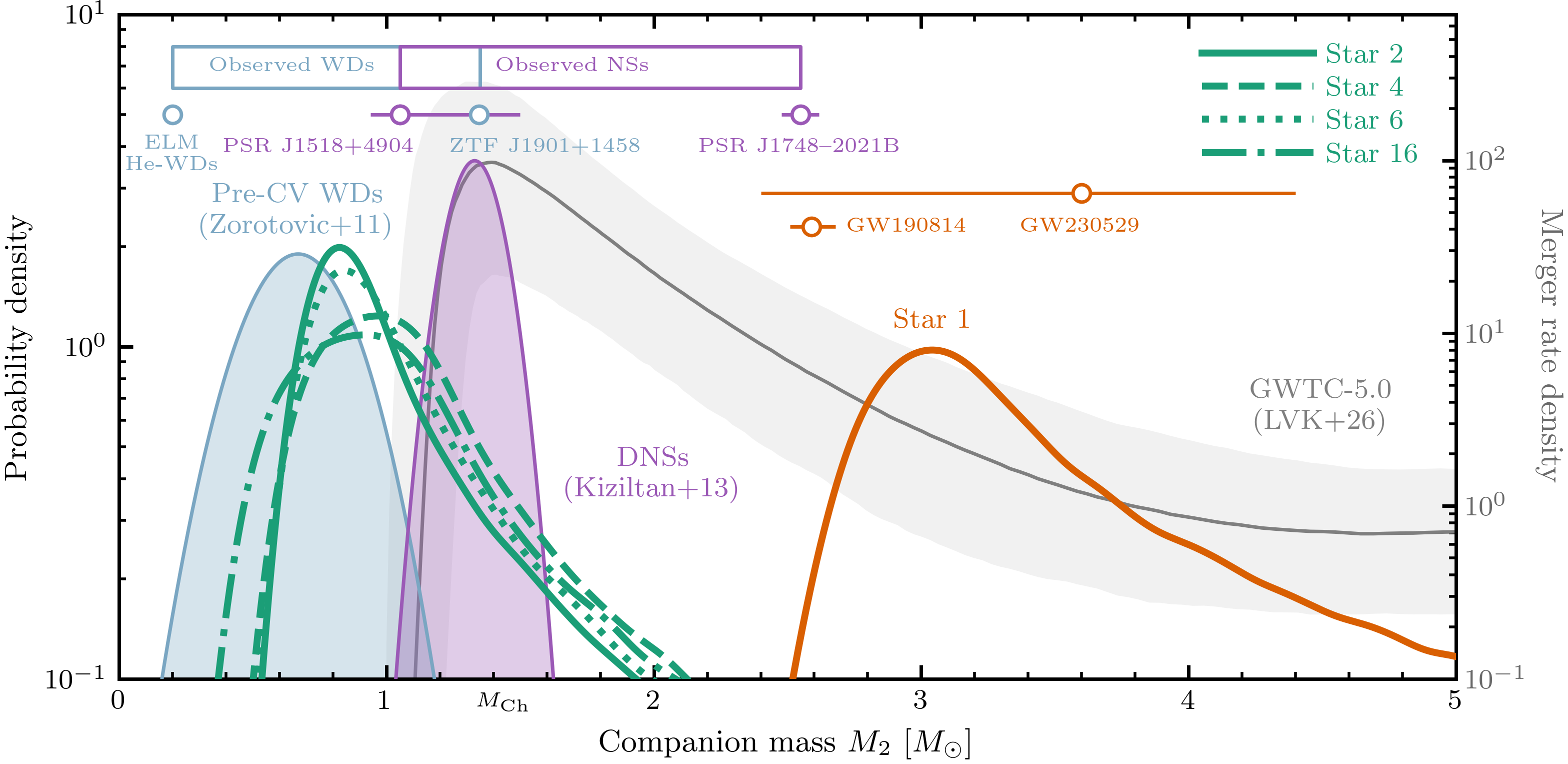}
    \caption{Companion mass distributions, while sampling over isotropic inclination angles ($\left<M_2\right>_{\rm inc}$, see \secref{sssec:M2inc}). We compare the mass distributions for Stars 1, 2, 4, 6 and 16 to: (i) the pre-cataclysmic variable WD mass distribution \citep[pre-CV WD, blue:][]{2011A&A...536A..42Z}, (ii) the double neutron star mass distribution \citep[DNS, purple:][]{2013ApJ...778...66K}, and (iii) the \textsc{FullPop} GW merger rate distribution \citep[GWTC-5.0, grey:][]{2026arXiv260527226T}. We also show observed mass ranges for WDs (blue) and NSs (purple), as well as mass gap objects (orange, see \secref{sec:co-comp}). Stars 2, 4, 6 and 16 could host WDs or NSs. Star 1's companion could lie in the putative lower mass gap between BHs and NS (orange). The Chandrasekhar mass is shown on the x-axis. Note that the GW merger rate axis (right, grey) has been scaled such that amplitude of the NS peak in the GW merger rate distribution matches the DNS distribution.}
    \label{fig:M2_co_prob}
\end{figure*}

While stellar companions to stripped stars are thought to be most common, some stripped stars are expected to orbit compact object companions \citep[e.g.][]{2005MNRAS.361..955B,
2016A&A...596A..58K,
2020ApJ...904...56G,
2021A&A...645A...5S,
2021ApJ...920L..36J,
2022MNRAS.511.3951F,
2024A&A...688A..87W}. For example, \cite{2017ApJ...842..125Z} estimated that $\sim$7\% of a stripped star population should have compact object companions, at the metallicities of the Magellanic Clouds. The UV excess detection method, used to identify the stripped star candidates in Swift-UVOT imagesc \citep{2026ApJ...999...73L}, was found to be particularly efficient at identifying this subset of the population, despite stripped star + compact object binaries being intrinsically rarer than stripped star + MS binaries \citep{2026PASP..138b4202B}. Compact object companions could therefore be overly represented in the sample of \citetalias{2023Sci...382.1287D}, relative to theoretical expectations. If a stripped star orbits a compact object, the binary is a direct progenitor of a double compact object binary. We will now consider the possibility that the optically dark companions for the confirmed binaries in our sample are compact objects. 

\subsection{Comparison to compact object binary mass distributions}\label{sssec:co-dist}

In \figref{fig:M2_co_prob} we present the mass distributions of $\left<M_2\right>_{\rm inc}$, obtained in \secref{ssec:m2}, for our five helium star binaries. We contextualize these distributions with three compact object mass distributions: 
\begin{enumerate}
    \item The mass distribution of WDs in short period binaries ($P_{\rm orb}\lesssim1$ d) with low-mass MS companions ($M_2\sim$ 0.1--1 $M_\odot$), with a mean WD mass of 0.67 $M_\odot$ and a standard deviation of 0.21 $M_\odot$ \citep{2011A&A...536A..42Z}. These are considered to be pre-cataclysmic varialble (CV) systems.\footnote{This comparison is not straightforward: a $\sim$0.7 $M_\odot$ WD did not necessarily come from a low-mass progenitor, since binary stripping of a more massive ($\sim$5--6 $M_\odot$) star can produce a WD of similar mass. Thus, a comparison of pre-CV WDs to WDs in binaries with stripped stars is complex. We nevertheless use this distribution as our fiducial comparison in the absence of a better-motivated, well-constrained alternative.}
    \item The neutron star mass distribution for double neutron star binaries \citep[DNS:][]{2013ApJ...778...66K}, with a mean mass of 1.33 $M_\odot$ and a (1.21, 1.43) $M_\odot$ 68\% confidence interval.
    \item The GW merger rate distribution as function of primary mass from the Gravitational-Wave Transient Catalog 5.0 \citep[GWTC-5.0:][]{2026arXiv260527226T}, which includes: NSs (first peak at $\sim$1.4 $M_\odot$), BHs (second peak at $\sim$10 $M_\odot$, not shown), and ambiguous objects (between NSs and BHs). Specifically, we visualize \textsc{FullPop}, a parametric hierarchical Bayesian model which accounts for all compact binary types. 
\end{enumerate}

The companion mass distributions for Stars 2, 4, 6 and 16 appear to be qualitatively consistent with both pre-CV WDs and DNSs. In contrast, the companion mass distribution of Star 1 is well above the DNS mass distribution and GWTC-5.0 NS peak, yet also below the GWTC-5.0 BH peak: the so-called “NS/BH mass gap,” which we will subsequently address.

In the following sections we discuss WD, NS and BH companions, making use of the following integral:
\begin{equation}\label{eq:int}
    P(X) = \int_{x_{\rm min}}^{\rm x_{\rm max}}
    P(M_2)\,dM_2,
\end{equation}
which estimates the probability that a companion mass lies between $[x_{\rm min},x_{\rm max}]$ for some compact object type $X$ given a companion mass distribution $P(M_2)$. These probabilities assume an isotropic inclination distribution and uniform mass ranges for compact object classes. We assume uniform distributions across the full observed compact object mass ranges to be as agnostic as possible, since a non-uniform distribution would decrease $P(X)$. We note that these probabilities assume that the companion is a compact object, since the purpose of this section is to consider this possibility. 

\subsection{White dwarf companions}
\label{ssec:wd}

To estimate the probability of WD companions, we evaluate \eqref{eq:int} over the mass range $0.2\leq M_2/M_\odot\leq 1.44$. The lower bound is motivated by extremely low-mass (ELM) helium WDs with $M_{\rm WD}\gtrsim0.2\,M_\odot$ \citep[e.g.][]{2016A&A...595A..35I}, while the upper bound corresponds to the Chandrashekar mass $M_{\rm Ch}$ \citep{1931ApJ....74...81C}. The most massive WD known, ZTF J1901+1458, with $M_{\rm WD}=1.327-1.365\,M_\odot$ \citep{2021Natur.595...39C} is below $M_{\rm Ch}$, but here we rely on the theoretical upper limit. This WD mass range is displayed in \figref{fig:M2_co_prob}. Stars 2, 4, 6 and 16 have $P(\rm WD)=0.81,0.76,0.79$ and 0.75, respectively. With this assumed WD mass range, WD companions are viable for Stars 2, 4, 6 and 16. A WD companion is ruled out for Star 1, with $P(\rm WD)=0$.

We can alternatively compare the companion masses to the pre-CV WD mass distribution for Stars 2, 4, 6 and 16. If these helium stars orbit WDs, they are more massive than the typical WD found in a low-mass MS+WD binary. This points to more massive WD progenitors, and thus to a formation channel distinct from standard low-mass star evolution.

The WD + stripped star scenario is intriguing for Stars 2, 4 and 6, since the helium stars are quite massive: $\sim3\,M_\odot$. While studying the formation of WD+NS binaries, \cite{2018A&A...619A..53T} identified the “(semi-)reversed WDNS” channels. In these scenarios, before the system has evolved to become a WD+NS binary, there is an intermediate evolutionary phase after two envelope-stripping episodes where the system becomes a WD + stripped star binary. \cite{2018A&A...619A..53T} found that a 6.5+5.9 (8.5+6.5) $M_\odot$ binary can evolve to become a $0.8\,M_\odot$ WD + $2.7\,M_\odot$ stripped star ($1.1\,M_\odot$ WD + $3.4\,M_\odot$ stripped star) in a $\sim$2 (1) day orbit, which after further binary interaction becomes a WD+NS binary. The WD + stripped star properties \cite[masses and periods of][]{2018A&A...619A..53T} are reasonably consistent with Stars 2, 4 and 6. To our knowledge, these are the only isolated binary channels capable of producing an intermediate-mass stripped star orbiting a WD. If Stars 2, 4 and 6 host WD companions, they could be systems in this intermediate evolutionary phase. We note that the WD + stripped star evolutionary pathway is very similar to the subdwarf + stripped star evolutionary scenario of \cite{2018A&A...619A..53T} (see \secref{sssec:sbd-comp}). It may even be possible for a WD + stripped star phase to follow a double stripped star phase in the same binary evolution channel, depending on the initial separation and mass ratio \citep[a NS + stripped star binary is also possible, see Figures 3 and 4 of][]{2018A&A...619A..53T}.

For Star 16, the WD + stripped star scenario is more straightforward. Given that Star 16 contains a lower mass $\sim0.7\,M_\odot$ helium star in a relatively tight $P_{\rm orb}\approx2$ day binary, such a system could originate from one of the subdwarf formation channels \citep[e.g.][]{2002MNRAS.336..449H, 2003MNRAS.341..669H}. In particular, common envelope ejection can lead to a short period binary where the secondary becomes a hot subdwarf, when the primary is already a white dwarf. The 2 day period of Star 16 is compatible with the inferred period distribution of subdwarf + WD binaries from the CE channel \citep[Fig. 7 in][]{2003MNRAS.341..669H}. Additionally, subdwarfs formed through this channel may be preferentially more massive than the canonical subdwarf mass of 0.46 $M_\odot,$ reaching subdwarf masses as high as $\sim$0.7 $M_\odot$ \citep[Fig. 12 in][]{2003MNRAS.341..669H}. 

\subsection{Neutron star companions}
\label{ssec:ns}

To estimate the probability of NS companions, we evaluate \eqref{eq:int} over the mass range $1\leq M_2/M_\odot\leq 3$. The lower bound is motivated by the NS companion in PSR J1518+4904, the lightest NS in the sample of \cite{2013ApJ...778...66K}, originally estimated by \cite{1999ApJ...512..288T} to have $M_{\rm NS}=1.05^{+0.45}_{-0.11}\,M_\odot.$ The upper bound is motivated by the theoretical upper limit for the neutron star mass \citep{1996ApJ...470L..61K}, while the most massive observed NS is PSR J1748--2021B \citep{2008ApJ...675..670F}, estimated by \cite{cliffordthesis} to have $M_{\rm NS}=2.548^{+0.047}_{
-0.078}\,M_\odot.$ This NS mass range is displayed at the top of \figref{fig:M2_co_prob}. Stars 1, 2, 4, 6 and 16 have $P(\rm NS)=0.21,0.34,0.52,0.40$ and 0.44, respectively. With this assumed NS mass range, NS companions are viable for Stars 2, 4, 6 and 16. Star 1 is unlikely to host a NS companion, unless the NS is much more massive than any observed NS. 

With respect the DNSs, the companion mass distributions of Stars 2, 4, 6 and 16 peak below the DNS mass range, while their high inclination tails overlap. As such, the inferred probabilities for DNS-like masses are small. For Star 1, the minimum companion mass is well above the DNS mass range. In other words, if the helium star binaries host NSs, their masses may be inconsistent with the observed DNS mass distribution \citep[e.g.][]{2017ApJ...846..170T}\footnote{However, we refer the reader to \cite{LudwigStar3} which examines Star 3 from the stripped star sample of \citetalias{2023Sci...382.1287D}.}. While Stars 2, 4 and 6 contain stripped stars that could be massive enough to reach core collapse and form NSs \citep{2015MNRAS.451.2123T}, the companion masses are low relative to NSs in DNS binaries, but they could be observed at low inclination angles. Star 16 is ruled out as a DNS progenitor due to its low helium star mass. Finally, if Star 1 contains a NS, it would be exceptionally massive relative to NSs in DNSs since $M_{2,\rm min}\sim2.9\,M_\odot$.

\subsection{Black hole companions}
\label{ssec:bh}

To estimate the probability of BH companions, we evaluate \eqref{eq:int} over the mass range $3\leq M_2/M_\odot\leq \infty$. We adopt this conservative threshold, rather than the mass of any observed system, to handle the (uncertain) transition between NSs and BHs, which we will discuss below. Stars 1, 2, 4, 6 and 16 have $P(\rm BH)=0.56,0.03,0.04,0.03$ and 0.04, respectively. With this assumed BH mass range, a BH companion is favoured for Star 1, while BH companions are effectively ruled out for Stars 2, 4, 6 and 16.

If Star 1 contains a BH, it is interesting to interpret its evolution within the context of the standard double compact object formation scenario \citep[e.g.][]{2018MNRAS.481.1908K}. In this scenario, the primary is stripped of its hydrogen envelope by the secondary, creating a helium star which would subsequently core collapse and form the BH. The secondary would then lose its hydrogen envelope in a common envelope phase, forming a BH + stripped star binary. It is unclear whether the relatively wide orbital period of Star 1 is consistent with this evolutionary pathway, as the common envelope phase is expected to significantly shrink the orbit. However, this channel is invoked to produce double compact object binaries that merge via GW-emission. In addition, we emphasize that wide, detached binaries containing a BH have been discovered, such as Gaia BH1/2 \citep{2023MNRAS.518.1057E,2023MNRAS.521.4323E}. It is worth considering whether the second envelope-stripping phase could have instead been stable mass transfer \citep[e.g.][]{2017MNRAS.471.4256V,2022ApJ...940..184V}, which need not shrink the orbit as substantially as CEE (although the orbital evolution is sensitive to the mass ratio). In either case, its orbital properties could provide empirical constraints on the natal kick imparted at BH formation \citep[e.g.][]{2025A&A...701L...3V,
2025PASP..137c4203N,
2026arXiv260910694L}.

\paragraph{Mass gap} Finally, we consider the putative mass-gap between NSs and BHs, from $\sim$3--5 $M_\odot$ (sometimes referred to as the “lower” mass gap). Mounting observational evidence from GW mergers \citep{2026arXiv260527226T}, X-ray binaries \citep{2012ApJ...757...36K} and microlensing \citep{2020A&A...636A..20W} have called into question the existence of a lower mass gap. To estimate the probability of a mass gap companion, we evaluate \eqref{eq:int} over the mass range $3\leq M_2/M_\odot\leq 5$, where we adopt the NS upper bound and BH lower bounds, as above. Stars 1, 2, 4, 6 and 16 have $P(\rm gap)=0.56,0.08,0.10,0.09$ and 0.10, respectively. With this assumed range, a mass gap companion is favored for Star 1, and strongly disfavored for Stars 2, 4, 6 and 16. Thus, if Star 1 contains a compact object, it is most likely in the lower mass gap. This would place it among a relatively small but growing number of systems occupying this uncertain regime. We highlight GW190814 \citep{2020ApJ...896L..44A} and GW230529 \citep{2024ApJ...970L..34A} which are GW merger events with one merger component in the mass gap: 2.50-2.67 $M_\odot$ for GW190814 and 2.5-4.5 $M_\odot$ for GW230529 (included in \figref{fig:M2_co_prob}). \cite{2026arXiv260212327S} demonstrated that asymmetric mass ratio mergers, like GW190814, may share an evolutionary pathway with asymmetric high mass X-ray binaries (HMXBs). Specifically, they outlined an evolutionary pathway involving a conservative first mass transfer phase, a large natal kick imparted to the first-born (lower mass) compact object, and a subsequent CE phase. \cite{2026arXiv260212327S} identified \mbox{HMXB 4U 1700-37}/ HD 153919 \citep{2015A&A...577A.130F} and GX 301-2 \citep{1995A&A...300..446K} as Galactic examples of such highly asymmetric HMXBs. The work of \cite{2026arXiv260212327S} highlights the importance of explaining anomalous GW mergers by identifying progenitor systems, but whether or not Star 1 is related to this evolutionary pathway, and whether it could form a GW source, is beyond the scope of this work.

If, as recent works suggest, the BH/NS mass gap is a consequence of observational biases as opposed to an intrinsic phenomenon, Star 1's companion may correspond to the low mass end of the BH distribution. Its inferred companion mass merits further study, and we will present additional observational constraints to discern the true nature of the companion beyond those presented here in follow-up work. This will be especially important for determining the mass gap progenitor, which would likely be a stripped star. For example, \cite{2021ApJ...916L...5V} found that compact object remnants of intermediate-mass stripped stars could populate the lower mass gap. The potential for infrared data to better constrain the absence of a low-mass MS companion orbiting Star 1, which would imply a compact object companion, is an intriguing possibility. 

\section{On the nature of intermediate-mass helium stars with non-detections of binarity}
\label{sec:singles}

We now turn our attention to possible explanations for the intermediate-mass helium stars without significant evidence for binary motion in Section~\ref{sec:binary_motion}. First, in \secref{sssec:bimod}, we will present indications that the non-detections are unlikely to arise from small RV shifts due to being viewed at low inclinations. We therefore consider evolutionary scenarios which could lead to apparently single intermediate-mass helium stars: wide orbits in \secref{ssec:single_wide}, runaway stripped stars in \secref{ssec:single_run}, or merger products in \secref{ssec:single_merge}.

\subsection{Surface composition suggests distinct sub-groups rather than inclination effects}\label{sssec:bimod}

\begin{figure}[t!]
    \centering
    \includegraphics
    [width=\columnwidth]{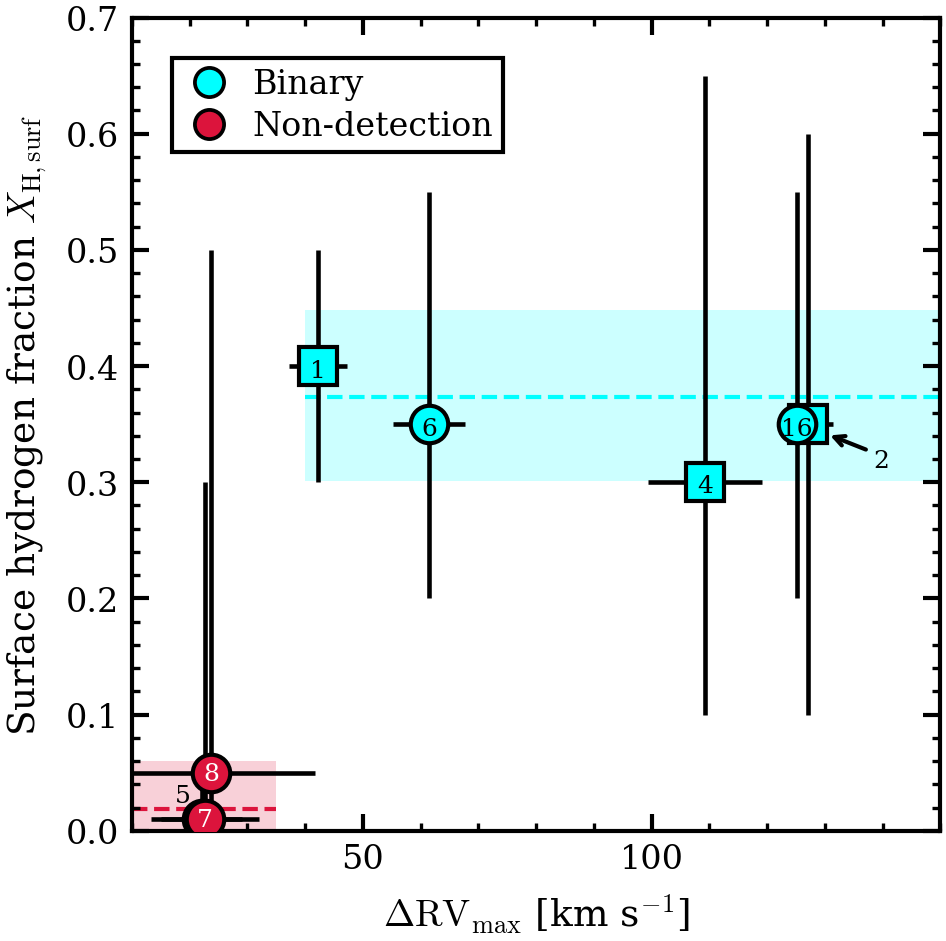}
    \caption{Surface hydrogen mass fractions \citepalias[$X_{\rm H,surf}$ from][]{2023ApJ...959..125G}, as a function of \dRVm. We observe a surprising bimodality: binary stripped stars (large RV shifts) are H-poor, whereas the helium stars with non-detections of binary motion (small RV shifts) are H-free. The error-weighted mean (standard deviation) for $X_{\rm H,surf}$ of the non-detections and binaries are displayed as dashed lines (filled regions).}
    \label{fig:bimod}
\end{figure}

The lower \dRVm~found in the three helium stars with non-detections of binary motion could---in principle---be due to inclination. However, we currently disfavor the interpretation that Stars 5, 7 and 8 have similar companions and orbital properties to the confirmed binaries in our sample but are simply viewed at low inclination. This is because of a stark difference in surface composition observed between the two groups: helium stars in confirmed binaries are H-poor ($X_{\rm H, surf}<0.4$), whereas the non-detections are consistent with being H-free ($X_{\rm H, surf}<0.05$). We present this bimodality in \figref{fig:bimod}, which shows the surface hydrogen mass fractions $X_{\rm H, surf}$ measured by \citetalias{2023ApJ...959..125G} as a function of \dRVm. While there are non-negligible individual system error bars, the error-weighted means for the $X_{\rm H, surf}$ for stripped stars in binaries and non-detections are very much inconsistent. Additionally, the CIII/IV multiplet near 4650 \AA~is visible in the spectra of the non-detections (see \appref{sapp:rv_stack}). This suggests their CNO surface abundances differ from the helium stars in binaries, for which the CIII/IV multiplet is absent.

There are two implications of this observed bimodality. First, the surface composition of the helium stars in binaries match theoretical expectations for the amount of leftover hydrogen from binary stripped star evolutionary models \citep[$X_{\rm H, surf}$$\sim$0.3-0.4:][]{2018A&A...615A..78G}. This implies that the core-envelope boundary for $M_{\rm 1,init}\sim8-25\,M_\odot$ stars is located inside the region where a strong chemical gradient develops, such that envelope stripping removes most, but not all, of the hydrogen. Second, this bimodality suggests that the stars in our sample with non-detections of binary motion may have a distinct origin from those in confirmed binaries. This is simply because surface composition should not depend on orbital inclination. It is therefore unlikely that the small RV shifts observed in Stars 5, 7 and 8 are due solely to inclination effects. As such, we will now consider other pathways to produce apparently single helium stars.

\subsection{The wide orbit scenario}\label{ssec:single_wide}

\begin{figure}[t!]
    \centering
    \includegraphics
    [width=\columnwidth]{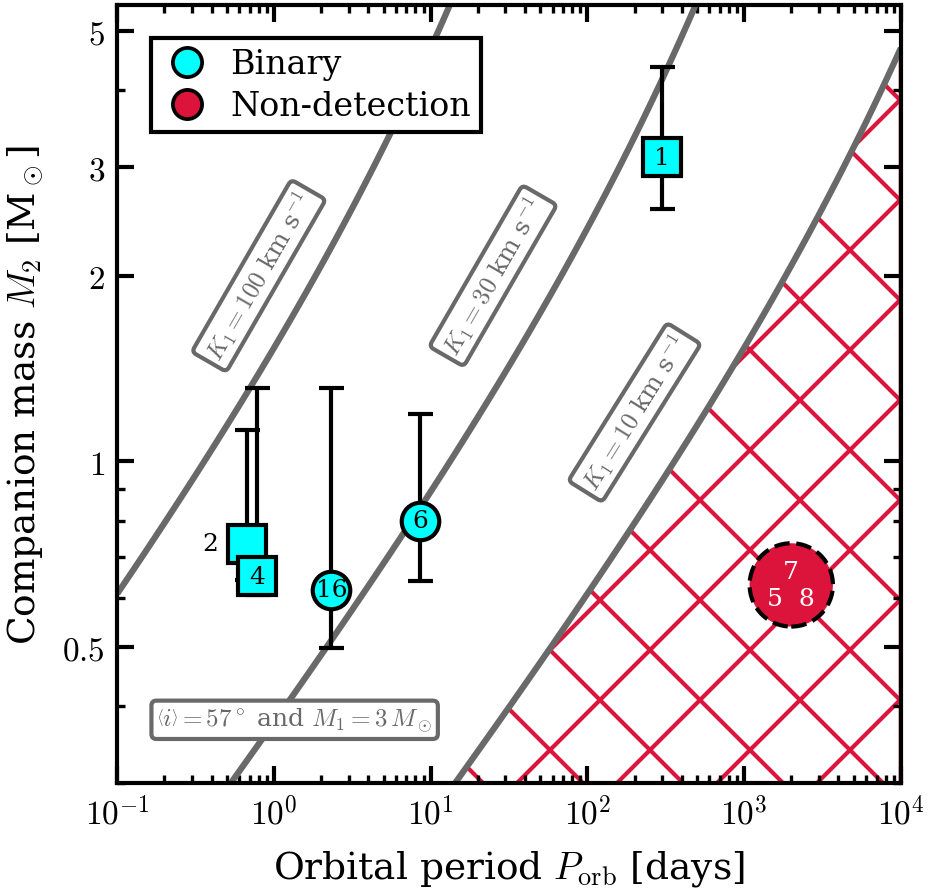}
    \caption{Helium star RV semi-amplitude $K_1$ as a function of orbital period $P_{\rm orb}$ and companion mass $M_2$ for the non-detections. We denote lines of constant $K_1=100,$ 30 and 10 km s$^{-1}$, assuming $M_1=3\,M_\odot$, $\left<i\right>=57^\circ$ and circular orbits. In comparison to binary stripped stars which display large RV shifts ($P_{\rm orb}$ and $\left<M_2\right>_{\rm inc}$, cyan), the non-detections (red) would have to be on wide orbits for our RV survey to not detect binary motion (hatched region).}
    \label{fig:non-detections}
\end{figure}

Lack of evidence for binary motion does not rule out that the non-detections are in binaries. As discussed above, we disfavor inclination effects due to surface composition. However, it is conceivable that the non-detections are in wide, undetectable  orbits. This would manifest as small RV shifts which could fall below our binary detection thresholds. In \figref{fig:non-detections}, we present a rough constraint on the viable binary parameter space which the non-detections could occupy yet fail our binary detection thresholds. Stars 5, 7 and 8 have helium star masses from $2\lesssim M_1/M_\odot\lesssim4$ (see \tabref{tab:orbit_props}). We therefore adopt a representative mass of $M_1=3\,M_\odot$ for the helium star, and calculate the RV semi-amplitude from the binary mass function (\eqref{eq:bmf}) for a range of $P_{\rm orb}$ and $M_2,$ at an average inclination angle of $\left<i\right>=57^\circ$ and $e=0.$ In order to exhibit shifts of $\dRVm\leq 20$ km s$^{-1}$ (i.e. $K_1\leq 10$ km s$^{-1}$) while in a binary, a 0.5 (1, 2, 3) $M_\odot$ companion would require $P_{\rm orb}\gtrsim50$ (500, 1000, 5000) days. Thus, there does exist a viable region of orbital parameter space for the non-detections that our RV survey is not sensitive to. 

Stripped stars in wide orbital periods could perhaps be consistent with a binary-evolution channel where mass transfer was stable and non-conservative. Depending on how angular momentum is lost from the system, orbital widening could potentially push the system below the range of detectable RV shifts, leaving a stripped star with a distant and optically faint companion (either a low mass star or a compact object). For analytic calculations of orbital widening as a function of MT conservation and angular momentum loss, see Sec. 3 of \cite{2019A&A...624A..66R}. However, it is unclear whether this scenario would naturally explain the low surface hydrogen mass fractions observed for this subset of the sample compared to those in confirmed binaries. Indeed, \citetalias{2023ApJ...959..125G} hypothesized that it is conceivable common envelope is more efficient at stripping a star, which would loosely imply that H-free stripped stars should be in short orbits ($P_{\rm orb}\lesssim1$ d). If the H-free stripped stars are in fact in wide orbits, the opposite would be true. With the current sample of orbital properties at our disposal, we cannot (yet) establish a relation between the amount of leftover hydrogen and the envelope-stripping mechanism. Lastly, if Roche lobe overflow was responsible for the envelope stripping, the orbits cannot be arbitrarily large. While \cite{2012Sci...337..444S} found that massive binaries with initial periods larger than $\sim$1200 days should not interact, an analysis mapping initial to final orbital periods after binary interaction is required to comment on this limit, quantitatively. 

\subsection{The walk/runaway scenario}\label{ssec:single_run}

Alternatively, truly single stripped stars can be produced through binary disruption \citep{1994A&A...290..119P}. In this scenario, it is thought that the initially more massive primary undergoes conservative Case A mass transfer. The central convective region, and therefore also the final helium core, is reduced and the core-hydrogen burning is slowed down. At the same time, efficient mass accretion can speed up the evolution of the secondary and in some cases the core-collapse order is reversed (i.e. the secondary star explodes first). Depending on the total mass lost during the resulting supernova and any natal kicks, the primary star (a stripped star at the time of explosion) will sometimes be ejected. \cite{2019A&A...624A..66R} performed extensive numerical studies for the evolution of massive runaway stars. $\sim$4\% of their simulated population were runaway stripped stars from disrupted binaries, with typical ejection velocities of $v_{\rm dis}\gtrsim 60$ \kms.

The production of single intermediate-mass stripped stars via the binary disruption scenario is typically thought to require efficient mass accretion in order to reverse the evolutionary order such that the secondary explodes first. This corresponds to a binary-evolution channel where mass transfer was stable and (at least partially) conservative. However, as above, further work is needed to understand whether such a channel could preferentially lead to a population of stripped stars that have very low surface hydrogen mass fractions. Future work analyzing the motions of the helium stars relative to their local environments with a joint population synthesis and galactic dynamics code such as \texttt{cogsworth} could help constrain the runaway scenario \citep{2025ApJS..276...16W}.

\subsection{The merger scenario}
\label{ssec:single_merge}

\begin{figure}[t!]
    \centering
    \includegraphics[width=\columnwidth]{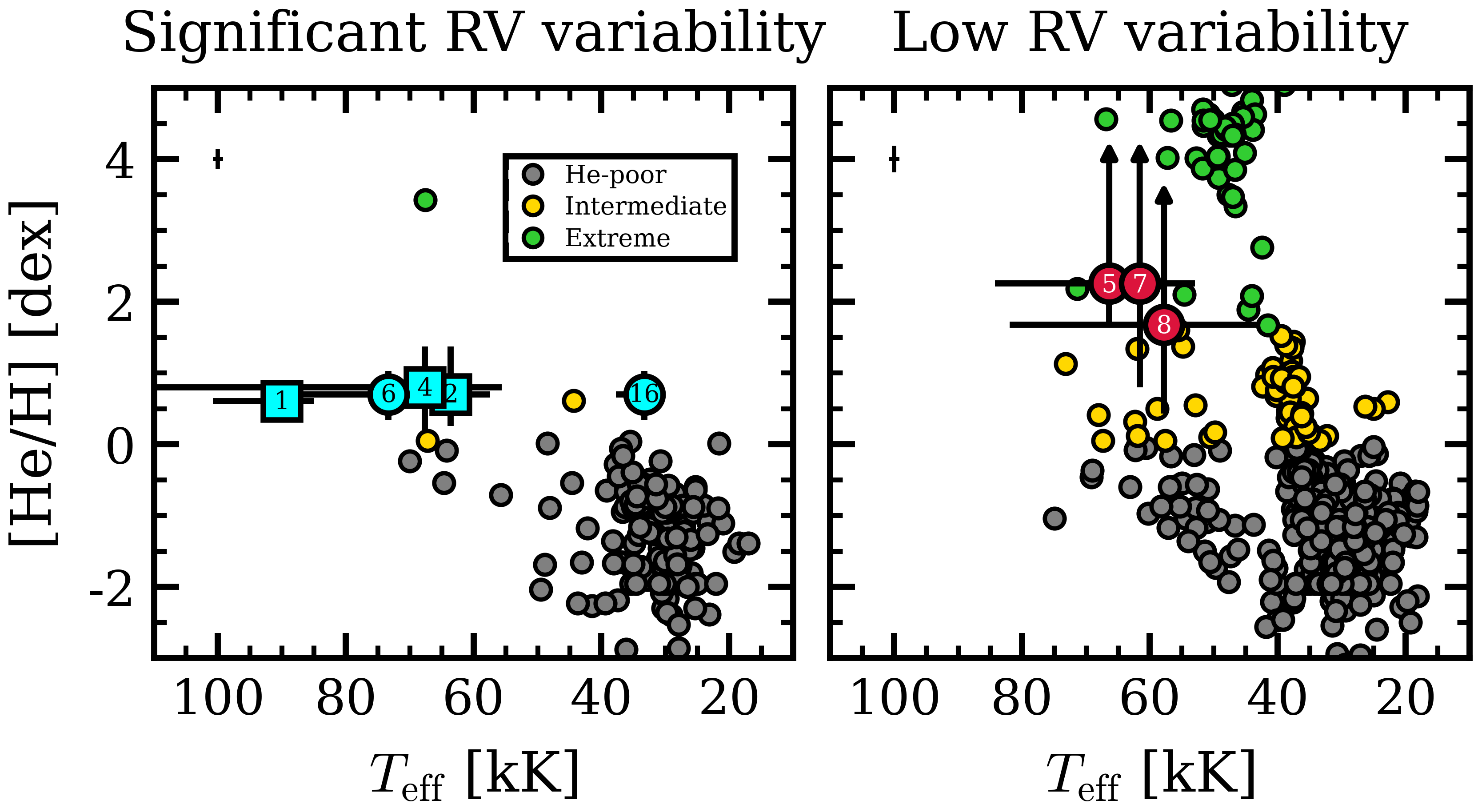}
    \caption{Helium stars in comparison to the hot subdwarf RV variability study of \cite{2022A&A...661A.113G}. Subdwarfs with significant RV variability (left column) are compared to the binary helium stars (numbered cyan), whereas subdwarfs with low RV variability (right column) are compared to the non-detections (numbered red). We present He abundance [He/H] as a function of surface temperature $T_{\rm eff}$. Grey, yellow, green colors for the hot subdwarfs correspond to poor, intermediate, extreme He abundance, respectively. Typical error bars are shown in the upper left of each panel.}
    \label{fig:sd-var}
\end{figure}

An additional evolutionary scenario we consider to produce a single helium star is a stellar merger. Formation of subdwarfs via mergers is thought to be common, in addition to binary stripping \citep[e.g.][]{2020A&A...642A.180P}. Population synthesis predicts that the fraction of subdwarfs from merger channels can range from $\sim$10--50\%, depending on environment and binary evolution assumptions \citep{2003MNRAS.341..669H}. Several hot subdwarf RV surveys indicate that many He-sdO/Bs and sdBs do not exhibit RV variability \citep[e.g.][]{2001MNRAS.326.1391M,
2004Ap&SS.291..321N,
2011MNRAS.415.1381C,
2015MNRAS.450.3514K}. As a result, several merger channels have been proposed \citep[e.g.][]{1984ApJ...277..355W,
1998AJ....116.1308S,
2008ApJ...687L..99P,
2011MNRAS.410..984J,
2011ApJ...733L..42C,
2020arXiv200700019K,
2022MNRAS.511L..66W,
2022MNRAS.511L..60M}. Moreover, fast rotation \citep{2008ApJ...687L..99P,
2026A&A...710A..99P} and magnetic fields \citep{2013ApJ...773..136J,
2015ApJ...806L...1Z} have been discovered in some subdwarfs as potential evidence for mergers \citep[e.g.][]{2011ApJ...733L..42C,
2011ApJ...733L..13G,
2013A&A...557A.122G,
2012A&A...543A.149G,
2019Natur.569..684G,
2022A&A...658L...9D,
2024A&A...691A.165D,
2024A&A...691A.179P}. R Coronae Borealis stars, which are hydrogen-deficient, carbon-rich, supergiants, with irregular variability, have also been proposed as WD merger products \citep[e.g.][]{2012MNRAS.427..190S, 2014MNRAS.445..660Z,
2012JAVSO..40..539C}.

Notably, when examining the surface properties of a sample of hot subdwarfs, \cite{2022A&A...661A.113G} found that He-rich subdwarfs almost exclusively show negligible RV variability. They use this to conclude that a merger origin is “very likely" for these systems. Quantitatively,  \cite{2022A&A...661A.113G} divide their sample of hot subdwarfs into three groups based on surface helium to hydrogen number ratio: He-poor sdOB/Os ($\log n(\text{He})/n(\text{H})\leq-1.0$), intermediate He-rich iHe-sdOB/Os ($-1.0\leq\log n(\text{He})/n(\text{H})\leq0.6$), and extremely helium-rich eHe-sdOB/Os ($\log n(\text{He})/n(\text{H})>0.6$). 
By these definitions, the intermediate-mass helium stars in our sample that are confirmed binaries would be categorized as intermediate He-rich while the non-detections have best-fit surface properties more consistent with the extremely He-rich regime.
 
In \figref{fig:sd-var}, we compare [He/H], the logarithmic ratio of helium to hydrogen relative to solar, as a function of effective temperature ($T_{\rm eff}$) for our binary helium stars (left column) and non-detections (right column) to subdwarfs with significant RV variability and low RV variability, respectively. Both similarities and differences between our sample and the hot subdwarfs are observed. For example, while the confirmed binaries in our sample have lower surface [He/H] ratios than the non-detections, they are still more He-rich than most of the hot subdwarfs with significant RV variability (perhaps due to e.g. differences in stellar wind strength). Conversely, the non-detections have [He/H] ratios very similar to the hot subdwarfs that \cite{2022A&A...661A.113G} argues are likely mergers. While the analogy between our non-detections and likely merger-produced subdwarfs is intriguing, we note that it may be challenging to reconcile the intermediate masses and luminosities of Stars 5, 7 and 8 ($M_1\approx2-4\,M_\odot$ and $\log_{10}(L_{\rm bol}/L_\odot)\approx3.6-4.4$) with many subdwarf merger channels. In particular, most of these hot subdwarfs are thought to arise from double He-WD mergers, which should produce helium stars with masses $\lesssim$0.9 $M_\odot$ \citep[He-WDs are at most 0.45 $M_\odot$, e.g.][]{2012MNRAS.419..452Z}. \cite{2026A&A...710A..99P} recently found that the extremely He-rich, rapidly rotating subdwarf HE 1518-0948, with $\log_{10}(L_{\rm bol}/L_\odot)\approx3.6,$ is consistent with a double He-WD merger. The resulting temperatures and luminosities from the models of \cite{2026A&A...710A..99P} are in good agreement with Stars 7 and 8, and their spectra are similar to HE 1518-0948. Lastly, a He-WD + CO-WD merger could perhaps explain the properties of Star 5 \citep{2002MNRAS.333..121S,
2019ApJ...885...27S}.

We can also compare to HD 45166: a $\sim$2 $M_\odot$ \citep{2025A&A...695L..20D} quasi-WR helium star with $(L_{\rm bol}/L_\odot)\approx3.8$ potentially formed through a stellar merger. \cite{2023Sci...381..761S} proposed that the $q$WR formed through the coalescence of the helium cores of two intermediate-mass stars ($M_1\sim5\,M_\odot$ and $M_2\sim3\,M_\odot$), with a wide tertiary ($M_3\sim3.4\,M_\odot$). Notably, the period of HD 45166 is very wide ($P_{\rm orb}=8200$ d). We showed in \secref{ssec:single_wide} that we are not sensitive to such wide period binaries. Thus, long-term RV monitoring of the non-detections (years to decades) may still reveal a distant companion, which if confirmed could support pathways for triple evolution to produce intermediate-mass helium stars \citep[e.g.][]{1968AJ.....73..190H,
2016ComAC...3....6T,
2019ApJ...872..119H,
2020A&A...640A..16T,
2021ApJ...907L..19V,
2022MNRAS.516.1406S,
2023A&A...678A..60K}. Additionally, very recent work by \cite{2026A&A...710L..12P} explored the “failed common envelope channel” which leads to the merger of a stripped star and a post-MS star, inspired by the proposed formation channel of HD 45166. They demonstrated that this channel can produce helium stars with similar properties to our sample \citep[see also][]{2024A&A...682A.169H}. We note that the spectra of Stars 5, 7 and 8 are very different from the $q$WR spectrum of HD 45166 (which is thought to be produced in part by trapping of the stellar outflow by strong magnetic field). It is also unclear whether the double helium star merger scenario can explain a broader population and whether it would lead to nearly H-free helium stars. Even if the double helium star merger rate is small relative to binary-stripping, the UV excess detection method should be efficient at detecting such systems. Future work to measure projected rotation rates could help provide evidence for or against a merger origin.

\section{Summary and Conclusions}\label{sec:conc}

In this manuscript, we present the results of a radial velocity survey for a subset of the intermediate-mass helium stars that were first unveiled in \citetalias{2023Sci...382.1287D}, with helium star surface properties measured by \citetalias{2023ApJ...959..125G}. Specifically, our sample consists of 8 helium stars with no signatures of a luminous companion in their UV/optical photometry or optical spectra. This includes most of the “Helium star-type” objects from \citetalias{2023Sci...382.1287D}. 

For our sample of 8 stars, we measure RVs from 5--24 epochs of optical spectroscopic data per star. Data was obtained with Magellan/MagE over a period of 2--5 years (2018--2023). We analyzed the RVs to find evidence for binarity. We detect clear evidence of binary motion in five of the helium stars in our sample (Stars 1, 2, 4, 6, 16). However, for the remaining three stars (Stars 5, 7 and 8) we do not find clear evidence for binary motion in our data.

We measured orbital properties for all five helium stars which displayed binary motion with \texttt{The Joker}. Their periods range from hours to hundreds of days, their RV semi-amplitudes span $\sim$20--80 \kms, and their eccentricities are consistent with zero. Their orbital properties broadly match expectations for post-interaction binaries, when compared to observations of other helium star binaries.
For these objects, we confirm the interpretation of \citetalias{2023Sci...382.1287D} and \citetalias{2023ApJ...959..125G} that these are stripped star in binaries with optically dark companions, since their RVs only show evidence of a helium star. 

We then combine the orbital properties with previous measurements of the helium star masses from \citetalias{2023ApJ...959..125G} to place constraints on the masses of the unseen companion stars. The minimum masses range from $M_{2,\rm min}\approx$ 0.6--2.9 $M_\odot.$ Assuming an isotropic distribution of inclinations, we derive a distribution of possible companion masses and find that the companions are likely $\left<M_2\right>_{\rm inc}\approx$ 0.7--3.1 $M_\odot$. In all cases, this establishes that the current mass ratios for the binaries are low, with $q=M_2/M_1\leq 1,$ unlike the vast majority of previously observed helium star binaries (helium star: $M_1$, companion: $M_2$).

We then use these constraints to discuss the possible nature of the optically dark companions to the confirmed helium stars in binaries in our sample. We consider two broad categories of companions: living stars and compact objects. For living stars, we first note that the small mass ratios imply that the envelope-stripping process must have been highly non-conservative. We use this as motivation to investigate whether low mass ‘unperturbed’ stellar companions (i.e. those that follow standard single star evolutionary tracks) are viable. We constrain stellar companions with a combination of age, optical flux, and radii. Given the system age constraints (based on the helium star masses), unperturbed stellar companions should be late pre-main-sequence or young main-sequence stars. We then find that in all cases an unperturbed stellar companion with mass $\left<M_2\right>_{\rm inc}$ would contribute $\lesssim$10\% of the optical flux of the overall binary system and would also fit within their Roche lobes. As such, late pre-MS and young MS companions are viable. In addition, we note that low-mass stripped star companions could remain optically hidden and fit within the Roche lobe of the binary orbits presented here. However, this would likely require a finetuned evolutionary history involving mass reversal of the binary components. 

To constrain possible compact object companions, we consider the distribution of possible companion masses for each of the confirmed binaries. We find that our sample could contain stripped stars orbiting a variety of compact object types. In particular, Star 1 could contain a high-mass NS or a low-mass BH. Star 2, 4 and 6 could contain low-mass NSs or high-mass WDs. Lastly, the binary properties of Star 16 match observations of subdwarf + WD binaries, but a NS star is still viable. We cannot at present make any definitive statements about the nature of the companions: they are consistent with both low-mass stars and compact objects.

Finally, we discuss the implications of our finding that three of the eight helium stars in our sample do not show significant evidence for binary motion. We identify a surface composition bimodality between the helium stars confirmed to be in binaries and the non-detections. While the stars stripped in binaries are H-poor, with $X_{\rm H, surf}\sim0.3-0.4,$ the helium stars with non-detections of binary motion are nearly H-free, with $X_{\rm H, surf}\lesssim0.05.$ This suggests that their low RV variations are not simply due to low inclinations, but rather that a separate evolutionary channel may be necessary for this sub-group. The hydrogen-free helium stars with non-detections of binary motion could be in (undetected) wide orbits, or they could be genuinely single as runaway stripped stars or merger products. 

Our RV survey of intermediate-mass helium stars conclusively demonstrates that several (five) of the intermediate-mass helium stars discovered by \citetalias{2023Sci...382.1287D} and characterized by \citetalias{2023ApJ...959..125G} \emph{are} in binaries. Our sample of orbital properties has the potential to further our understanding of the progenitors of stripped-envelope supernovae, orbital evolution during mass transfer, and compact object binary formation. Future work will present additional constraints on the nature of the unseen companions in the confirmed binary sample as well as implications for their evolutionary past and future. Finally, our sample also offers the possibility to explore a novel formation channel for intermediate-mass helium stars, since some may be (currently) single.

\begin{acknowledgments}
We especially thank all the staff at Las Campanas Observatory for making our large spectroscopic survey smooth and efficient. 
We also acknowledge the Lorentz Center and the workshop ``From discovery to a population: benchmarking stripped stars and companions'' held in July 2024, and the ``Stable Mass Transfer in Binaries'' workshops hosted at the Center for Computational Astrophysics of the Flatiron Institute, in March 2024 and May 2025. 

A.L. acknowledges support from the NSERC and is funded through a NSERC Canada Graduate Scholarship—Doctoral. A.L. is also supported by the Data Sciences Institute at the University of Toronto through grant number DSIDSFY3R1P02. M.R.D. acknowledges support from the NSERC through grant RGPIN-2019-06186, the Canada Research Chairs Program, and the Dunlap Institute at the University of Toronto. B.L. has received financial support from the Flemish Government under the long-term structural Methusalem funding program by means of the project SOUL: Stellar evolution in full glory, grant METH/24/012 at KU Leuven. The Dunlap Institute is funded through an endowment established by the David Dunlap family and the University of
Toronto. M.R. acknowledge support from NASA (ATP: 80NSSC24K0932). A.O. acknowledges the support of the McWilliams Postdoctoral Fellowship in the McWilliams Center for Cosmology and Astrophysics. Support for this work was provided by NASA through the NASA Hubble Fellowship Program grant \#HST-HF2-51457.001-A awarded by the Space Telescope Science Institute, which is operated by the Association of Universities for Research in Astronomy, Inc., for NASA, under contract NAS5-26555.
Y.G.\ appreciates the support from the Observatories of the Carnegie Institution for Science. 

The authors advocate for the judicious and ethical use of artificial intelligence (AI) based tools in science. In accordance with the \href{https://baas.aas.org/pub/2023i016/release/1}{guidelines} of the American Astronomical Society on the appropriate use of AI-based writing tools, the authors confirm that AI assistance \citep[with Claude and ChatGPT;][]{claude, chatgpt} was used to edit code. The authors retain the full responsibility for the accuracy, integrity, and content of this work.

\end{acknowledgments}

\facilities{We acknowledge the use of the MagE instrument on the Magellan/Baade 6.5m telescope at Las Campanas Observatory (LCO) in Chile \citep{2008SPIE.7014E..54M}.}

\software{\texttt{astroNN} \citep{2019MNRAS.483.3255L}, \texttt{astropy} \citep{astropy:2013,astropy:2018,astropy:2022}, \texttt{astroquery} \citep{astroquery:2019}, \texttt{Carpy} \footnote{https://code.obs.carnegiescience.edu/mage-pipeline}\citep{2000ApJ...531..159K, 2003PASP..115..688K}, \texttt{lmfit} \citep{2016ascl.soft06014N}, \texttt{matplotlib} \citep{Hunter:2007}, \texttt{numpy} \citep{harris2020array}, \texttt{pandas} \citep{mckinney-proc-scipy-2010,reback2020pandas}, \texttt{PyRAF} \citep{2012ascl.soft07011S}, \texttt{SciPy} \citep{2020SciPy-NMeth}, \texttt{Simbad} \citep{2000A&AS..143....9W}, \texttt{The Joker} \citep{2017ApJ...837...20P},
\texttt{Vizier} \citep{2000A&AS..143...23O}.}

\bibliographystyle{aasjournalv7}
\bibliography{references_bin.bib} 

\appendix

\section{Radial Velocity Details}
\label{app:rv}
\restartappendixnumbering

Here, we provide details on RV measurements (\appref{sapp:tab_rv}), spectral line selection (\appref{sapp:rv_lines}), comparison of \HeII~and \HeII-H RVs to demonstrate that we do not find evidence for RV motion from a companion (\appref{sapp:rv_compare}), systematic errors (\appref{sapp:rv_sys_err}) and updated stacked spectra (\appref{sapp:rv_stack}). 

\subsection{Radial Velocity Measurements}
\label{sapp:tab_rv}

\begin{center}
\begin{minipage}[t!]{\textwidth}
\begin{minipage}[t!]{0.45\textwidth}
\centering
\footnotesize
\hypertarget{tab:all_rv}{}
\textbf{Table A1:} Overview of observations.\\[0.3em] 
\renewcommand{\arraystretch}{.5}
\begin{tabular}{ccccccc}
\toprule\midrule
Name & Time & Rel. RV & Abs. RV & $\sigma_{\rm RV}$ & S/N & Flag \\
& [HJD] & [km s$^{-1}$] & [km s$^{-1}$] & 
[km s$^{-1}$] & & \\
\midrule
Star 1 & 2458866.5349 & 0.84 & 164.64 & 1.57 & 41 & 0 \\
Star 1 & 2458838.5348 & 14.66 & 178.46 & 1.56 & 41 & 0 \\
Star 1 & 2460281.5309 & 11.99 & 175.79 & 1.80 & 35 & 0 \\
Star 1 & 2458836.5402 & 15.41 & 179.21 & 2.09 & 34 & 0 \\
Star 1 & 2459203.5692 & -10.88 & 152.92 & 2.06 & 33 & 0 \\
Star 1 & 2458867.5191 & 5.29 & 169.09 & 2.29 & 32 & 0 \\
Star 1 & 2459575.5397 & -15.34 & 148.46 & 2.39 & 30 & 0 \\
Star 1 & 2458485.5448 & 16.04 & 179.84 & 2.67 & 30 & 0 \\
Star 1 & 2459202.6389 & -7.00 & 156.80 & 2.37 & 30 & 0 \\
Star 1 & 2459249.5206 & -19.95 & 143.85 & 2.43 & 29 & 0 \\
Star 1 & 2459621.5247 & -6.48 & 157.32 & 2.29 & 29 & 0 \\
Star 1 & 2459204.6882 & -7.92 & 155.88 & 2.48 & 29 & 0 \\
Star 1 & 2459865.8234 & -20.59 & 143.21 & 2.43 & 28 & 0 \\
Star 1 & 2460149.9250 & -16.96 & 146.84 & 2.63 & 27 & 0 \\
Star 1 & 2458836.7026 & 17.84 & 181.64 & 3.06 & 26 & 0 \\
Star 1 & 2459251.5176 & -37.32 & 126.48 & 3.42 & 24 & 2 \\
Star 1 & 2459203.7211 & -11.25 & 152.55 & 2.80 & 24 & 0 \\
Star 1 & 2460265.7820 & 0.60 & 164.40 & 2.60 & 24 & 0 \\
Star 1 & 2459246.5378 & -13.75 & 150.05 & 3.26 & 22 & 0 \\
Star 1 & 2459929.7083 & 0.15 & 163.95 & 3.74 & 21 & 0 \\
Star 1 & 2459248.5237 & -24.38 & 139.42 & 3.91 & 20 & 0 \\
Star 1 & 2459586.5227 & -18.68 & 145.12 & 4.11 & 18 & 0 \\
Star 1 & 2459590.5401 & -10.30 & 153.50 & 4.94 & 14 & 0 \\
Star 1 & 2459877.8623 & -12.68 & 151.12 & 5.83 & 13 & 1 \\
\midrule
Star 2 & 2458839.5559 & 12.06 & 224.43 & 3.20 & 27 & 0 \\
Star 2 & 2459575.5850 & -114.94 & 97.43 & 3.09 & 25 & 0 \\
Star 2 & 2459249.5854 & -88.72 & 123.65 & 3.61 & 23 & 0 \\
Star 2 & 2459941.6218 & -43.15 & 169.22 & 3.36 & 22 & 0 \\
Star 2 & 2459251.5993 & -95.64 & 116.73 & 4.40 & 20 & 0 \\
Star 2 & 2459576.6956 & -0.58 & 211.78 & 4.60 & 20 & 0 \\
Star 2 & 2459940.6098 & -59.28 & 153.09 & 4.79 & 18 & 0 \\
Star 2 & 2459247.5313 & -99.12 & 113.25 & 3.89 & 18 & 0 \\
Star 2 & 2459587.5481 & -58.79 & 153.58 & 4.75 & 17 & 0 \\
Star 2 & 2459246.5872 & -33.41 & 178.96 & 6.26 & 15 & 0 \\
Star 2 & 2459586.6346 & -7.81 & 204.55 & 5.32 & 15 & 0 \\
Star 2 & 2459248.5424 & 1.35 & 213.72 & 5.86 & 13 & 0 \\
Star 2 & 2459622.5653 & -33.02 & 179.35 & 7.05 & 12 & 0 \\
Star 2 & 2459590.6151 & -63.20 & 149.17 & 11.37 & 6 & 1 \\
\midrule
Star 4 & 2460281.5516 & 5.66 & 147.67 & 4.85 & 25 & 0 \\
Star 4 & 2459576.5556 & 87.25 & 229.26 & 4.50 & 24 & 0 \\
Star 4 & 2458840.5453 & 65.23 & 207.24 & 4.70 & 23 & 0 \\
Star 4 & 2459251.5360 & -18.02 & 123.99 & 6.30 & 20 & 0 \\
Star 4 & 2459960.5975 & -9.83 & 132.18 & 4.67 & 20 & 0 \\
Star 4 & 2460283.7218 & -19.91 & 122.10 & 5.03 & 19 & 0 \\
Star 4 & 2459940.6557 & 17.79 & 159.80 & 6.41 & 16 & 0 \\
Star 4 & 2459621.5491 & 67.71 & 209.72 & 8.07 & 16 & 0 \\
Star 4 & 2458836.6051 & 76.48 & 218.49 & 7.10 & 16 & 0 \\
Star 4 & 2459247.5917 & 1.37 & 143.38 & 6.54 & 16 & 0 \\
Star 4 & 2459975.5314 & -2.68 & 139.32 & 9.47 & 15 & 0 \\
Star 4 & 2459587.6250 & 87.87 & 229.87 & 5.97 & 13 & 0 \\
Star 4 & 2459929.6618 & 9.98 & 151.99 & 7.55 & 13 & 0 \\
Star 4 & 2459976.5345 & 89.35 & 231.35 & 8.43 & 13 & 0 \\
Star 4 & 2459248.5947 & 58.24 & 200.25 & 14.37 & 12 & 1 \\
Star 4 & 2459586.5895 & 56.32 & 198.33 & 9.27 & 11 & 0 \\
\midrule
Star 5 & 2458840.6852 & 0.87 & 287.95 & 2.13 & 32 & 0 \\
Star 5 & 2459202.7168 & -6.34 & 280.74 & 2.31 & 28 & 0 \\
Star 5 & 2459621.6242 & 3.87 & 290.95 & 3.10 & 24 & 0 \\
Star 5 & 2459576.7312 & -7.43 & 279.66 & 3.14 & 24 & 0 \\
Star 5 & 2459251.6825 & -0.03 & 287.05 & 2.98 & 24 & 0 \\
Star 5 & 2459960.8069 & 2.16 & 289.24 & 2.76 & 23 & 0 \\
\vdots & \vdots & \vdots & \vdots & \vdots & \vdots & \vdots \\
\bottomrule
\vspace{0.3em}
\end{tabular}
\end{minipage}
\hspace{0.05\textwidth}
\raisebox{.195ex}[0pt][0pt]{
\begin{minipage}[t!]{0.45\textwidth}
\centering
\footnotesize
\hypertarget{tab:lmc_rv}{}
\textbf{Table A1:} continued.\\[0.3em] 
\renewcommand{\arraystretch}{.5}
\begin{tabular}{ccccccc}
\toprule\midrule
Name & Time & Rel. RV & Abs. RV & $\sigma_{\rm RV}$ & S/N & Flag \\
& [HJD] & [km s$^{-1}$] & [km s$^{-1}$] & 
[km s$^{-1}$] & & \\
\midrule
\vdots & \vdots & \vdots & \vdots & \vdots & \vdots & \vdots \\
Star 5 & 2459203.7733 & -4.60 & 282.48 & 3.02 & 23 & 0 \\
Star 5 & 2459869.8669 & -3.84 & 283.24 & 3.57 & 23 & 0 \\
Star 5 & 2459249.6756 & -5.61 & 281.47 & 3.11 & 22 & 0 \\
Star 5 & 2459976.6488 & 9.06 & 296.14 & 3.92 & 22 & 0 \\
Star 5 & 2459975.7019 & -10.13 & 276.95 & 3.18 & 21 & 0 \\
Star 5 & 2459622.6135 & 3.55 & 290.63 & 4.26 & 19 & 0 \\
Star 5 & 2458836.8078 & -3.20 & 283.88 & 4.54 & 19 & 0 \\
Star 5 & 2459940.7869 & 3.16 & 290.25 & 3.79 & 18 & 0 \\
Star 5 & 2459246.7137 & -13.08 & 274.00 & 5.82 & 15 & 0 \\
Star 5 & 2459586.7903 & 18.62 & 305.70 & 6.23 & 12 & 2 \\
Star 5 & 2459590.8235 & -7.82 & 279.26 & 8.38 & 8 & 1 \\
\midrule
Star 6 & 2458483.6440 & 0.62 & 296.42 & 2.68 & 29 & 0 \\
Star 6 & 2459621.5988 & -30.82 & 264.98 & 3.70 & 24 & 0 \\
Star 6 & 2459202.5374 & -50.60 & 245.19 & 3.18 & 22 & 0 \\
Star 6 & 2459203.6208 & -35.89 & 259.91 & 3.61 & 22 & 0 \\
Star 6 & 2459251.6332 & -20.72 & 275.08 & 3.77 & 21 & 0 \\
Star 6 & 2458866.7199 & -50.99 & 244.81 & 4.31 & 20 & 0 \\
Star 6 & 2459204.8375 & -25.21 & 270.59 & 4.36 & 20 & 0 \\
Star 6 & 2459246.6352 & -39.58 & 256.22 & 5.20 & 19 & 0 \\
Star 6 & 2458836.7281 & 10.46 & 306.26 & 4.39 & 19 & 0 \\
Star 6 & 2459576.5334 & 31.79 & 327.58 & 5.07 & 18 & 2 \\
Star 6 & 2458485.7375 & 15.28 & 311.07 & 4.50 & 18 & 2 \\
Star 6 & 2459203.8234 & -36.25 & 259.54 & 4.67 & 17 & 0 \\
Star 6 & 2459249.7625 & 2.97 & 298.76 & 5.77 & 16 & 0 \\
Star 6 & 2458836.5715 & 4.19 & 299.99 & 5.11 & 15 & 0 \\
Star 6 & 2459587.7985 & -37.89 & 257.91 & 6.97 & 13 & 0 \\
Star 6 & 2459586.8115 & 4.35 & 300.15 & 8.16 & 12 & 0 \\
Star 6 & 2459248.6675 & -0.91 & 294.89 & 8.27 & 10 & 0 \\
Star 6 & 2459590.6611 & -26.78 & 269.02 & 8.69 & 10 & 0 \\
\midrule
Star 7 & 2460283.7725 & -1.82 & 283.84 & 4.94 & 20 & 0 \\
Star 7 & 2459941.6693 & -8.96 & 276.71 & 4.83 & 19 & 0 \\
Star 7 & 2459869.8010 & 5.51 & 291.17 & 5.70 & 19 & 0 \\
Star 7 & 2459975.5794 & 5.08 & 290.74 & 5.32 & 19 & 0 \\
Star 7 & 2459960.6831 & -2.86 & 282.81 & 4.39 & 16 & 0 \\
Star 7 & 2458866.6823 & -4.29 & 281.37 & 6.18 & 15 & 0 \\
Star 7 & 2459203.7418 & -3.41 & 282.25 & 7.09 & 14 & 0 \\
Star 7 & 2458840.8009 & -11.27 & 274.39 & 6.23 & 12 & 0 \\
Star 7 & 2459976.6791 & -11.14 & 274.52 & 7.43 & 12 & 0 \\
Star 7 & 2459587.7207 & -6.40 & 279.26 & 8.71 & 12 & 0 \\
Star 7 & 2459204.7398 & -12.47 & 273.19 & 8.97 & 11 & 0 \\
Star 7 & 2459246.6644 & -32.55 & 253.11 & 9.59 & 11 & 2 \\
Star 7 & 2459202.7950 & -8.57 & 277.09 & 6.87 & 10 & 0 \\
Star 7 & 2459247.6410 & -17.19 & 268.48 & 7.42 & 9 & 0 \\
\midrule
Star 8 & 2458867.6262 & -16.48 & 279.68 & 8.64 & 17 & 0 \\
Star 8 & 2458838.6605 & -0.32 & 295.84 & 8.42 & 16 & 0 \\
Star 8 & 2459576.7901 & -7.41 & 288.75 & 9.60 & 14 & 0 \\
Star 8 & 2459622.6862 & -19.57 & 276.59 & 12.97 & 14 & 0 \\
Star 8 & 2459586.6828 & 4.14 & 300.29 & 12.49 & 13 & 0 \\
\midrule
Star 16 & 2458839.7130 & -0.43 & 278.53 & 1.56 & 26 & 0 \\
Star 16 & 2459202.8273 & -62.76 & 216.20 & 1.93 & 23 & 0 \\
Star 16 & 2459204.7715 & -0.55 & 278.41 & 2.15 & 22 & 0 \\
Star 16 & 2459203.6476 & 3.13 & 282.09 & 1.93 & 21 & 0 \\
Star 16 & 2458866.7901 & 62.45 & 341.41 & 2.23 & 21 & 0 \\
Star 16 & 2459202.5749 & -19.23 & 259.73 & 2.03 & 20 & 0 \\
Star 16 & 2459204.6619 & 14.73 & 293.69 & 2.47 & 19 & 0 \\
Star 16 & 2459203.7954 & 35.81 & 314.77 & 2.39 & 18 & 0 \\
Star 16 & 2458867.8320 & -45.71 & 233.25 & 2.99 & 17 & 0 \\
Star 16 & 2459576.8453 & -10.69 & 268.27 & 2.94 & 15 & 0 \\
Star 16 & 2459248.6906 & 57.73 & 336.69 & 7.85 & 8 & 1 \\
\bottomrule
\vspace{0.3em}
\end{tabular}
\end{minipage}}
\footnotesize
\\
\textbf{Notes.} Columns list source name, observing dates, relative RV, absolute RV, RV uncertainty, estimated SNR, and quality flag.
\end{minipage}
\end{center}

\begin{figure}[ht!]
    \centering
    \includegraphics[width=\textwidth]{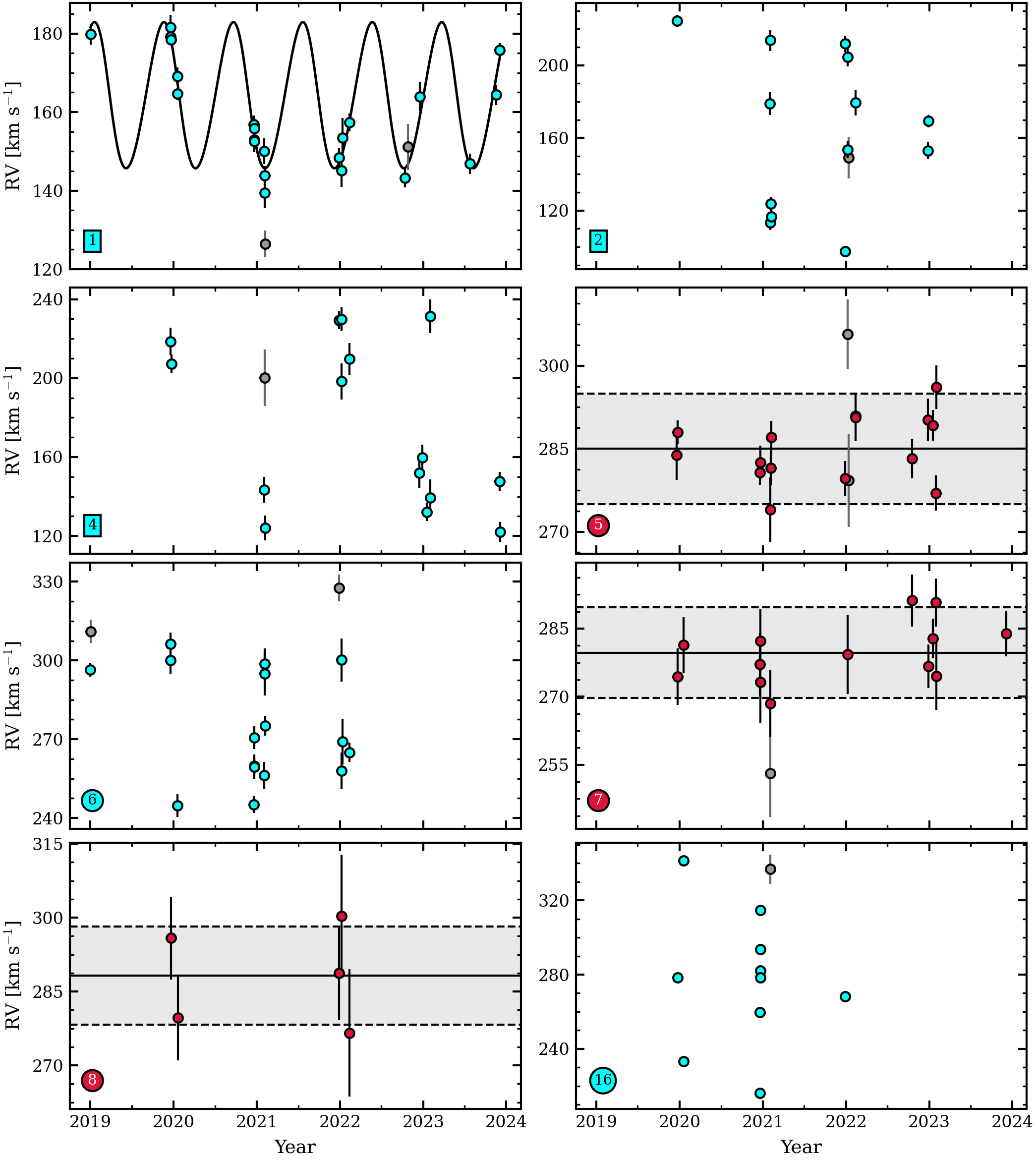}
    \caption{Measured absolute RVs as a function of time, given in years. SB1 stripped stars are cyan, whereas non-detections are red, while all stars are numbered in the bottom left of each panel. Unreliable RVs which we discard from subsequent analysis are shown in grey. For the non-detections, we present the mean $\mu$ (black line) and $\mu\pm10$ km s$^{-1}$ (dashed black lines) of the RVs to visualize our binary detection threshold of 20 km s$^{-1}$. Uniquely, for Star 1 we also show the best-fit orbit (MAP, \tabref{tab:detailed_orbits}), since it has a wide enough period to observe the orbit over multi-year timescales. Finally, we highlight the magnitude of this work, which began in late 2018 and ended in late 2023: 6 years of observations.}
    \label{fig:rv_jd}
\end{figure}

\noindent
In \tabref{sapp:tab_rv} we present a complete list of the RVs measured in this work, and will provide the table on Zenodo. We include the name of the source \citepalias[star number, as in][]{2023Sci...382.1287D,2023ApJ...959..125G}, relative and absolute RVs (where we both exclude and include the systemic velocities given in \tabref{tab:orbit_props}, respectively), uncertainties, estimated SNRs (averaged across all orders within an epoch) and quality flags. \texttt{Flag=0} are reliable RVs, whereas \texttt{Flag=1} or \texttt{Flag=2} correspond to unreliable RVs (see \secref{ssec:cuts} for details). Then, in \figref{fig:rv_jd}, we present these RVs as a function of observing time, where colored (grey) RVs are reliable (unreliable). For the non-detections, we visualize our RV shift threshold of 20 km s$^{-1}$ in shaded grey (\eqref{eq:binary_bool}).

\subsection{Line Selection}
\label{sapp:rv_lines}

The hot temperatures of our stars ($T_{\rm eff} \sim $60-100 kK) limit the number of available spectral lines. All epoch spectra for each source were visually inspected to identify spectral lines strong enough to be present in individual epochs. This does eliminate some lines which are only robustly visible in the high-SNR stacked spectra: some weaker metal lines are present, but they are not strong enough for robust RV measurements. We therefore settled on only including the strongest \HeII\,and \HeII-H blends, and, only in the case of Star 16, \HeI~(see below for details).  The seven spectral lines which are (largely) present across our sample include: four \HeII\,lines (\HeII\,4200, 4542, 4686, 5412 \AA) and three \HeII-H lines (\HeII-H 4100, 4339, 4860 \AA). We exclude the H$\alpha$ blend because of its sensitivity to wind features and surrounding nebula. We will now justify our spectral line selection for each star in our sample: 

For Stars 1, 2 and 5, we use all seven of the aforementioned spectral lines to measure RVs, since all lines are strong. For Star 4, we only consider \HeII\,lines for RV measurements. One epoch we observed displayed strong \HeII-H absorption (weaker in another epoch). The origin of these transient Balmer features is unclear, perhaps: scattered light from a nearby source, eclipses, or chance alignment. To avoid any potential contamination of our RV measurements, not only do we only fit \HeII\,lines, but we also discard both epochs displaying strong Balmer features. For Star 6, we exclude \HeII-H 4860. The 2D spectra of Star 6 show evidence for surrounding ionized gas, and a narrow nebular emission feature is present in the center of H$\beta.$ This leads to an over-corrected sky subtraction which makes \HeII-H 4860 ill-suited for RV measurements. For Star 7, we exclude \HeII-H 4100, whose line strength is too weak across epochs for accurate RV measurements. For Star 8, we exclude \HeII\,4686 and \HeII-H 4100, since similarly to Star 7 these lines are too weak. For Star 16, we use a set of \HeI\,absorption features: \HeI\,4388, 4713 and 5016 \AA. These lines are absent from the rest of our sample, as their properties match expectations for core-He burning. Rather, these \HeI\,features are symptomatic of the cooler temperature ($T_{\rm eff}\sim$ 35 $k$K) and larger radius of Star 16 (twice as large as expected relative to core-helium burning). 

\subsection{Comparison of \HeII\,and \HeII-H lines}
\label{sapp:rv_compare}

\citetalias{2023Sci...382.1287D} classified our sample as “Helium-star type” according to the optical dominance of an intermediate-mass helium star, while \citetalias{2023ApJ...959..125G} demonstrated that a MS companion could contribute at most $\sim$10\% of the optical flux through spectral analysis. Nevertheless, it is still a potential concern that a low-mass star could contribute just enough optical flux to bias RV measurements. In particular, while we do not expect a low-mass MS companion to exhibit \HeII\,lines: they are simply not hot enough, it is conceivable that a low-mass MS companion could impact \HeII-H features through Balmer absorption. 

\begin{figure}[t!]
    \centering
    \includegraphics[width=\textwidth]{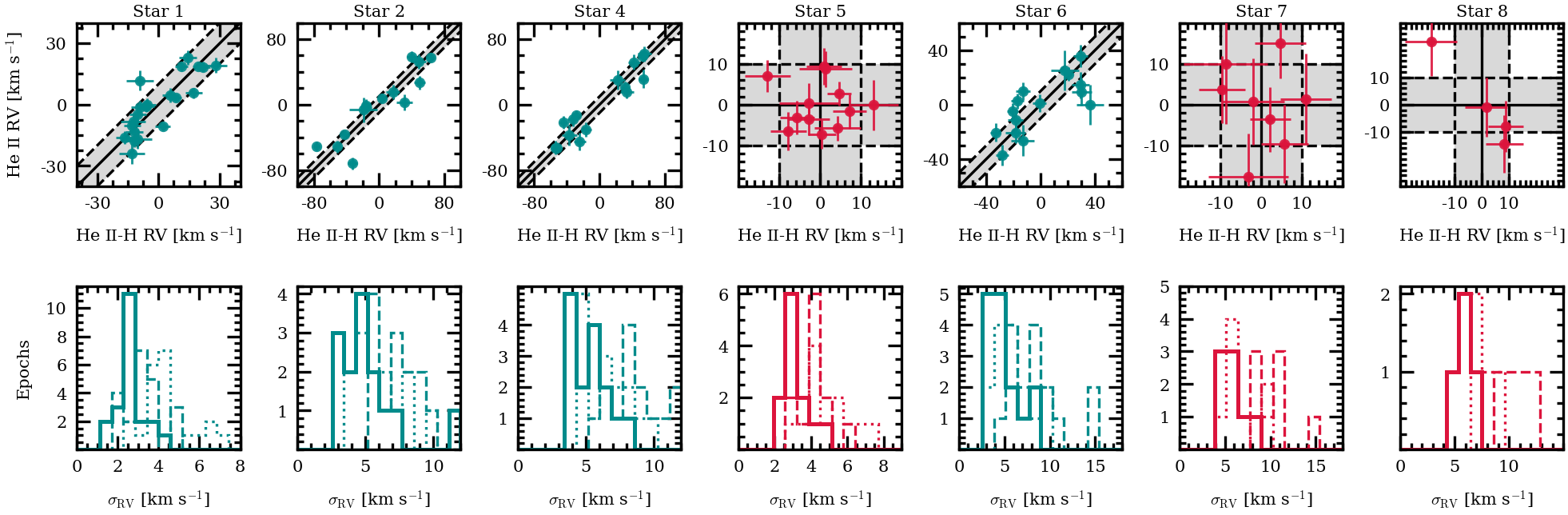}
    \caption{A comparison of \HeII-H and \HeII~RVs for Stars 1, 2, 4, 5, 6, 7 and 8, where dark cyan (red) corresponds to binaries (non-detections). In the top row, for the binaries, we present \HeII-H and \HeII~RVs, while also showing the 1:1 line (solid black) and $\pm 10$ km s$^{-1}$ deviations from 1:1 (dashed), whose range is shaded in grey, since $\pm 10$ km s$^{-1}$ is our typical RV residual from orbital fitting (see \figref{fig:orbits}). Stars 1, 2, 4 and 6 all exhibit directly proportional relationships between \HeII-H and \HeII~RVs, demonstrating they are SB1s. For the non-detections, we show $\pm 10$ km s$^{-1}$ scatter about zero, which corresponds to our RV shift threshold (20 km s$^{-1}$, see \eqref{eq:binary_bool}). For Stars 5, 7 and 8, both \HeII-H and \HeII~RVs display stochastic shifts which are consistent with single star variability. Note that both the \HeII-H and \HeII~RVs are centered about zero since we subtract their means. Then, in the bottom row, we present the RV error distributions for \HeII-H (dashed), \HeII~(dotted), and \HeII-H and \HeII~combined (solid). In all cases, the combination of \HeII-H and \HeII~ yields the smallest uncertainties. We clarify that: $(i)$ while we do show \HeII-H RVs for Star 4, we do not use them for RV measurements and orbital properties,  $(ii)$ Star 16 is excluded since we use \HeI~lines to measure RVs, where we justify both of these decisions in \secref{sapp:rv_lines} and $(iii)$ we only consider RVs which pass our quality cuts in \secref{ssec:cuts}.}
    \label{fig:rv_compare}
\end{figure}

In \figref{fig:rv_compare}, we compare the \HeII-H and \HeII~RVs for all stars in our sample (except Star 16, since we use \HeI\,lines). More specifically, we repeat our RV measurement procedure while only including either set of lines. In the top row, we observe that for all of the binary helium stars, the \HeII-H and \HeII~RVs are directly proportional. There is of course some scatter about the one-to-one line; inclusion of fewer spectral lines degrades RV accuracy. Overall, the direct proportionality is confirmation that the optically invisible companions are dim enough for our purposes: unbiased RV measurements. For the non-detections, we observe that RVs are stochastically distributed, with a scatter of $\sim20$ km s$^{-1}$, while no trend across \HeII-H and \HeII~RVs is present. Then, in the bottom row we present RV uncertainty distributions for \HeII-H, \HeII~and their combination (dashed, dotted, and thick solid, respectively). The combination of \HeII-H and \HeII~lines clearly improves RV accuracy, since their error distributions are smaller than both \HeII-H and \HeII\, alone. We conclude that the binary helium stars are SB1s. Morevoer, we find that combining as many spectral lines as possible produces the best RV measurements. In summary, the line selections we detailed in \secref{sapp:rv_lines} are optimal for this work.

\begin{figure}[t!]
    \centering
    \includegraphics[width=.44\textwidth]{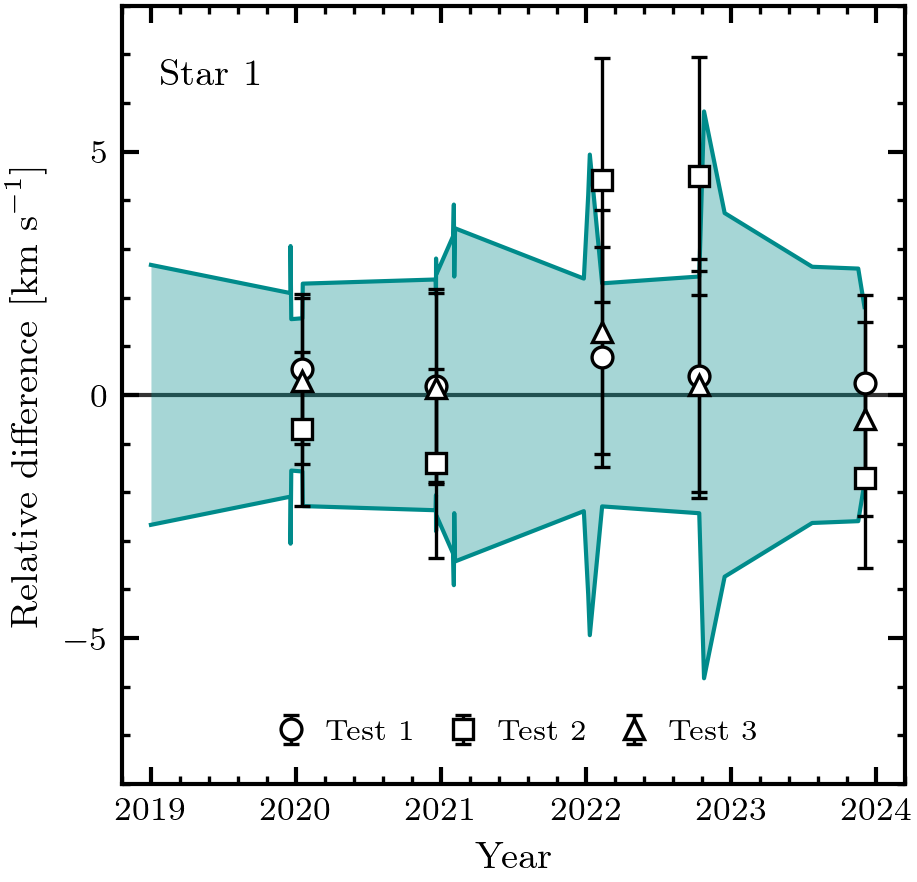}
    \hspace{2em}
    \includegraphics[width=.45\textwidth]{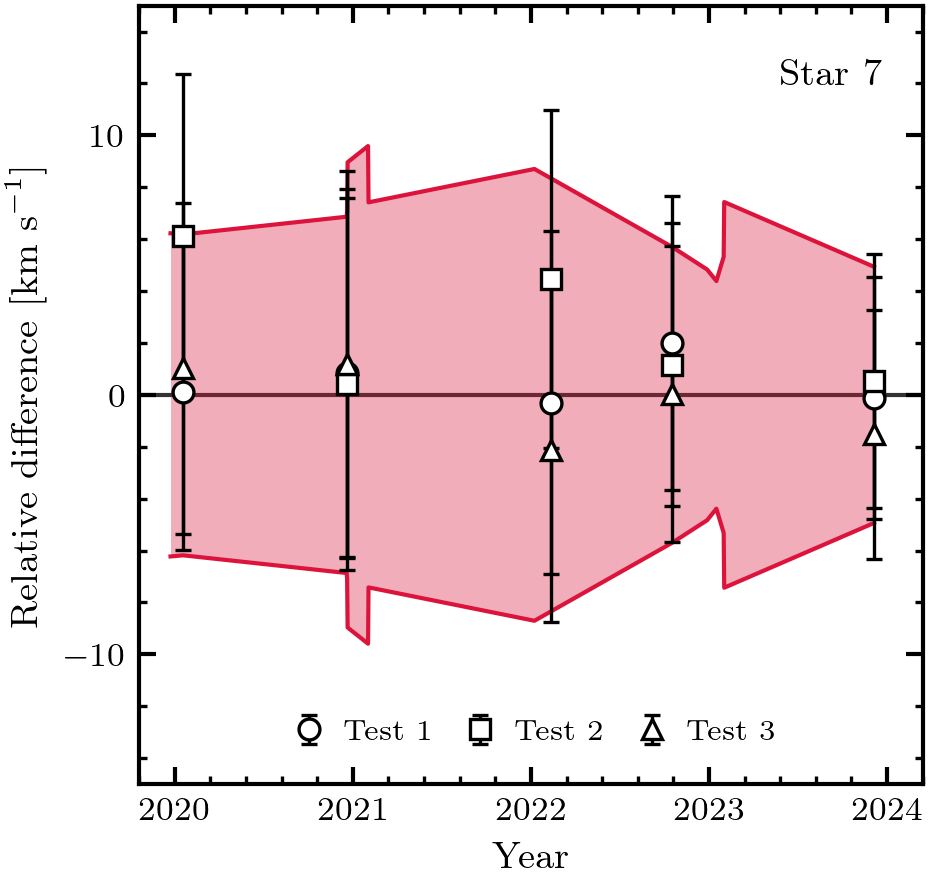}
    \caption{Analysis of the sources of systematic error presented in \appref{sapp:rv_sys_err}, for Stars 1 and 7. The relative difference of the RVs between the tests we perform (scatter points) and the fiducial value (zero line) is presented as a function of observing time. We also color the region spanning our fiducial RV measurement uncertainties for all measured epochs.}
    \label{fig:rv_tests}
\end{figure}

\subsection{Systematic error}
\label{sapp:rv_sys_err}

To investigate the importance of systematic error in our RV measurements, we perform a series of tests. Specifically, we modify the \texttt{CarPy} reduction pipeline which outputs the normalized epoch spectra in three ways: In Test 1, we re-flatten the epoch spectra with a different spline order. If the continuum level is not accurately estimated, we would expect to see a significant difference in an RV measurement after re-flattening. In Test 2, we use a different ThAr exposure than the default for wavelength calibration, that is, the ThAr exposure taken closest to the observation. This test accounts for temperature changes, pressure changes, and telescope rotation that can occur during a night, causing the wavelength solution to shift, which in turn could bias RVs. Finally, in Test 3, we adjust a flag within  CarPy that impacts how the rectifications of the different (curved) orders are done during image reduction. 

We apply Tests 1, 2 and 3 to five observations spanning our observing window (2020-2023) for Stars 1 and 7, which are representative of the binaries and non-detections, respectively. We then measure RVs from the new resulting observations and compare them to the fiducial values. We visualize this comparison in \figref{fig:rv_tests}, where we present the relative difference between test and fiducial as a function of time. Since the variation is contained within the 1$\sigma$ interval of our measured RV uncertainties (coloured regions), we infer that RV systematics related to the MagE instrument and/or our reduction procedure are not significantly biasing our RV measurements. However, it is worth noting that Test 2, using a different ThAr exposure for wavelength calibration, induces the largest systematic RV shift. Despite the larger relative difference between Test 2 and the fiducial value, Test 2 is still consistent with our measured RV uncertainty. In particular, RV errors are larger during 2022-2023 for Star 1, which corresponds to the largest deviation for Test 2. We conclude that our RV systematic uncertainties are typically $\sim$0.5 km s$^{-1}$ and at most $\sim$4 km s$^{-1},$ which is sub-dominant and contained within our measurement uncertainty, respectively.

\subsection{Updated stacked spectra}
\label{sapp:rv_stack}

\begin{figure*}
\centering
\includegraphics[width=\textwidth]{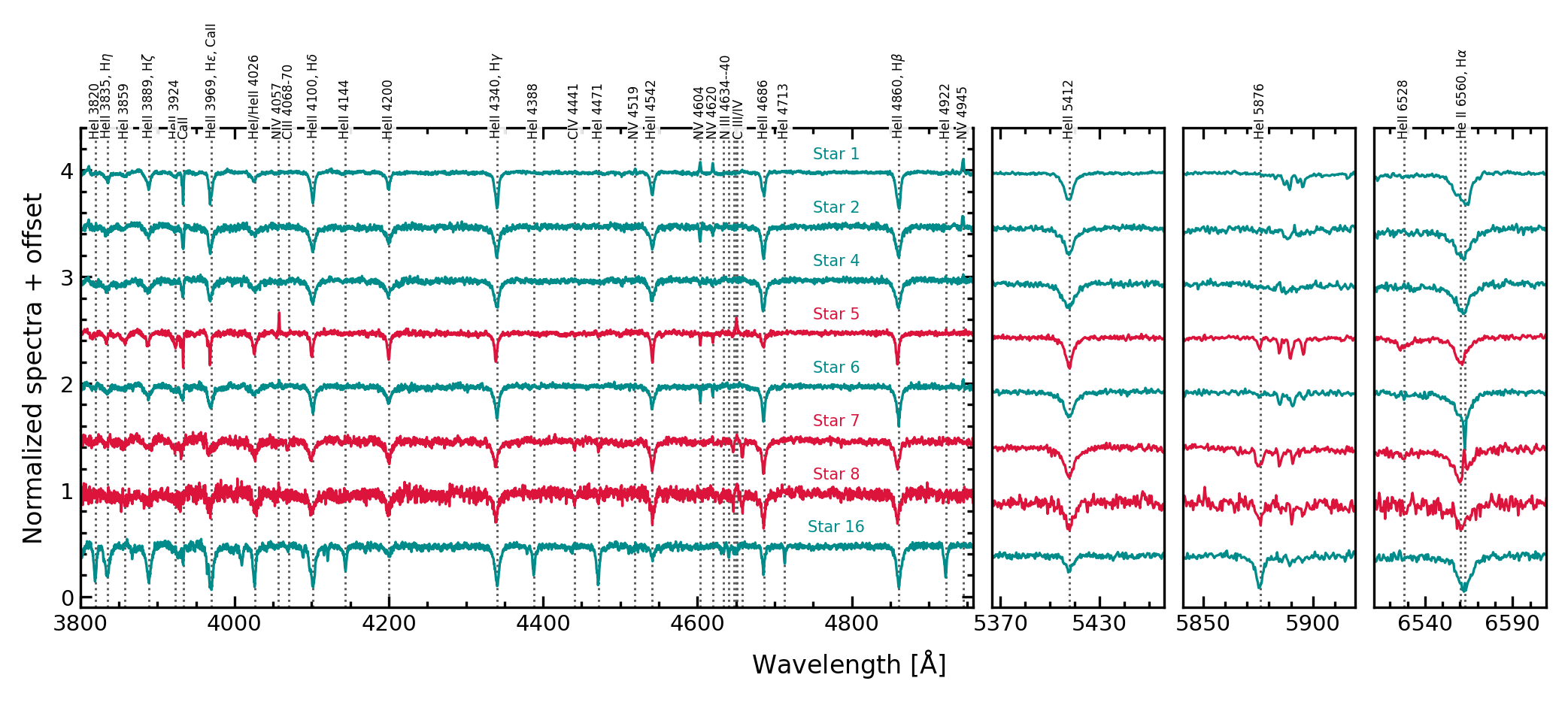}
\caption{The updated normalized spectra for the sample we analyze in our RV survey, described in \secref{ssec:stack} and \appref{sapp:rv_stack}. We reproduce Figure 1 from \citetalias{2023ApJ...959..125G}. Each spectrum is colored according to class: cyan for binary, and red for non-detection.}
\label{fig:stacks}
\end{figure*}

Here, we present the updated stacked optical spectra for all stars in our sample, which will be publicly available on Zenodo. In \tabref{tab:stack}, we give analogous information to Table 1 in \citetalias{2023ApJ...959..125G}, which summarizes the observations we obtained to create our high S/N stacks. Two main changes have occurred. First, since 
\citetalias{2023ApJ...959..125G}, we have gathered more epoch spectra for our sample. Second, we are now shifting the epoch spectra by our new measured RVs. Hence, our S/N are comparable or higher than \citetalias{2023ApJ...959..125G}. The updated stacked optical spectra for our 8 stars are visualized in \figref{fig:stacks}.

\begin{table}[t!]
\centering
\footnotesize
\caption{Overview of observations used for updated stacked optical spectra.}
\label{tab:stack}
\renewcommand{\arraystretch}{1}
\begin{tabularx}{0.95\textwidth}{lcccccccc}
\toprule
\midrule
Star & Location & RA & DEC & Dates of observation & $N_{\rm RV}/N_{\rm obs}$ & Approximate exposure times & S/N & Continuous $\lambda$ coverage \\
\midrule
1 & SMC & 01:00:59.70 & -72:37:13.7 & 2019-2023 & 22/24 & 2$\times$600s & 140 & 3390-8810 \AA \\
2 & SMC & 00:57:01.56 & -72:36:03.3 & 2019-2022 & 13/14 & 2$\times$1200s & 70 & 3540-8110 \AA \\
4 & SMC & 01:04:00.48 & -72:16:42.7 & 2019-2023 & 15/16 & 3$\times$1200s & 70 & 3390-8000 \AA \\
\midrule
5 & LMC & 05:08:49.38 & -69:05:29.8 & 2019-2023 & 15/17 & 2$\times$800s & 90 & 3390-8130 \AA \\
6 & LMC & 05:04:46.68 & -69:02:25.3 & 2018-2022 & 16/18 & 2$\times$850s & 80 & 3390-8130 \AA \\
7 & LMC & 05:28:01.15 & -69:59:48.7 & 2019-2023 & 13/14 & 2$\times$1200s & 50 & 3540-8170 \AA \\
8 & LMC & 05:47:28.01 & -69:06:07.6 & 2019-2022 & 5/5 & 3$\times$1200s & 30 & 3760-8210 \AA \\
16 & LMC & 05:35:33.63 & -70:19:06.1 & 2019-2021 & 10/11 & 2$\times$900s & 60 & 3390-8030 \AA \\
\bottomrule
\end{tabularx}
\vspace{0.3em}\\
\tablecomments{S/N is calculated per pixel from the stacked spectra, averaged and rounded by 10 over the wavelength ranges: 4230-4300, 4400-4430, 4730-4820, 5030-5250 \AA. The continuous wavelength coverage is rounded to the highest (lowest) 10 \AA~for the lower (upper) bound.}
\end{table}

\section{Justification of binary motion thresholds}
\label{app:binary-motion} 
\restartappendixnumbering

Previous multiplicity studies have implemented various thresholds on measured RV shifts to determine binary motion, which differ depending on the stellar population. \cite{2013A&A...550A.107S} adopted a threshold of $20\,\rm km\,s^{-1}$ for O-type stars, motivated by the variability of single blue supergiants (BSGs). \cite{2009A&A...507.1585R} found that observed RV shifts of $\sim$15-25 km s$^{-1}$ in BSGs could arise from either orbital or photospheric motion. A $20\,\rm km\,s^{-1}$ limit was also utilized by \cite{2023MNRAS.521.4473C} when studying O-type hypergiants and WNLha-type WR stars, since they deemed their sample variability similar to the \cite{2013A&A...550A.107S} sample. \cite{2015A&A...580A..93D} selected $16\,\rm km\,s^{-1}$ for B-type stars due to an observed break at $16\,\rm km\,s^{-1}$ for the B-type supergiants in their sample, whereas \cite{2022A&A...660A..89R} adopted $25\,\rm km\,s^{-1}$ to mitigate the larger pulsations of early B-type supergiants. For Galactic WR stars, different limits have been imposed: $10\,\rm km\,s^{-1}$ for carbon-rich WR (WC), up to $50\,\rm km\,s^{-1}$ for nitrogen-rich WR (WN)
\citep{2020A&A...641A..26D,2022A&A...664A..93D, 2023A&A...674A..88D}. Most recently, the BLOeM survey \citep{2024A&A...690A.289S} adopted a threshold of $20\,\rm km\,s^{-1}$ for Oe/Be stars \citep{2025arXiv250202641B}, early-B type dwarfs \citep{2025arXiv250321936V}, early-B type supergiants \citep{2025A&A...698A..40B} and BAF-type supergiants \citep{2025A&A...698A..39P}, motivated by the lack of an observed $16\,\rm km\,s^{-1}$ break in their sample as well as several previous works \citep[e.g.][]{2021A&A...652A..70B,2022A&A...658A..69B}. Regardless of minimum RV shift to detect binarity, the aforementioned studies used a shift significance limit of $\sigma_{\rm detect}>4$ (the ratio of observed RV shift to error) which equates to 1 false variability detection per 1000 stars. 

In this work, we adopt the recent RV threshold implemented by the BLOeM survey of $20\,\rm km\,s^{-1}$ since (i) our stellar samples are in the same environment--the Magellanic Clouds (impact of metallicity on variability), (ii) we expect our stripped star sample to exhibit RV variability more analogous to OB-stars than the extreme variability of WR stars since they have absorption-line spectra, (iii) $20\,\rm km\,s^{-1}$ is the prevailing shift threshold adopted in the literature, and (iv) we do not currently have an observational understanding of the variability of intermediate-mass helium stars, since no previous studies exist.

\section{Orbital Property Details}
\label{app:orbits} 
\restartappendixnumbering

Here, we provide additional details on orbital solutions (\appref{sapp:orbits}), companion mass constraints for a variety of assumed helium star masses (\appref{sapp:m2}), and posterior/rejection sampling distributions (\appref{sapp:corner_plots}). In doing so, we make our orbital solutions and orbit dependent companion masses as transparent and reproducible as possible.

\subsection{Reproducible Orbital Solutions}
\label{sapp:orbits}

In \tabref{tab:detailed_orbits}, we present the MAP orbit parameters for each of our binary helium stars, which can yield the radial velocity $v$ at time $t$ \citep{1609anov.book.....K,2010exop.book...15M}:
\begin{equation}\label{eq:rv_eq}
    v(t;\theta) = v_{\rm sys} 
    + K_1 \left[\cos(\omega+f)+e\cos\omega\right],
\end{equation}
where $v_{\rm sys}$ is the systemic velocity, $K_1$ is the stripped star RV semi-amplitude, $\omega$ is the pericenter argument, and $f$ is the true anomaly:
\begin{equation}\label{eq:f_eq}
    \cos f = \frac{\cos E - e}{1 - e\cos E},
\end{equation}
where $E$ is the eccentric anomaly, which is related to the mean anomaly $M(t)$:
\begin{equation}\label{eq:M_eq}
    M(t) = \frac{2\pi(t-t_{\rm ref})}{P} - M_0;
    M(t) = E(t) - e\sin E(t),
\end{equation}
where $t_{\rm ref}$ is the reference time (date of first observation), and $M_0$ is the mean anomaly constant reported in \tabref{tab:detailed_orbits}. Since we find that all orbits are consistent with circular, we also provide orbital solutions assuming $e=0,$ which is the MAP orbital solution from posterior samples which have near-zero eccentricities. We transform each orbit in \tabref{tab:detailed_orbits} into \texttt{The Joker} compatible objects, specifically \texttt{twobody.KeplerOrbit}, which will be publicly available on Zenodo.

\begin{table}[ht!]
\centering
\footnotesize
\caption{Overview of orbital solutions.}
\label{tab:detailed_orbits}
\renewcommand{\arraystretch}{1}
\begin{tabularx}{.65\textwidth}{lcccccccc}
\toprule\midrule
Star & Type & $P_{\rm orb}$ & $e$ & $\omega$ & $M_0$ & $K_1$ & $v_{\rm sys}$ & $t_{\rm ref}$ \\
& & [d] & & [rad] & [rad] & 
[km s$^{-1}$] & [km s$^{-1}$] & [MJD] \\
\midrule
Star 1 & MAP & 300.3819 & 0.1610 & 0.2938 & 0.8048 & 18.4100 & 161.8651 & 58485.0448 \\
Star 1 & Circular & 301.6352 & 0 & 0 & 0.7171 & 18.1297 & 162.4455 & 58485.0448 \\
Star 2 & MAP & 0.6738 & 0.0892 & 1.3740 & 0.9410 & 65.2493 & 160.1130 & 58839.0559 \\
Star 2 & Circular & 0.6738 & 0 & 0 & 1.0089 & 64.5760 & 162.0916 & 58839.0559 \\
Star 4 & MAP & 0.7766 & 0.1255 & 1.0374 & 0.1986 & 64.3284 & 183.3477 & 58836.1051 \\
Star 4 & Circular & 0.7766 & 0 & 0 & 2.2455 & 61.5952 & 182.5111 & 58836.1051 \\
Star 6 & MAP & 8.6022 & 0.3062 & 1.1275 & 1.4773 & 33.7799 & 272.4136 & 58483.1440 \\
Star 6 & Circular & 8.5956 & 0 & 0 & 2.1532 & 28.6674 & 273.0017 & 58483.1440 \\
Star 16 & MAP & 2.3252 & 0.0984 & -1.7057 & 3.0516 & 78.7011 & 273.9140 & 58839.2130 \\
Star 16 & Circular & 2.3252 & 0 & 0 & 0.6000 & 75.4476 & 277.9904 & 58839.2130 \\
\bottomrule
\end{tabularx}
\end{table}

\subsection{Companion Mass Constraints}
\label{sapp:m2}

\begin{figure*}[t!]
\centering
\includegraphics[width=\textwidth]{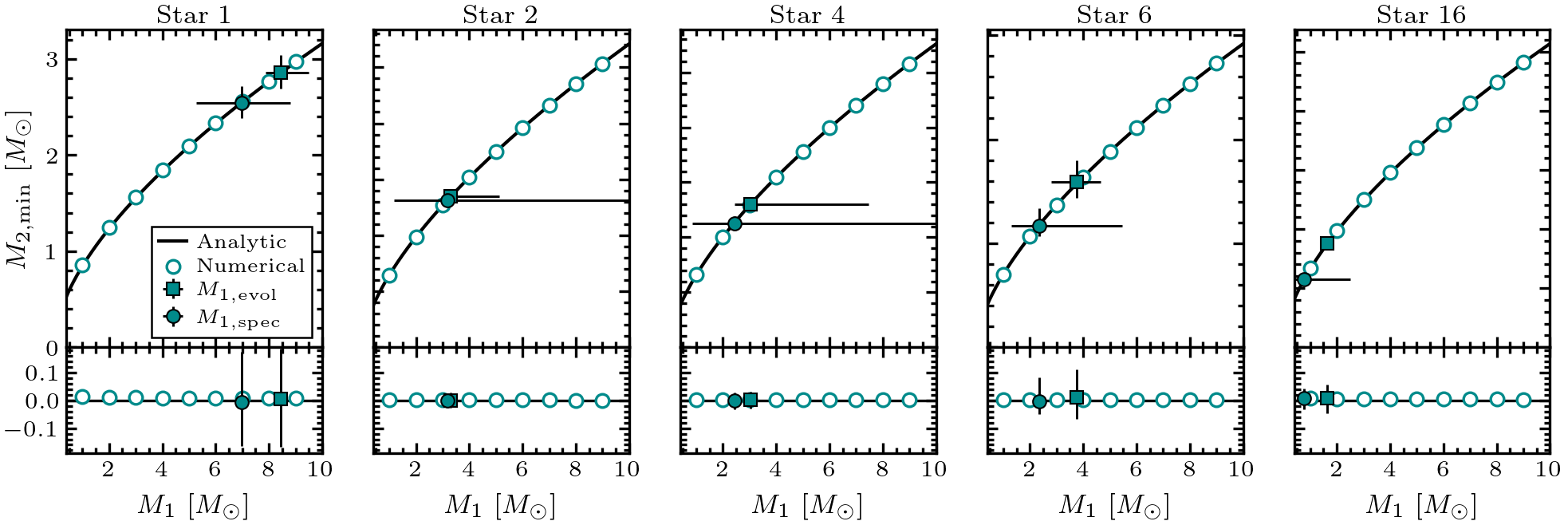}
\caption{Analytic approximations for $M_{2,\rm min}$ given in \tabref{tab:detailed_m2} (black lines) in comparison to numerical solutions for $M_{2,\rm min}$ assuming a broad range of stripped star masses (dark cyan circles), as well as $M_{2,\rm min}(M_{1,\rm spec})$ and $M_{2,\rm min}(M_{1,\rm evol})$ (bright cyan markers) for each SB1 stripped star. Residuals are shown below each $M_1$-$M_{2,\rm min}$ panel, and are typically $10^{-2}$ $M_\odot$ (at most $10^{-1}$ $M_\odot$). Note that in all cases $M_{2,\rm min}(M_{1,\rm evol})>M_{2,\rm min}(M_{1,\rm spec})$ since $M_{1,\rm evol} > M_{1,\rm spec}.$}
\label{fig:m2-approx}
\end{figure*}

Although we took considerable care in selecting the helium star mass $M_1$ which we assume when deriving companion mass constraints for our binary helium stars (see \secref{ssec:m2} and \tabref{tab:orbit_props}), we acknowledge that the helium star mass is uncertain. Therefore, in \tabref{tab:detailed_m2} we present minimum masses $M_{2,\rm min}$ and inclination-marginalized masses $\left<M_2\right>_{\rm inc}$ assuming both the spectroscopic $M_{1,\rm spec}$ and evolutionary $M_{1,\rm evol}$ stripped star mass from \citetalias{2023ApJ...959..125G} (see \secref{ssec:m2} for definitions). All values and uncertainties quoted in \tabref{tab:detailed_m2} are analogous to those detailed in \secref{ssec:m2}. Even more agnostically, we present an analytic approximation for $M_{2,\rm min}(f,M_1)$:
\begin{equation}\label{eq:M2_approx}
    M_{2,\rm min}(f,M_1) = 
    \left(\frac{f}{M_\odot}\right)^{1/3}
    \left(\frac{M_1}{M_\odot}\right)^{2/3}
    + \frac{2}{3}\left(\frac{f}{M_\odot}\right)^{2/3}
    \left(\frac{M_1}{M_\odot}\right)^{1/3}
    + \frac{1}{3}\left(\frac{f}{M_\odot}\right),
\end{equation}
where $f$ is the binary mass function (computed from the orbital parameters, \eqref{eq:bmf}) and $M_1$ is an arbitrary stripped star mass. \eqref{eq:M2_approx} is a third-order truncated series solution for the mass function (\eqref{eq:bmf}). For ease of use, in \tabref{tab:detailed_m2}, we compute the $f$ dependent coefficients and present explicit formulae for each star as a function of $x=M_1/M_\odot.$ In \figref{fig:m2-approx}, we compare our analytic formulae to numerical solving for $M_{2,\rm min}$ from $M_1=1$-10 $M_\odot$, while also comparing to $M_{2,\rm min}(M_{1,\rm spec})$ and $M_{2,\rm min}(M_{1, \rm evol}).$ Analytic formulae errors relative to numerical $M_{2,\rm min},$ $M_{2,\rm min}(M_{1, \rm spec})$ and $M_{2,\rm min}(M_{1, \rm evol})$ are typically $\sim10^{-2}\,M_\odot$.

\begin{table}[t!]
\centering
\footnotesize
\caption{Overview of companion mass constraints.}
\label{tab:detailed_m2}
\renewcommand{\arraystretch}{1}
\begin{tabularx}{0.91\textwidth}{lccccccc}
\toprule
\midrule
Star & $M_{1,\rm spec}$ & $M_{2,\rm min}(M_{1,\rm spec})$ & $\left<M_2(M_{1,\rm spec})\right>_{\rm inc}$ & $M_{1,\rm evol}$ & $M_{2,\rm min}(M_{1,\rm evol})$ & $\left<M_2(M_{1,\rm evol})\right>_{\rm inc}$ & $M_{2,\rm min}(x),\,x=M_1/M_\odot$ \\
& [$M_\odot$] & [$M_\odot$] & [$M_\odot$] & [$M_\odot$] & [$M_\odot$] & [$M_\odot$] & [$M_\odot$] \\
\midrule
\midrule
Star 1 & 6.97$_{-1.70}^{+1.84}$ & 2.54$_{-0.16}^{+0.18}\left(_{-0.25}^{+0.30}\right)$ & 2.8$_{-0.4}^{+0.9}\left(_{-0.6}^{+1.1}\right)$ & 8.45$_{-0.58}^{+1.04}$ & 2.86$_{-0.17}^{+0.18}\left(_{-0.17}^{+0.25}\right)$ & 3.1$_{-0.6}^{+1.2}\left(_{-0.6}^{+1.2}\right)$ & 
$0.570 x^{2/3} + 0.217 x^{1/3} + 0.062$ \\
Star 2 & 3.18$_{-2.00}^{+6.96}$ & 0.66$_{-0.03}^{+0.03}\left(_{-0.06}^{+0.55}\right)$ & 0.7$_{-0.0}^{+0.3}\left(_{-0.1}^{+0.8}\right)$ & 3.31$_{-0.18}^{+1.82}$ & 0.67$_{-0.03}^{+0.03}\left(_{-0.03}^{+0.17}\right)$ & 0.7$_{-0.1}^{+0.3}\left(_{-0.1}^{+0.5}\right)$ & 
$0.269 x^{2/3} + 0.048 x^{1/3} + 0.006$ \\
Star 4 & 2.43$_{-1.59}^{+11.17}$ & 0.56$_{-0.03}^{+0.03}\left(_{-0.03}^{+0.91}\right)$ & 0.6$_{-0.0}^{+0.2}\left(_{-0.0}^{+1.2}\right)$ & 3.04$_{-0.59}^{+4.43}$ & 0.65$_{-0.03}^{+0.03}\left(_{-0.03}^{+0.40}\right)$ & 0.7$_{-0.0}^{+0.3}\left(_{-0.0}^{+0.6}\right)$ & 
$0.274 x^{2/3} + 0.050 x^{1/3} + 0.007$ \\
\midrule
Star 6 & 2.34$_{-1.04}^{+3.11}$ & 0.58$_{-0.05}^{+0.09}\left(_{-0.05}^{+0.34}\right)$ & 0.6$_{-0.1}^{+0.3}\left(_{-0.2}^{+0.5}\right)$ & 3.74$_{-0.94}^{+0.91}$ & 0.79$_{-0.08}^{+0.10}\left(_{-0.11}^{+0.12}\right)$ & 0.8$_{-0.1}^{+0.4}\left(_{-0.2}^{+0.4}\right)$ & 
$0.296 x^{2/3} + 0.058 x^{1/3} + 0.009$ \\
Star 16 & 0.76$_{-0.35}^{+1.73}$ & 0.58$_{-0.04}^{+0.04}\left(_{-0.04}^{+0.43}\right)$ & 0.6$_{-0.1}^{+0.3}\left(_{-0.1}^{+0.7}\right)$ & 1.63$_{-0.15}^{+0.21}$ & 0.88$_{-0.05}^{+0.05}\left(_{-0.06}^{+0.07}\right)$ & 0.9$_{-0.1}^{+0.4}\left(_{-0.1}^{+0.5}\right)$ & 
$0.475 x^{2/3} + 0.151 x^{1/3} + 0.036$ \\
\bottomrule
\end{tabularx}
\end{table}

\subsection{Orbital Property Rejection and Posterior Distributions}
\label{sapp:corner_plots}

\begin{figure*}
\centering
\includegraphics[width=\textwidth]{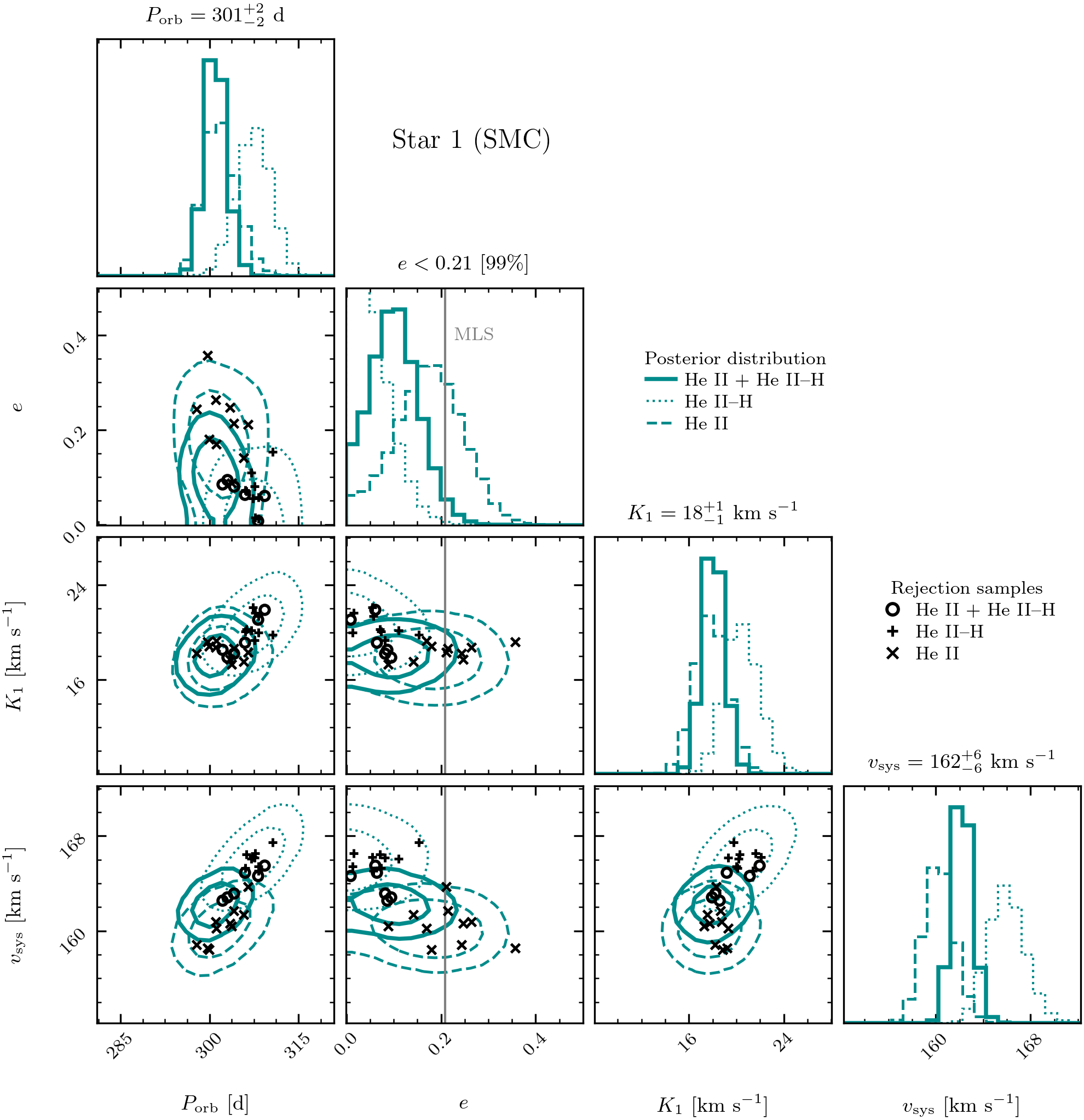}
\caption{Corner plot visualizing the orbital properties we measure for Star 1, after applying \texttt{The Joker} to our RVs. Rejection samples are shown as black points, whereas the posterior distributions are dark cyan contours (1$\sigma$ and 2$\sigma$). Black circles and solid contours correspond to the posterior distributions from which we estimate our orbital properties: \HeII~and \HeII-H lines. Additionally, we re-run \texttt{The Joker} with RVs computed from only \HeII-H (+ and dotted) and \HeII~(x and dashed). Regardless of the line combination used for RV measurements, the resulting posterior distributions are consistent within error. We implement the modified Lucy-Sweeney test \citep[MLS:][]{2013A&A...551A..47L}, and find that the orbit is consistent with circular, while we are 99\% confident that $e<0.21.$}
\label{fig:star1-corner}
\end{figure*}

\begin{figure*}
\centering
\includegraphics[width=\textwidth]{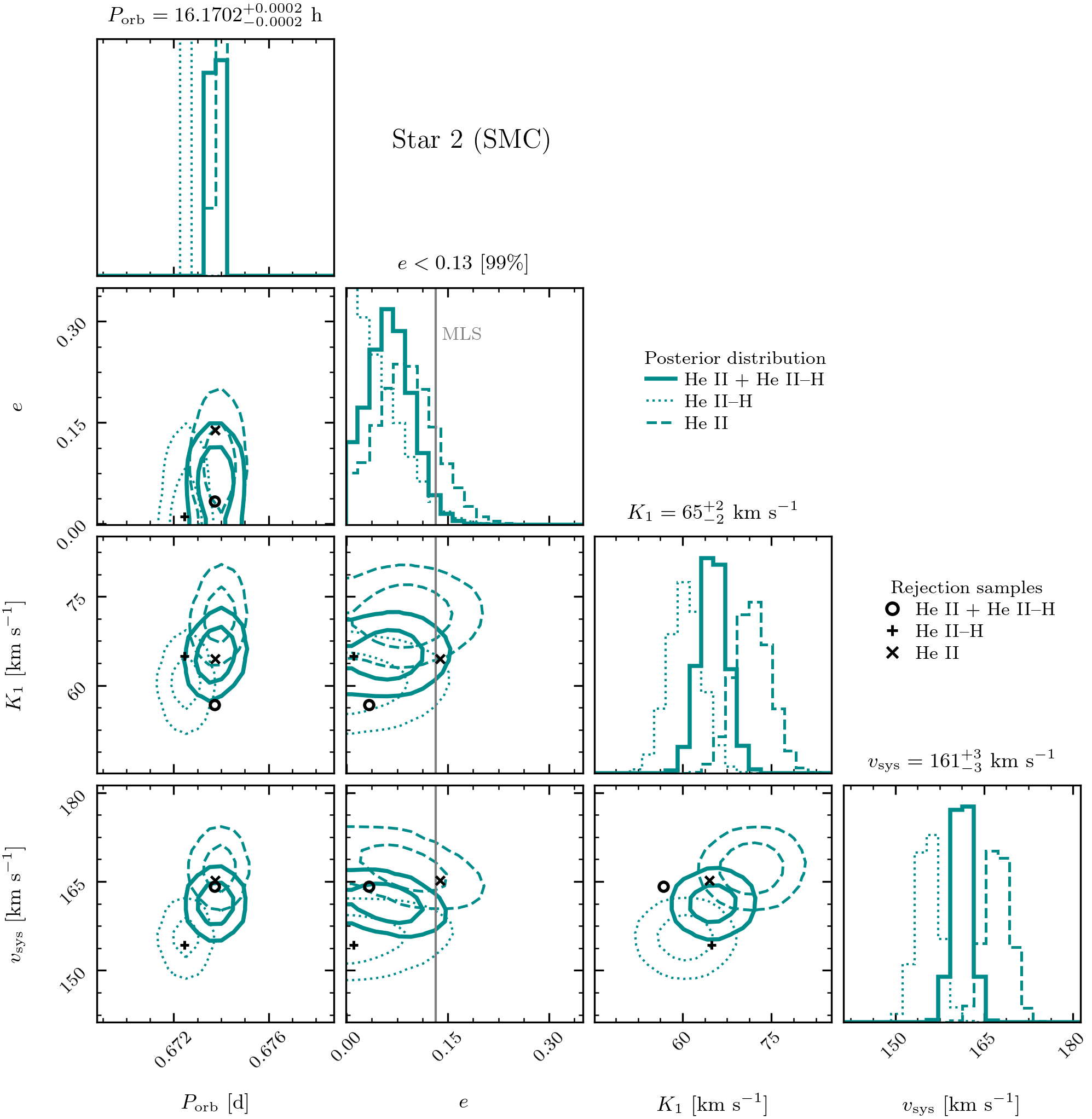}
\caption{Similar to \figref{fig:star1-corner}, for Star 2. We find that $e<0.13$ with 99\% confidence.}
\label{fig:star2-corner}
\end{figure*}

\begin{figure*}
\centering
\includegraphics[width=\textwidth]{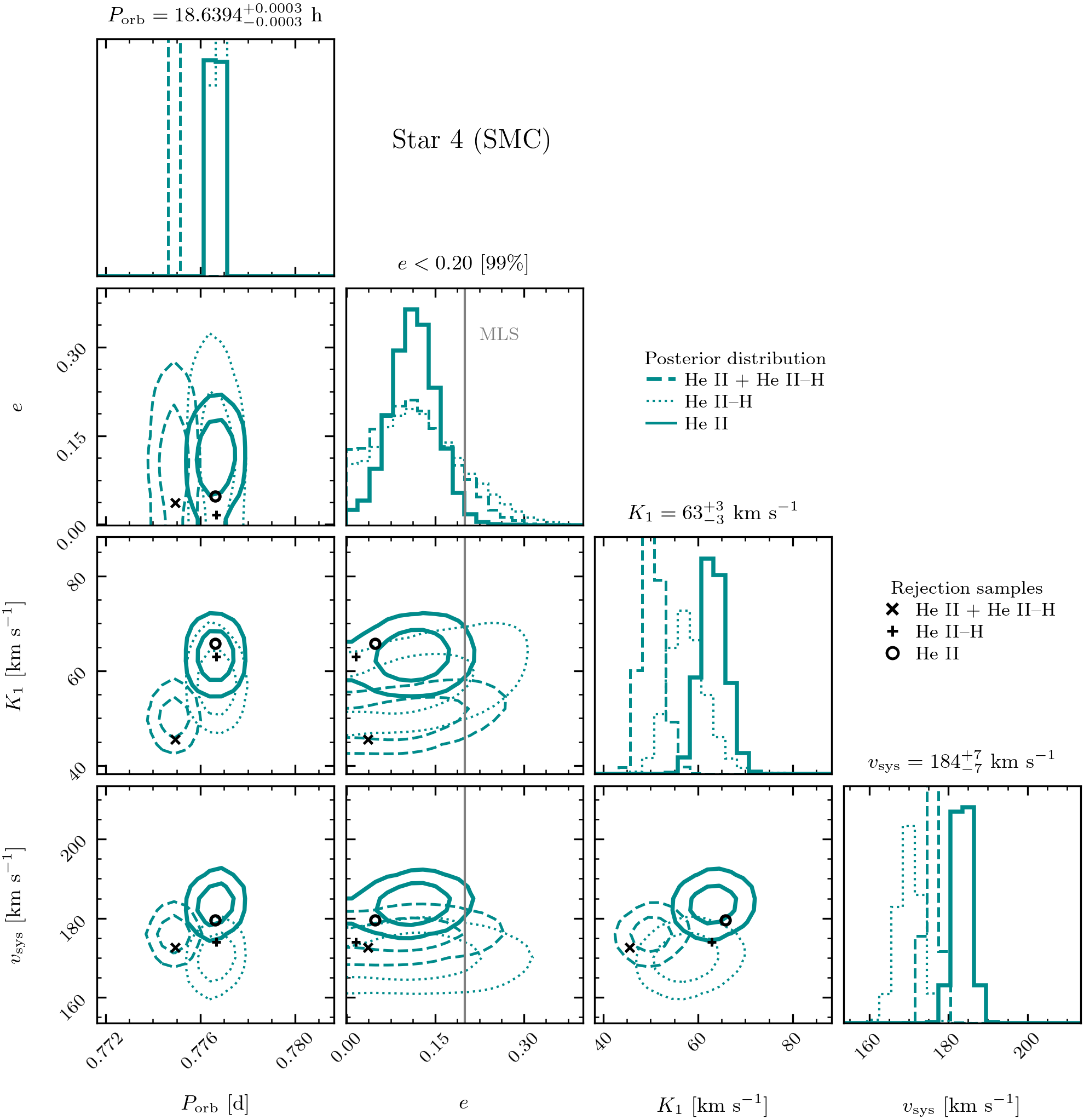}
\caption{Similar to \figref{fig:star1-corner}, for Star 4. We find that $e<0.20$ with 99\% confidence.}
\label{fig:star4-corner}
\end{figure*}

\begin{figure*}
\centering
\includegraphics[width=\textwidth]{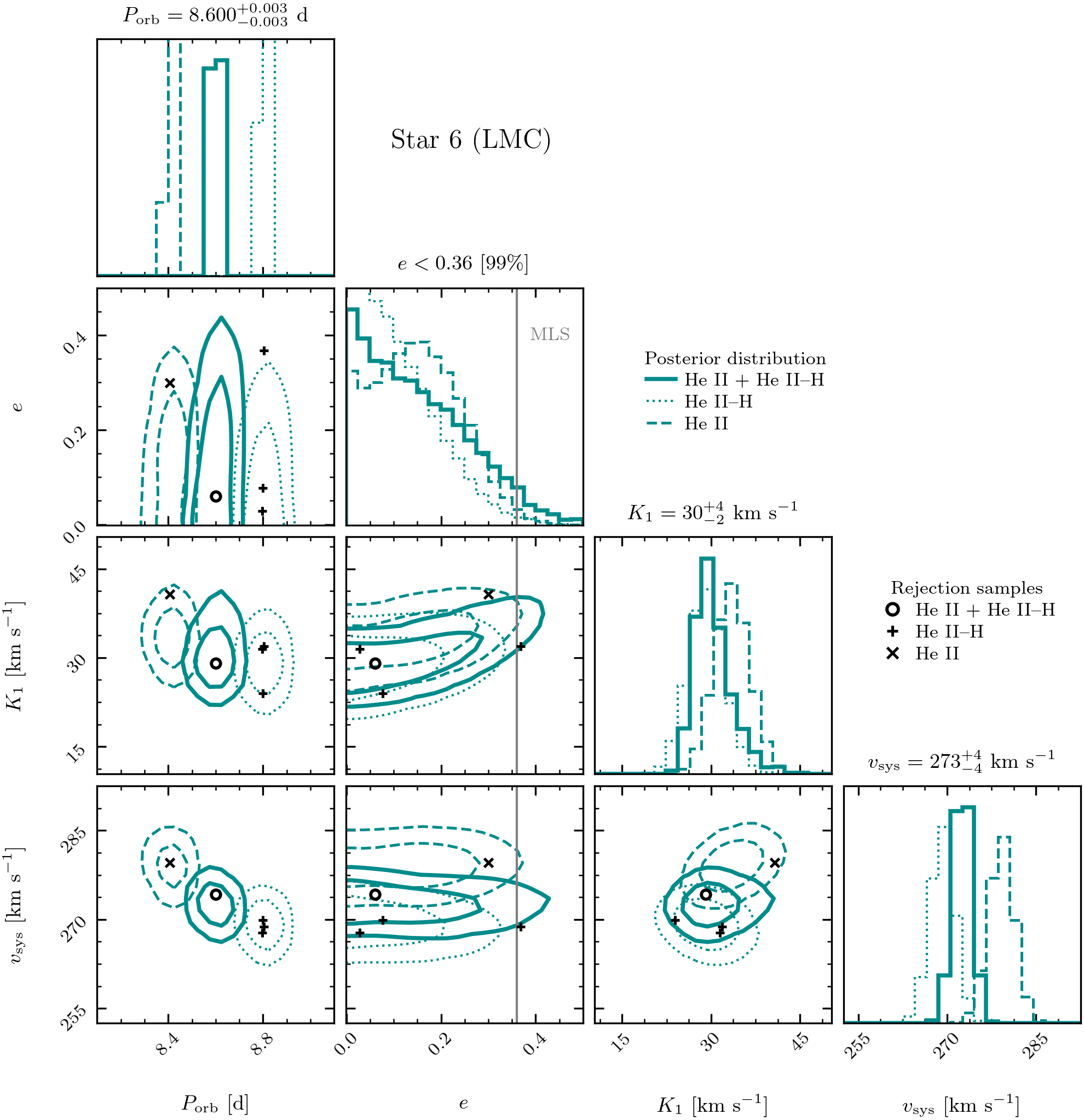}
\caption{Similar to \figref{fig:star1-corner}, for Star 6. We find that $e<0.36$ with 99\% confidence.}
\label{fig:star6-corner}
\end{figure*}

\begin{figure*}
\centering
\includegraphics[width=\textwidth]{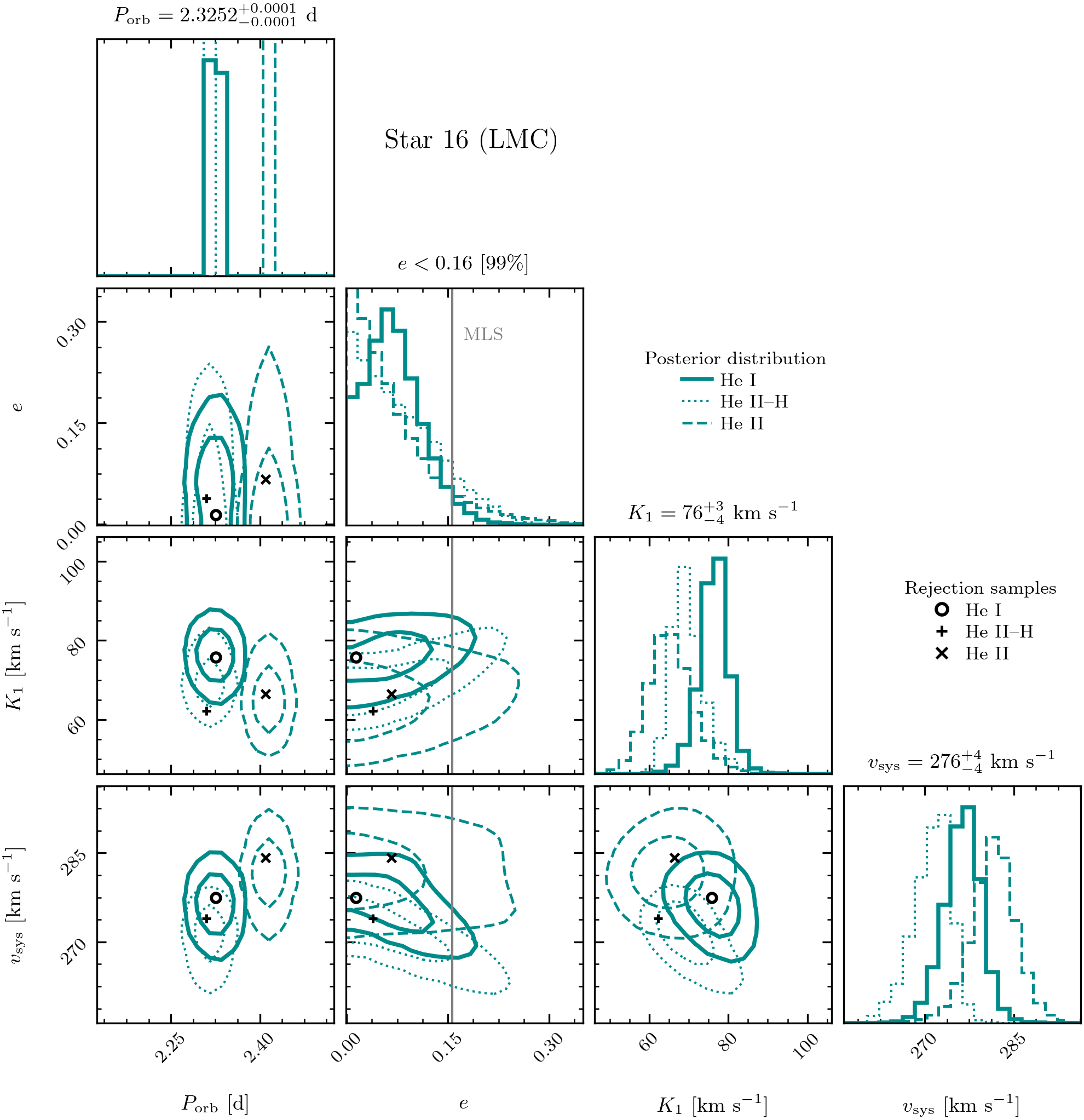}
\caption{Similar to \figref{fig:star1-corner}, for Star 16. We find that $e<0.16$ with 99\% confidence.}
\label{fig:star16-corner}
\end{figure*}

In Figs. \ref{fig:star1-corner}, \ref{fig:star2-corner},\ref{fig:star4-corner}, \ref{fig:star6-corner}, and \ref{fig:star16-corner}, we present the orbital property rejection and posterior distributions resulting from applying \texttt{The Joker} (see \secref{sec:orbits}) to our RVs (see \secref{sec:rv}) for Stars 1, 2, 4, 6, and 16, respectively. For each star, not only do we present the rejection and posterior distributions resulting from the RVs we use to estimate the binary properties we publish in this work, but we also present the rejection and posterior distributions for subsets of spectral lines. We note that very few rejection samples survive, which demonstrates that the orbits are unimodal and sharply constrained. Additionally, we varied the number of rejection samples and found that the rejection sample distribution remains unchanged. This is in contrast to the non-detections, whose rejection samples are numerous and severely multi-modal (see below).

In all cases, we use the RV measurements from the spectral lines specified in \tabref{tab:rv_summary} to estimate an orbit. However, as a sanity check we also implement \texttt{The Joker} for subsets of these spectral lines to convince ourselves that we are not suffering from invisible spectral line contamination from an optically faint companion. In particular, while we are confident \HeII\,lines are originating from the stripped star in the binary, it is conceivable that a secondary could contribute to \HeII-H blends and in so biasing the RVs. However, if all line sets produce the same orbital solution, we conclude that unseen flux contribution is not a concern for our purposes. In all cases, regardless of the spectral line combination used for RV measurements, the derived orbital posterior distributions are consistent within error. This is additional evidence that our sample contains genuine SB1s with negligible, if any, optical flux contribution from a companion. Lastly, we observe that $K_1$ and $v_{\rm sys}$ can vary somewhat depending on the line combination used to measure the RVs. First, varying $K_1$ by the mean shift in distributions for different line combinations changes our inferred $M_{2,\rm min}$ by at most $\sim$0.1 $M_\odot$. Second, differences in $v_{\rm sys}$ largely arise from relative RVs being converted to absolute RVs, which are nevertheless consistent within $\sim$10 km s$^{-1}$ (see \ref{ssec:orb_params}).

\begin{figure*}
\centering
\includegraphics[width=\textwidth]{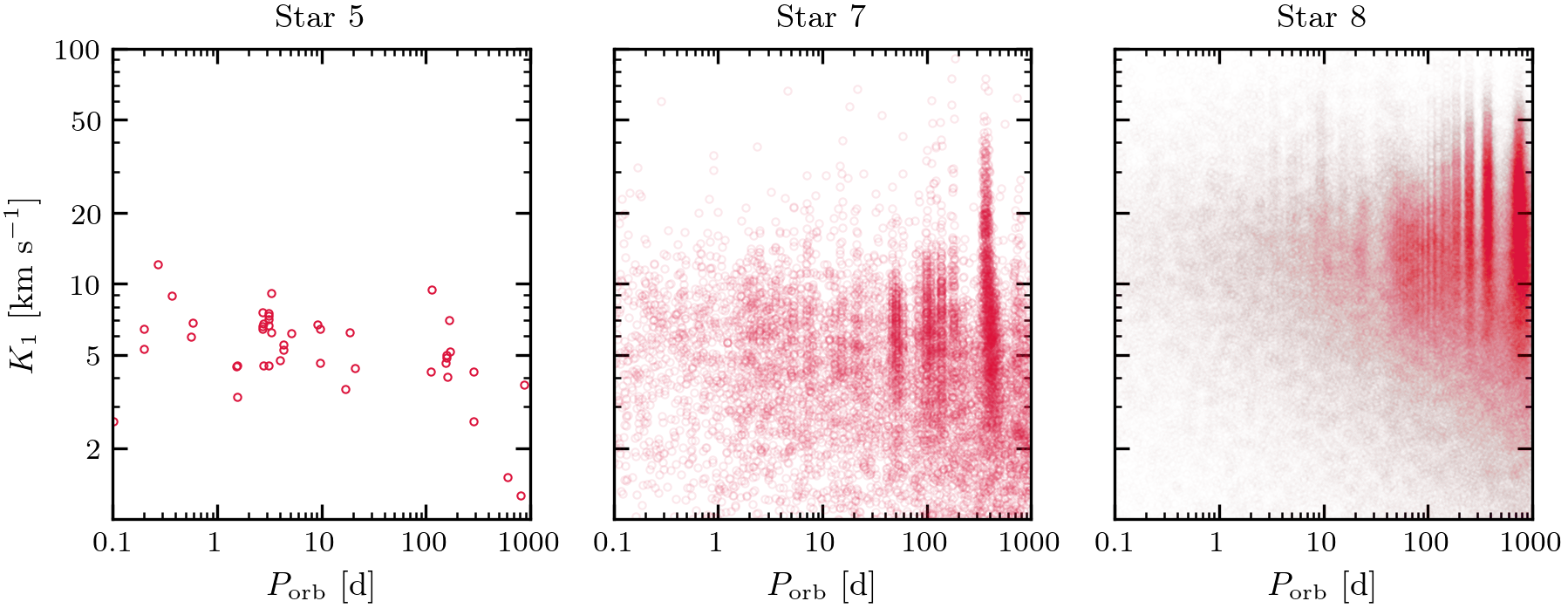}
\caption{Rejection samples in $P_{\rm orb}$-$K_1$ space for the non-detections: Star 5, 7 and 8. In contrast to the strongly peaked, unimodal rejection sample distributions of Stars 1, 2, 4, 6 and 16 (see Figs. \ref{fig:star1-corner}--\ref{fig:star16-corner}), their $P_{\rm orb}$ and $K_1$ are entirely unconstrained. Thus, no orbital solutions are preferred, supporting our initial finding that the non-detections do not exhibit binary motion (see \secref{sec:binary_motion}).}
\label{fig:star578-reject}
\end{figure*}

Additionally, in \figref{fig:star578-reject}, for the helium stars where we do not detect binary motion (Stars 5, 7, and 8), we present the rejection sampling distributions in $P_{\rm orb}$-$K_1$ space. For the non-detections, the orbital parameter space is completely unconstrained (strongly degenerate). We conclude that there is no evidence for a robust orbital solution in any of the non-detections. 

\section{Details for Observed Helium Star Binaries}
\label{app:obs} 
\restartappendixnumbering

\subsection{Observations of helium star binaries}\label{sapp:he_obs}

We summarize our compilation of 142 helium star binaries in the literature, which we separate into subdwarfs, Wolf-Rayets, intermediate-mass helium stars and puffed-up helium stars in \appref{ssapp:sd_obs}, \ref{ssapp:wr_obs}, \ref{ssapp:int_obs} and \ref{ssapp:puff_obs}), respectively. 

\subsubsection{Hot Subdwarfs}\label{ssapp:sd_obs}

On the low mass-end, we identified 16 sdOB+Be binaries with measured binary properties \citep{2008ApJ...686.1280P,
2013ApJ...765....2P,
2015A&A...577A..51M,
2018ApJ...865...76C,
2022ApJ...940...86K,
2023AJ....165..203W,
2024ApJ...962...70K,
2025A&A...694A.208K}, which \cite{2025arXiv250514780L} demonstrated likely underwent conservative mass transfer. 58 confirmed or strong candidate sdB+WD binaries were compiled by \cite{2022A&A...666A.182S}, who derived $M_{2,\rm min}$ assuming a canonical subdwarf mass for core-helium burning $M_1=0.47\,M_\odot.$ \cite{2023A&A...673A..90S} measured $M_2$ for 19 short period sdB+MS binaries ($<$1 day, assuming $M_1=0.47\,M_\odot$), while \cite{2017A&A...605A.109V} measured and/or tabulated binary properties for 11 long period sdB+MS binaries \citep[$>$500 days:][]{2012MNRAS.421.2798D,
2012A&A...548A...6V,
2013ApJ...771...23B,
2013A&A...559A..54V}. \cite{2026arXiv260827673D} presented a 500 pc volume-limited sample of short-period sdO/B+MS and sdO/B+WD binaries. They derived miminum companion masses assuming the SED-fit subdwarf masses of \cite{2026A&A...707A...6D}, except for eight systems where periods and masses were derived from light curve fitting. When the samples of \cite{2022A&A...666A.182S,2023A&A...673A..90S} and \cite{2026arXiv260827673D} overlap, we retain the periods
and minimum companion masses of \citet{2026arXiv260827673D} and subdwarf masses of
\citet{2026A&A...707A...6D}.
HD 49798 is a well-studied low-mass X-ray binary containing an sdO orbiting a compact object whose nature is debated \citep{2017ApJ...847...78B,2023MNRAS.523.3043R}. Finally, two double subdwarf binaries have been discovered: a double He-rich sdB binary \citep[PG 1544+488:][]{2004A&A...418..275A,
2014MNRAS.440.2676S}, and a double He-rich sdO binary \citep[HE 0301$-$3039:][]{2004Ap&SS.291..351L,
2007A&A...462..269S}. However, only PG 1544+488 has measured binary properties, with minimum binary masses since it is a double-lined spectroscopic binary (SB2). Due to the anomalously low minimum mass constraints for PG 1544+488, we exclude it from our main comparisons ($(M_1,M_2)\sin^3 i\sim(0.16,0.15)$).

\subsubsection{Wolf-Rayets}\label{ssapp:wr_obs}

On the high mass end, there are a 18 likely binary-stripped WR+O systems in the Magellanic Clouds, with measured binary properties. There are 4 in the SMC: SMC AB 3, 6, 7 and 8 \citep{1979A&A....75..243A,1980A&AS...39...19A,2003MNRAS.338..360F,2016A&A...591A..22S,2018A&A...616A.103S,2002MNRAS.333..347N, 2016A&A...591A..22S,1990ApJ...348..232M, 2005ApJ...628..953S}, and 15 in the LMC: BAT99 19, 28, 29, 38, 39, 43, 49, 59, 64, 71, 77, 92, 103, 126 and 129 \citep{1999A&AS..137..117B,2003MNRAS.338.1025F,2019A&A...627A.151S,1990ApJ...348..232M,2008MNRAS.389..806S,2006A&A...447..667F}. All aforementioned WR binaries have O-type companions, except for BAT99 29 which has an early-B type companion. Four of the LMC systems have measured spin rates which support the binary-stripping interpretation \cite{2020MNRAS.492.4430S}. Lastly, there is Cyg X-3: a Galactic X-ray binary with a WR donor and either a NS or BH companion \citep{2013MNRAS.429L.104Z,
2002A&A...392..161S}. In addition to Cyg X-3, only SMC AB 6, 7, 8 and BAT99 19, 129 have binary mass estimates, whereas the remaining WRs have minimum mass constraints. As such, we only include these WRs in our main comparisons. Finally, \cite{2017MNRAS.464.2066S} measured the spin rates of eight Galactic WR+O binaries \citep[compiling their masses, along with 3 additional systems]{1990A&A...240..105B,
1994ApJ...422..810M,
1996AJ....112.2227L,
2011MNRAS.411..635D}, and demonstrated that they were likely the result of envelope stripping. We include a subset of six of these systems whose inclinations are well constrained: WR 21, 30, 42, 113, 11 and 139.

\subsubsection{Hot, intermediate-mass helium stars}\label{ssapp:int_obs}

Two hot, compact, intermediate-mass helium stars have been observed, with measured orbital properties. First, the Galactic $q$WR HD 45166 now has precise binary properties which constrain the stripped star mass to the lower end of the intermediate-mass regime \citep[$\sim$2 $M_\odot$:][]{2025A&A...695L..20D}. Second, HD 96670 was recently suggested to contain an intermediate-mass stripped star \citep[$\sim4.5\,M_\odot$:][]{2025A&A...696A..84N}. While observed eclipses disfavor a black hole companion, HD 96670 requires further analysis to conclusively establish the stripped star interpretation, since its spectral morphology is not typical for a stripped star. As such, we only include HD 45166 in our main comparisons.

\subsubsection{Puffed-up stripped stars}
\label{ssapp:puff_obs}

We identify 11 puffed-up/bloated stripped stars with binary property estimates. The black hole impostors are: HR 6819 \citep[$Y_{\rm H, surf}\approx0.55\to X_{\rm H, surf}\approx0.45$:][]{2021MNRAS.502.3436E,2025A&A...694A.208K}, NGC 1850 BH1 \citep[$X_{\rm H, surf}$ unknown:][]{2022MNRAS.511L..24E}, and LB1 \citep[$n(\text{He})/n(\text{H})=0.21\to X_{\rm H, surf}=0.54$:][]{2020A&A...639L...6S}, while other examples include: VFTS 291 \citep[$Y_{\rm H, surf}\approx0.29\to X_{\rm H, surf}\approx0.71$:][]{2023MNRAS.525.5121V}, AzV 476 \citep[$X_{\rm H, surf}=0.73$:][]{2022A&A...659A...9P}, HIP 15429 \citep[$X_{\rm H, surf}\approx0.3$:][]{2025A&A...701A...9M}, SMCSGS-FS 69 \citep[$X_{\rm H, surf}=0.59_{-0.10}^{+0.10}$:][]{2023A&A...674L..12R}, 2dFS 163, 2dFS 2553 and Sk -71 35 \citep[$X_{\rm H, surf}=0.33_{-0.05}^{+0.10},0.60_{-0.05}^{+0.10},0.70_{-0.10}^{+0.04}$:][]{2024A&A...692A..90R}. Most recently, \cite{2026arXiv260805276M} argued that the Galactic short-period binary WR 2-1 contains a $\sim$4 $M_\odot$ puffed-up stripped star with $X_{\rm H, surf}\gtrsim0.5.$

The surface properties of our core-helium burning stripped stars appear to be different from the majority of the puffed-up stripped stars. Whereas our stripped star binary sample has $X_{\rm H, surf}\lesssim0.4,$ most of the puffed-up stripped stars have higher measured $X_{\rm H, surf}$. In order to make an informative, fair comparison  we only include puffed-up stripped stars with measured $X_{\rm H, surf}<0.5$ in our main comparisons: HR 6819, 2dFS 163 and HIP 15429. However, we include the entire puffed-up stripped star sample displayed in \figref{fig:M1_M2_He_all}. Finally, there is some ambiguity between partially stripped stars \citep{2022A&A...662A..56K} and puffed-up stripped stars \citep{2024A&A...687A.215D} in the literature, which \cite{2025A&A...693A..10S} attempt to clarify.

\subsection{Additional observations of helium star binaries}
\label{sapp:obs-all}

\begin{figure*}[ht!]
    \centering
    \includegraphics[width=.45\textwidth]{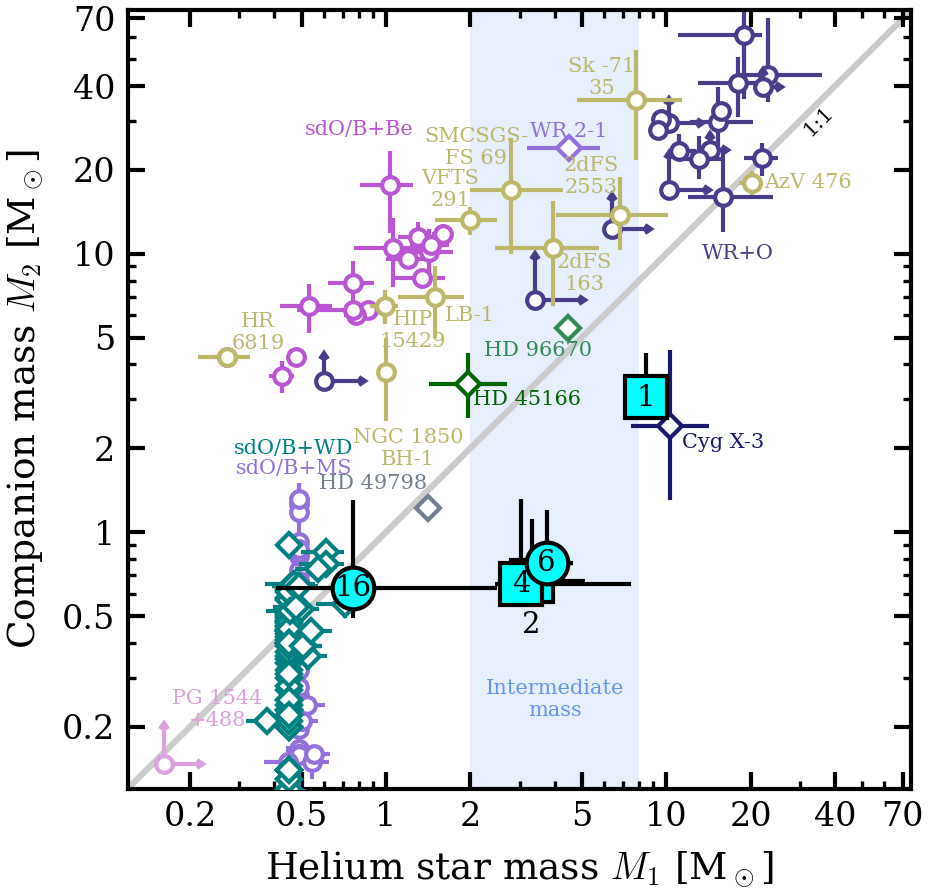}
    \includegraphics[width=.47\textwidth]{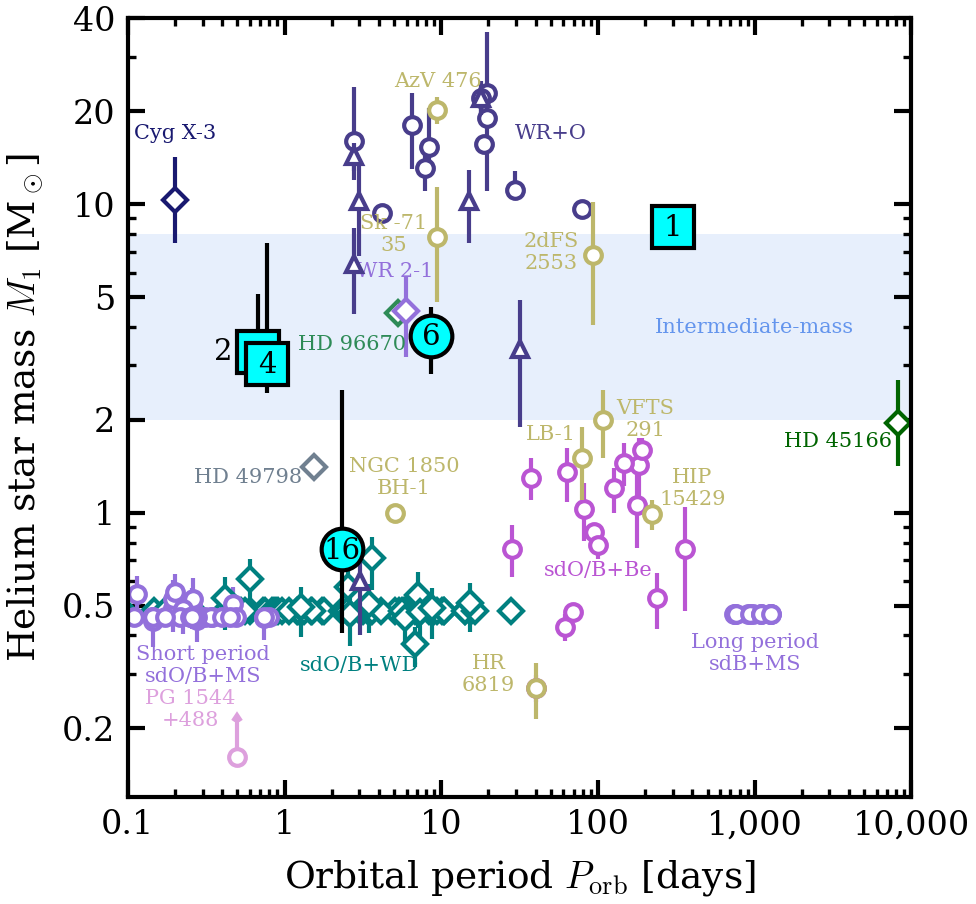}
    \caption{Analogous to \figref{fig:M1_M2_He} (left) and \figref{fig:M1_Porb_He} (right) while also including: PG 1544+488, WR+O binaries with minimum masses, NGC 1850 BH1, LB-1, VFTS 291, SMCSGS-FS 69, 2dFS 2553, Sk -71 35, AzV 476, HD 96670 and WR 2-1.}
    \label{fig:M1_M2_He_all}
\end{figure*}

When comparing our binary stripped stars to observed samples of helium star binaries, we omitted some observed systems, justified above. In \figref{fig:M1_M2_He_all}, we reproduce these comparisons, while including the observed systems we excluded: $M_1$-$M_2$ and $P_{\rm orb}$-$M_1$ for helium star binaries.

\section{Details for Stellar Companions}
\label{app:stellar-comp}
\restartappendixnumbering

Here, we present additional results for our constraints on stellar companions (see \secref{sec:stellar-comp}). When discussing the stellar companion constraints which result from optical fluxes, in \secref{sssec:opt-flux}, and radii relative to their Roche lobes, in \secref{sssec:roche-lobe}, we included exemplary constraints for Star 2. Here, in \secref{sapp:opt-flux}, we present optical flux constraints for Stars 1, 4, 6 and 16. Then, in \secref{sapp:roche-lobe}, we present Roche lobe constraints for the same stars. 

\subsection{Optical flux constraints}
\label{sapp:opt-flux}

\begin{figure*}[ht!]
    \centering
    \includegraphics[width=.4\textwidth]{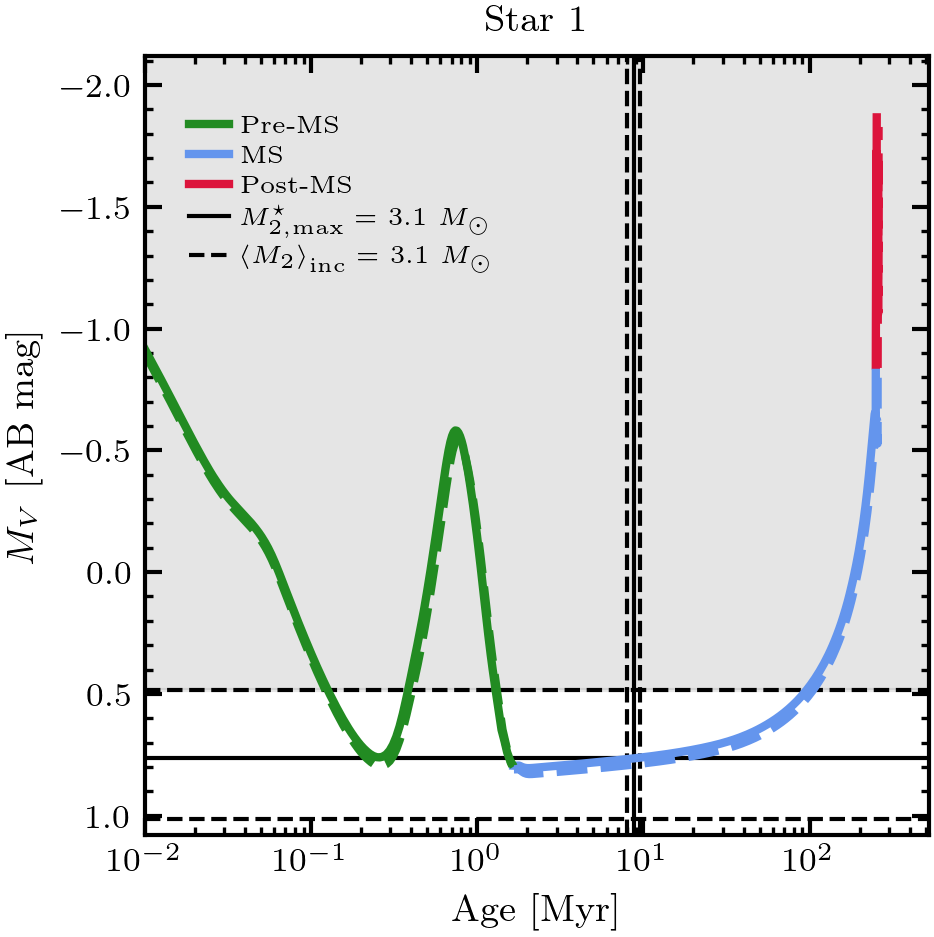}
    \hspace{2em}
    \includegraphics[width=.39\textwidth]{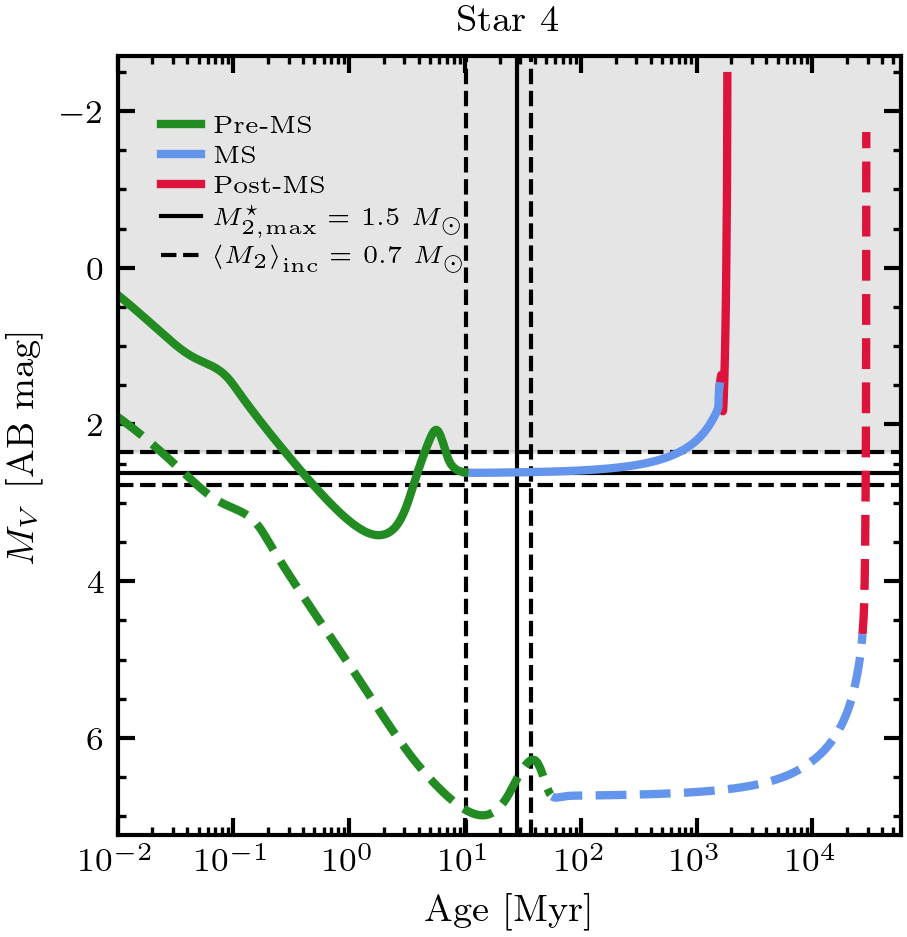}\\
    \vspace{2em}
    \includegraphics[width=.4\textwidth]{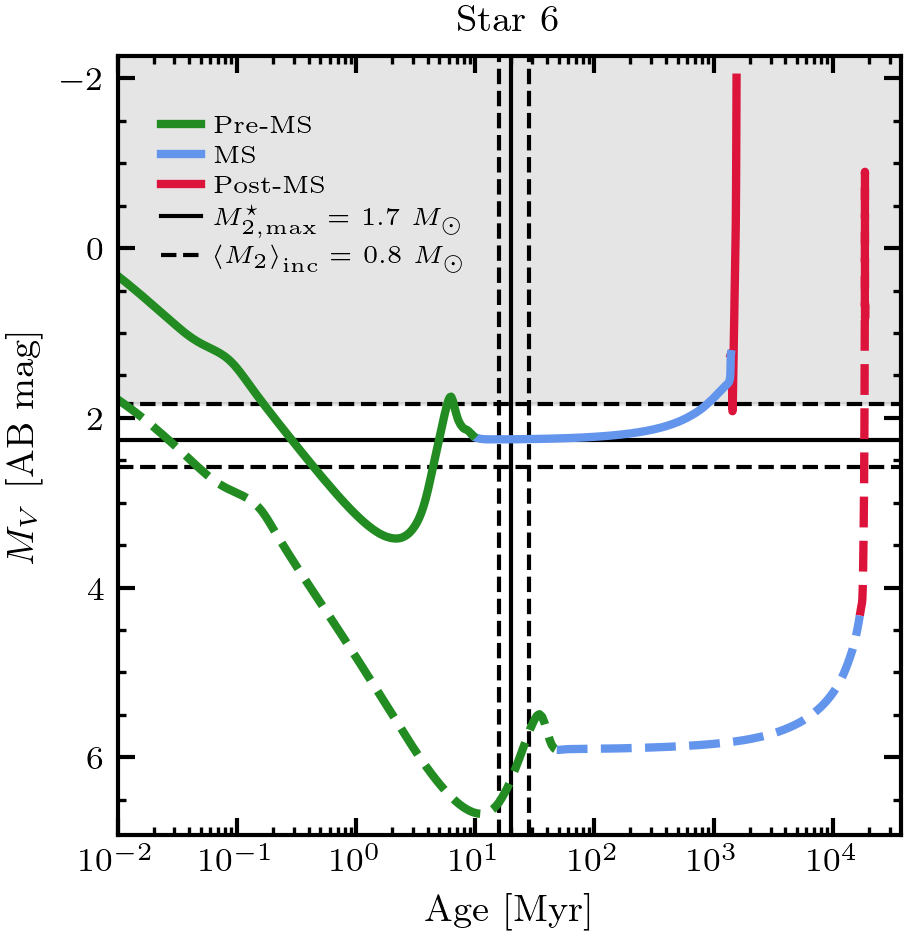}
    \hspace{2em}
    \includegraphics[width=.4\textwidth]{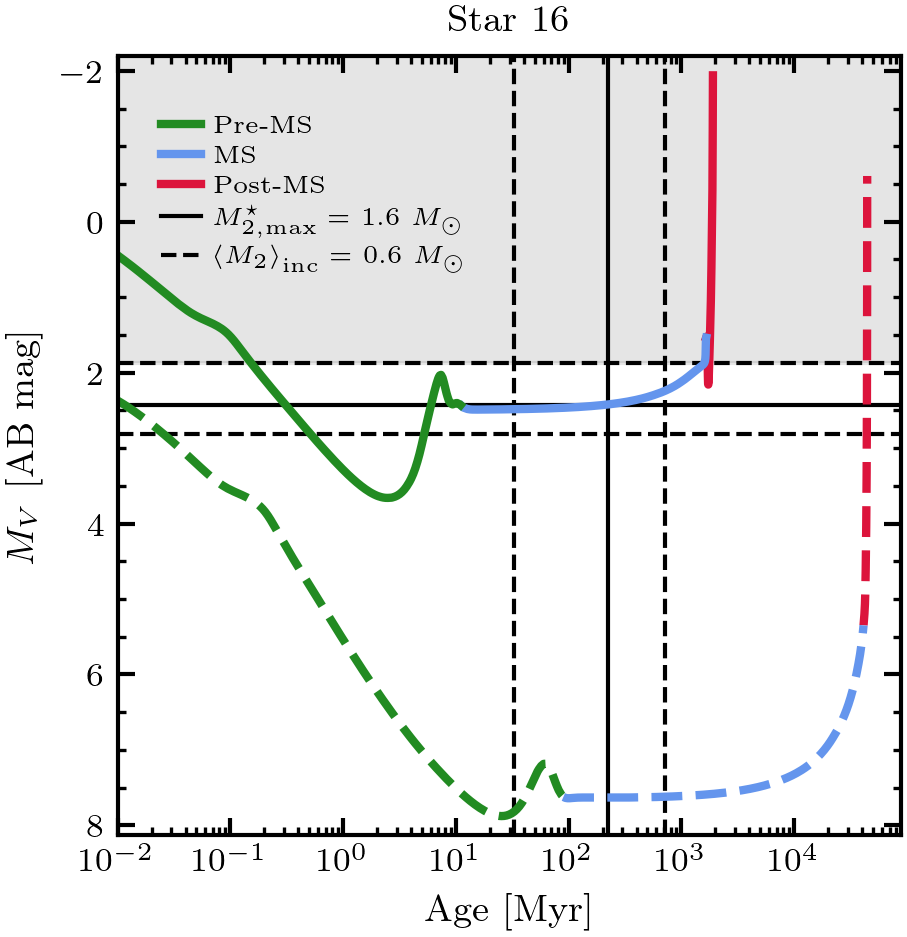}
    \caption{Optical flux contribution of a stellar companion for Star 1 (top left), Star 4 (top right), Star 6 (bottom left) and Star 16 (bottom right). System ages (vertical) and 10\% V-band flux contributions (horizontal) are shown (solid lines), with uncertainties (dashed lines). A stellar companion should not be more massive than $M_{\rm 2,max}^\star$, otherwise we would see it in the optical spectrum. In all cases, a stellar companion of mass $\left<M_{2}\right>_{\rm inc}$ is optically hidden, while young pre-MS and post-MS stars are too bright. The post-MS extends to the end of the red giant branch.}
    \label{fig:opt-flux-14616}
\end{figure*}

In \figref{fig:opt-flux-14616}, we present optical flux constraints for Stars 1, 4, 6 and 16. As described in \secref{sssec:opt-flux}, according to the analysis of \citetalias{2023ApJ...959..125G}, a stellar companion which contributes more than 10\% of the optical flux should be detectable. Thus, we can set upper limits on the mass of a stellar companion with the $V$-band photometry from \cite{2026ApJ...999...73L}, given by $M_{2,\rm max}^\star$ (see \tabref{tab:orbit_props}). Simultaneously, we can check whether the optical brightness of a stellar companion of mass $\left<M_{2}\right>_{\rm inc}$ (the inclination-averaged mass, see \secref{sssec:M2inc}) is consistent with our 10\% optical flux limit. For Star 1, $M_{2,\rm min}=2.9\,M_\odot,$ whereas $M_{2,\rm max}^\star=3.1\,M_\odot.$ There is a narrow 0.2 $M_\odot$ mass range from 2.9--3.1 $M_\odot$ where a stellar companion is viable. Conversely, for Stars 4, 6 and 16, $\left<M_{2}\right>_{\rm inc}$ is well below $M_{2,\rm max}^\star$. As such, a stellar companion could be optically hidden. 

\subsection{Roche lobe constraints}
\label{sapp:roche-lobe}

\begin{figure*}[ht!]
    \centering
    \includegraphics[width=.4\textwidth]{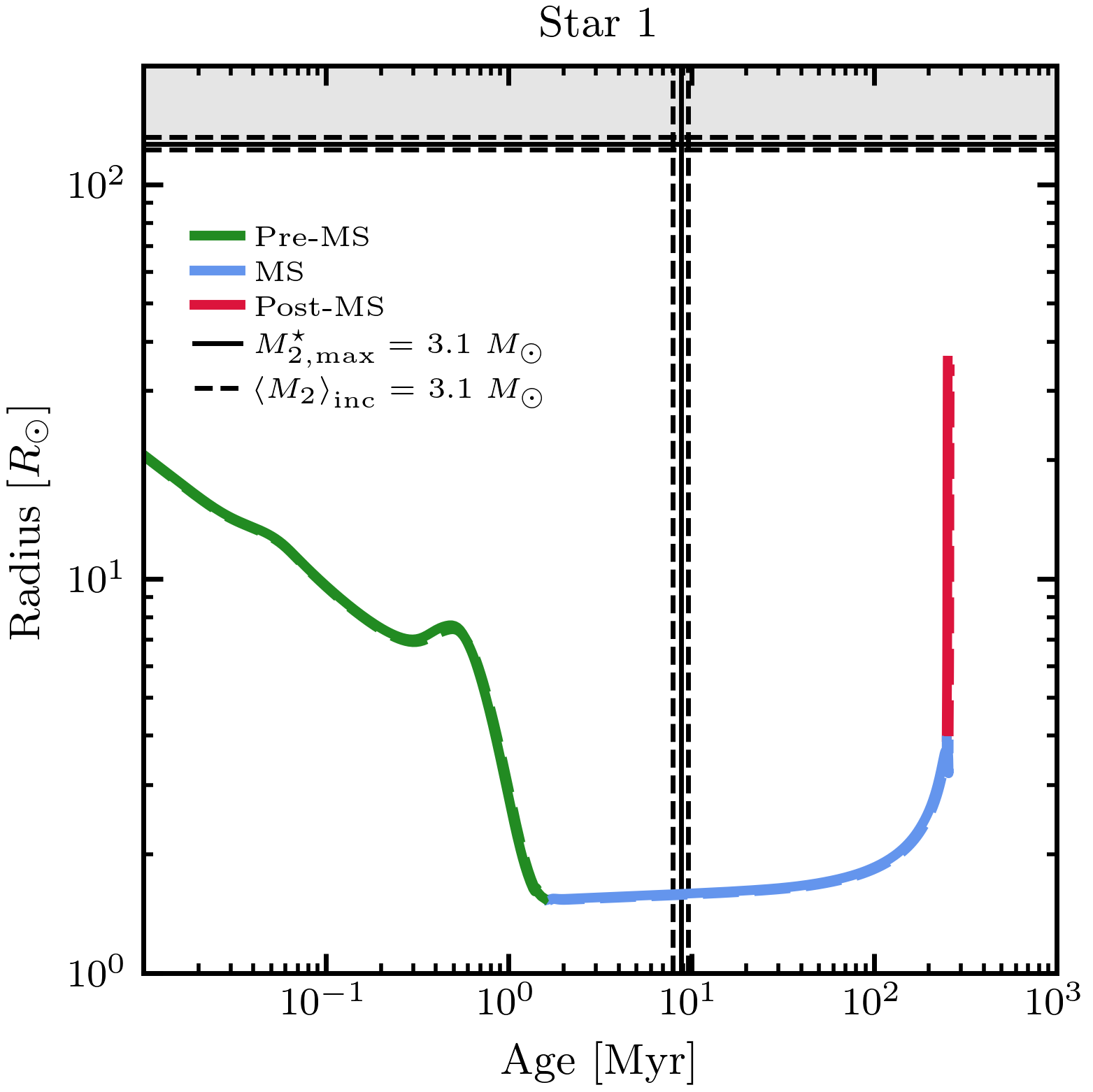}
    \hspace{2em}
    \includegraphics[width=.37\textwidth]{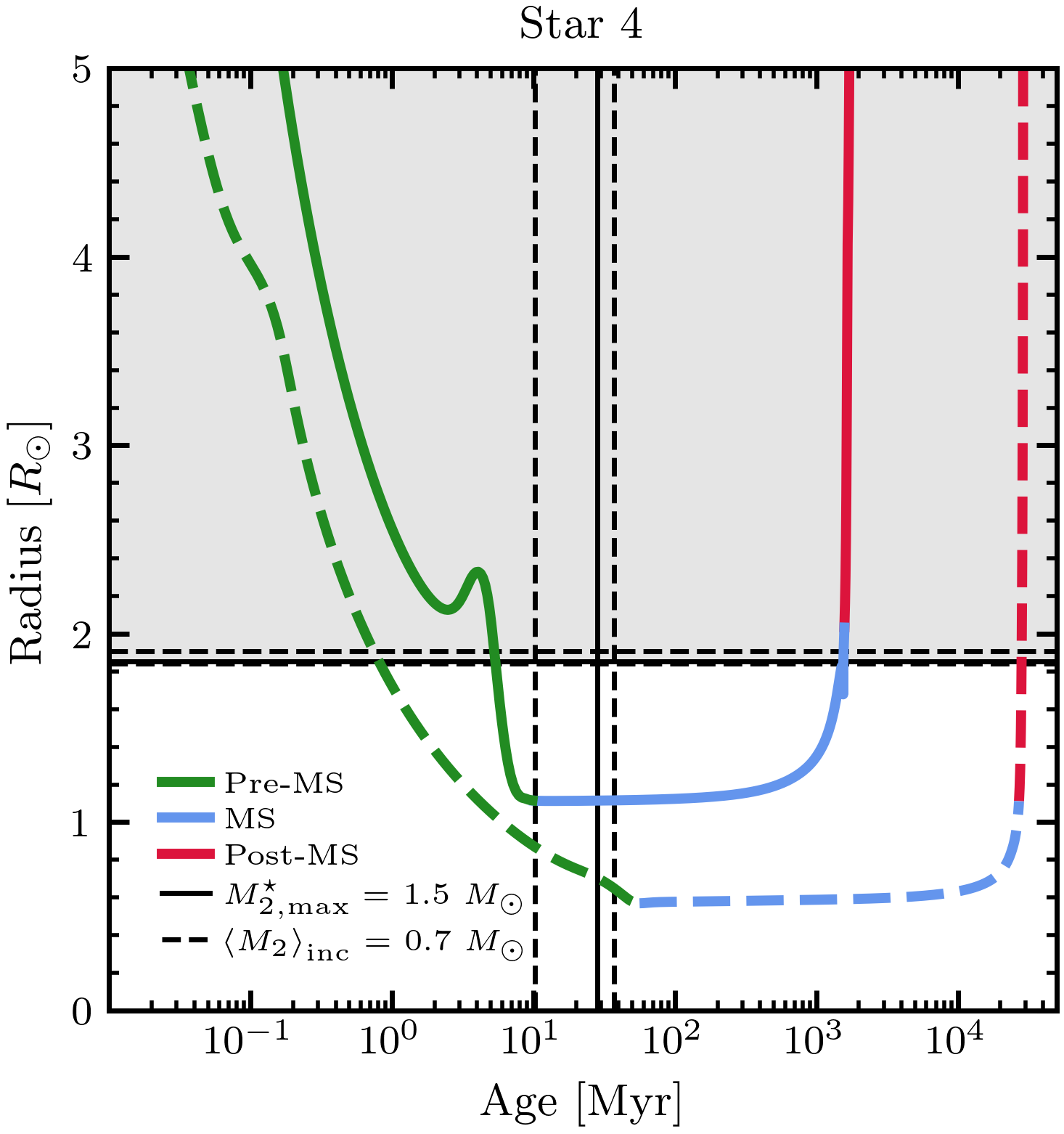}\\
    \vspace{2em}
    \includegraphics[width=.4\textwidth]{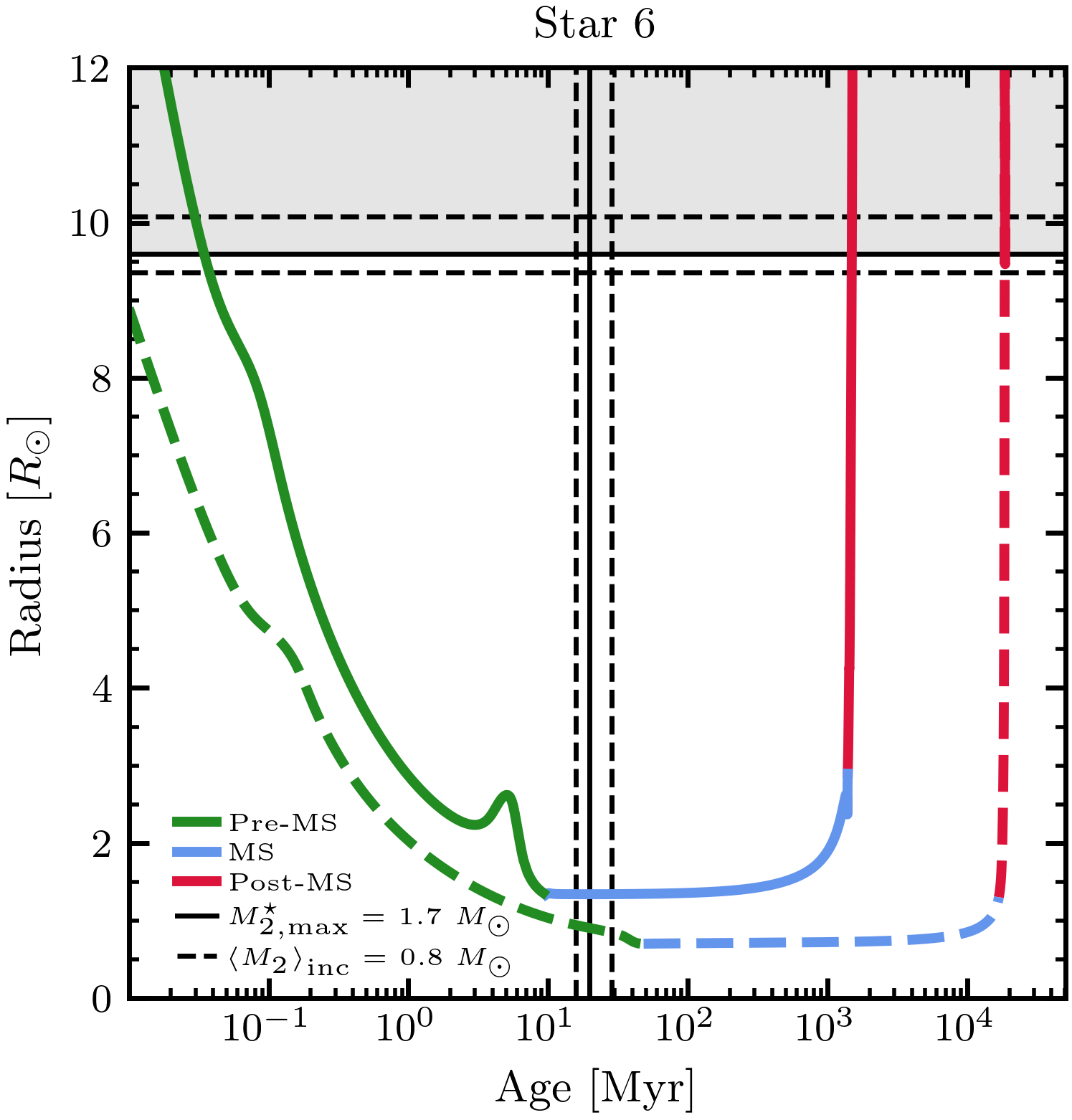}
    \hspace{2em}
    \includegraphics[width=.39\textwidth]{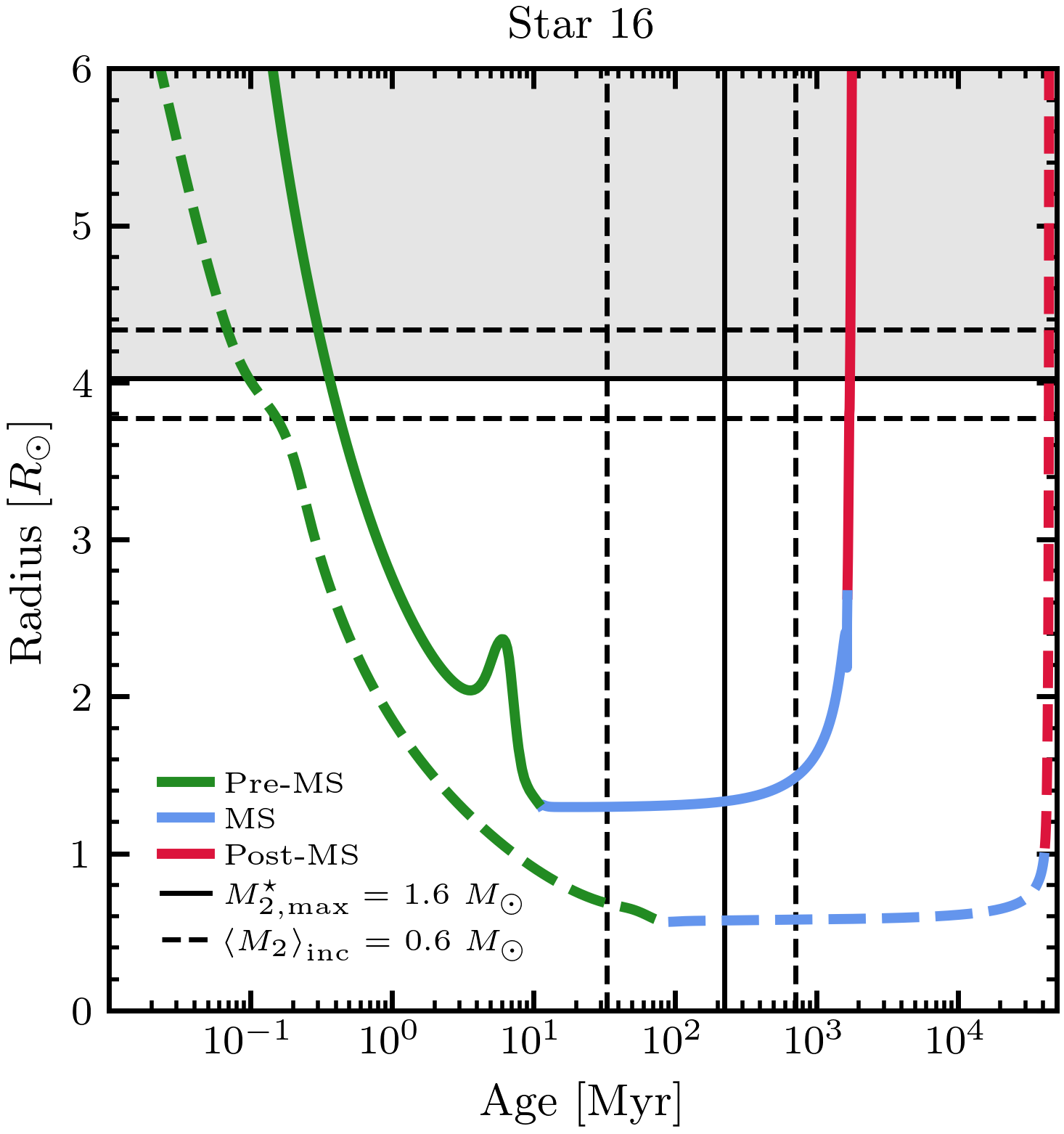}
    \caption{Radii relative to the Roche lobe of a stellar companion for Star 1 Star 1 (top left), Star 4 (top right), Star 6 (bottom left) and Star 16 (bottom right). System ages (vertical) and Roche lobe radii (horizontal) are shown (solid lines), with uncertainties (dashed lines). A stellar companion should not overflow its Roche lobe, since our systems are not actively mass transferring. In all cases, we find that a stellar companion would be contained within its Roche lobe, while young pre-MS and post-MS stars are too large (except for Star 1). The post-MS extends to the end of the red giant branch.}
    \label{fig:radius-14616}
\end{figure*}

In \figref{fig:radius-14616}, we present radius constraints for Stars 1, 4, 6 and 16. As described in \secref{sssec:roche-lobe}, a stellar companion should not overflow its Roche lobe. Thus, we can check whether the radius of a stellar companion is consistent with our companion mass constraints. In all cases, we find that a stellar companion would be contained within its Roche lobe.  For Stars 4, 6 and 16, young pre-MS and post-MS companions would overflow their Roche lobes and are thus disfavored. The wide period of Star 1 is such that any stellar companions, regardless of evolutionary stage, would be contained within its Roche lobe.

\end{document}